\documentclass[useAMS,usenatbib,breaklinks=true]{mnras}
\usepackage{graphicx}
\usepackage{times}
\usepackage{mn2e-breakabs}
\usepackage{amssymb}
\usepackage{amsmath}
\usepackage{hyperref}
\hypersetup{
    colorlinks=true,
    linkcolor=blue,
    citecolor=blue,
}

\def\be{\begin{equation}}
\def\ee{\end{equation}}
\def\ba{\begin{eqnarray}}
\def\ea{\end{eqnarray}}

\def\M{{\rm M}}

\begin{document}

\title[DR14 eBOSS Quasar RSD Measurements] 
{The clustering of the SDSS-IV extended Baryon Oscillation Spectroscopic Survey DR14 quasar sample: measurement of the growth rate of structure from the anisotropic correlation function between redshift 0.8 and 2.2}

\author[P. Zarrouk et al.]{\parbox{\textwidth}{
Pauline Zarrouk$^{1}$\thanks{E-mail: pauline.zarrouk@cea.fr},
Etienne Burtin$^{1}$,
H\'ector Gil-Mar\'in$^{2,3}$,
Ashley J. Ross$^{4,5}$,
Rita Tojeiro$^{6}$,
Isabelle P\^aris$^{7}$,
Kyle S. Dawson$^{8}$,
Adam D. Myers$^{8}$,
Will J. Percival$^{5}$,
Chia-Hsun Chuang$^{9,10}$,
Gong-Bo Zhao$^{11,5}$,
Julian Bautista$^{5}$,
Johan Comparat$^{12}$,
Violeta Gonz\'alez-P\'erez$^{5}$,
Salman Habib$^{13}$,
Katrin Heitmann$^{13}$,
Jiamin Hou$^{12}$,
Pierre Laurent$^{1}$,
Jean-Marc Le Goff$^{1}$,
Francisco Prada$^{14,15,16}$,
Sergio A. Rodr\'iguez-Torres$^{14,15,17}$,
Graziano Rossi$^{18}$,
Rossana Ruggeri$^{5}$,
Ariel G. S\'anchez$^{12}$,
Donald P. Schneider$^{19,20}$,
Jeremy L. Tinker$^{21}$,
Yuting Wang$^{11}$,
Christophe Y\`eche$^{1}$,
Falk Baumgarten$^{9,22}$,
Joel R. Brownstein$^{8}$,
Sylvain de la Torre$^{7}$,
H\'elion du Mas des Bourboux$^{8}$,
Jean-Paul Kneib$^{23}$,
Nathalie Palanque-Delabrouille$^{1}$,
John Peacock$^{24}$,
Patrick Petitjean$^{25}$,
Hee-Jong Seo$^{26}$,
Cheng Zhao$^{27}$
 } \vspace*{4pt} \\ 
\scriptsize $^{1}$ IRFU,CEA, Universit\'e Paris-Saclay, F-91191 Gif-sur-Yvette, France\vspace*{-2pt} \\  
\scriptsize $^{2}$ Sorbonne Universit\'es, Institut Lagrange de Paris (ILP), 98 bis Boulevard Arago, 75014 Paris, France\vspace*{-2pt} \\ 
\scriptsize $^{3}$ Laboratoire de Physique Nucl\'eaire et de Hautes Energies, Universit\'e Pierre et Marie Curie, 4 Place Jussieu, 75005 Paris, France\vspace*{-2pt} \\ 
\scriptsize $^{4}$ Center for Cosmology and Astro-Particle Physics, Ohio State University, Columbus, Ohio, USA\vspace*{-2pt} \\ 
\scriptsize $^{5}$ Institute of Cosmology \& Gravitation, Dennis Sciama Building, University of Portsmouth, Portsmouth, PO1 3FX, UK\vspace*{-2pt} \\
\scriptsize $^{6}$ School of Physics and Astronomy, University of St Andrews, St Andrews, KY16 9SS, UK\vspace*{-2pt} \\ 
\scriptsize $^{7}$ Aix-Marseille Universit\'e, CNRS, LAM (Laboratoire d’Astrophysique de Marseille), 38 rue F. Joliot-Curie 13388 Marseille Cedex 13, France\vspace*{-2pt} \\
\scriptsize $^{8}$ Department Physics and Astronomy, University of Utah, 115 S 1400 E, Salt Lake City, UT 84112, USA\vspace*{-2pt} \\ 
\scriptsize $^{9}$ Leibniz-Institut f\"{u}r Astrophysik Potsdam (AIP), An der Sternwarte 16, D-14482 Potsdam, Germany\vspace*{-2pt} \\ 
\scriptsize $^{10}$ Kavli Institute for Particle Astrophysics and Cosmology \& Physics Department, Stanford University, Stanford, CA 94305, USA\vspace*{-2pt} \\
\scriptsize $^{11}$ National Astronomy Observatories, Chinese Academy of Science, Beijing, 100012, P.R. China\vspace*{-2pt} \\ 
\scriptsize $^{12}$ Max-Planck-Institut f\"ur Extraterrestrische Physik, Postfach 1312, Giessenbachstr., 85748 Garching bei M\"unchen, Germany\vspace*{-2pt} \\
\scriptsize $^{13}$ HEP and MCS Divisions, Argonne National Laboratory, Lemont, IL 60439, USA\vspace*{-2pt} \\ 
\scriptsize $^{14}$ Instituto de F\'isica Te\'orica (UAM/CSIC), Universidad Aut\'onoma de Madrid, Cantoblanco, E-28049 Madrid, Spain\vspace*{-2pt} \\ 
\scriptsize $^{15}$ Campus of International Excellence UAM+CSIC, Cantoblanco, E-28049 Madrid, Spain\vspace*{-2pt} \\ 
\scriptsize $^{16}$ Instituto de Astrof\'isica de Andaluc\'ia (CSIC), E-18080 Granada, Spain\vspace*{-2pt} \\
\scriptsize $^{17}$ Departamento de F\'isica Te\'orica M8, Universidad Aut\'onoma de Madrid, E-28049 Cantoblanco, Madrid, Spain\vspace*{-2pt} \\ 
\scriptsize $^{18}$ Department of Physics and Astronomy, Sejong University, Seoul 143-747, Korea\vspace*{-2pt} \\ 
\scriptsize $^{19}$ Department of Astronomy and Astrophysics, The Pennsylvania State University, University Park, PA 16802, USA\vspace*{-2pt} \\ 
\scriptsize $^{20}$ Institute for Gravitation and the Cosmos, The Pennsylvania State University, University Park, PA 16802, USA\vspace*{-2pt} \\ 
\scriptsize $^{21}$ Center for Cosmology and Particle Physics, New York University, New York, NY 10003, USA\vspace*{-2pt} \\ 
\scriptsize $^{22}$ Humboldt-Universit\"{u}t zu Berlin, Institut f\"{u}r Physik, Newtonstrasse 15, D-12589 Berlin, Germany\vspace*{-2pt} \\
\scriptsize $^{23}$ Institute of Physics, Laboratory of Astrophysics, Ecole Polytechnique F\'ed\'erale de Lausanne (EPFL), Observatoire de Sauverny, 1290 Versoix, Switzerland\vspace*{-2pt} \\
\scriptsize $^{24}$ Institute for Astronomy, University of Edinburgh, Royal Observatory, Edinburgh EH9 3HJ, UK\vspace*{-2pt}  \\
\scriptsize $^{25}$ Institut d'Astrophysique de Paris, Universit\'e Paris 6 et CNRS, 98bis Boulevard Arago, 75014 Paris, France\vspace*{-2pt} \\
\scriptsize $^{25}$ School of Physics and Astronomy, University of St Andrews, St Andrews, KY16 9SS, UK\vspace*{-2pt} \\ 
\scriptsize $^{26}$ Department of Physics and Astronomy, Ohio University, Clippinger Labs, Athens, OH 45701\vspace*{-2pt} \\ 
\scriptsize $^{27}$ Tsinghua Center for Astrophysics and Department of Physics, Tsinghua University, Beijing 100084, China\vspace*{-2pt} \\ 
}

\date{To be submitted to MNRAS} 

\pagerange{\pageref{firstpage}--\pageref{lastpage}} \pubyear{2017}
\maketitle
\label{firstpage}

\begin{abstract}
We present the clustering measurements of quasars in configuration space based on the Data Release 14 (DR14) of the Sloan Digital Sky Survey IV extended Baryon Oscillation Spectroscopic Survey. This dataset includes 148,659 quasars spread over the redshift range $0.8\leq z \leq 2.2$ and spanning 2112.9 square degrees. We use the Convolution Lagrangian Perturbation Theory (CLPT) approach with a Gaussian Streaming (GS) model for the redshift space distortions of the correlation function and demonstrate its applicability for dark matter halos hosting eBOSS quasar tracers.
At the effective redshift $z_{\rm eff} = 1.52$, we measure the linear growth rate of structure $f\sigma_{8}(z_{\rm eff})= 0.426 \pm 0.077$, the expansion rate $H(z_{\rm eff})= 159^{+12}_{-13}(r_{s}^{\rm fid}/r_s){\rm km.s}^{-1}.{\rm Mpc}^{-1}$, and the angular diameter distance $D_{A}(z_{\rm eff})=1850^{+90}_{-115}\,(r_s/r_{s}^{\rm fid}){\rm Mpc}$, where $r_{s}$ is the sound horizon at the end of the baryon drag epoch and $r_{s}^{\rm fid}$ is its value in the fiducial cosmology. The quoted errors include both systematic and statistical contributions.
The results on the evolution of distances are consistent with the predictions of flat $\Lambda$-Cold Dark Matter ($\Lambda$-CDM) cosmology with Planck parameters, and the measurement of $f\sigma_{8}$ extends the validity of General Relativity (GR) to higher redshifts($z>1$)
This paper is released with companion papers using the same sample. The results on the cosmological parameters of the studies are found to be in very good agreement, providing clear evidence of the complementarity and of the robustness of the first full-shape clustering measurements with the eBOSS DR14 quasar sample.
\end{abstract}


\newpage
\section{Introduction}
\label{sec:intro}

The discovery of the accelerating expansion of the Universe from the cosmic distance-redshift relation using Type 1a supernovae~\citep{Riess+98,Perlmutter+99}  led to the conclusion that matter alone is not sufficient to describe the energy content of the universe.
With the cosmic microwave background measurements~\citep[e.g.][]{wmap9,planck15} and the use of the imprint of Baryon Acoustic Oscillations (BAO) on the spatial distribution of galaxies~\citep{Eisenstein+2005,Cole+2005,boss-dr12} obtained from spectroscopic galaxy surveys, all data are converging towards a standard model, $\Lambda$CDM, where the universe is described by the theory of general relativity (GR) and is composed of collisionless Cold Dark Matter (CDM), baryons, photons, neutrinos, and a fluid of negative pressure designated as ``Dark Energy". In its simplest form which fits the data, Dark Energy can be described as a cosmological constant ($\Lambda$) entering Einstein's equation and accounting for the late-time acceleration of the expansion of the Universe. One of the quests of the next decades is to achieve enough precision in cosmological data to constrain $\Lambda$CDM.

The nature of dark energy can be probed by measuring the growth of cosmic structure. In linear theory, one defines the linear growth rate of structure as:
\begin{equation}
f(a)=\frac{d \ln(D(a))}{d \ln(a)}
\end{equation}
where $D$ is the linear growth function of density perturbations and $a$ is the scale factor at a given epoch. In general relativity, $f$ is related to the matter contribution to the energy content of the universe~\citep{Peebles1980} through the following approximation: $f(z) \simeq (\Omega_{\rm m}(z))^\gamma$. The exponent $\gamma$ depends weakly on the energy content of the universe and is predicted to be $\gamma_{\rm GR}\simeq0.55$~\citep{LinderCahn07}. Measuring the evolution of $f$ with redshift becomes an important test for the $\Lambda$CDM+GR concordance model and it is a key observable for constraining dark energy or modified gravity models~\citep{Guzzo+08}. Indeed, alternative scenarios of gravity that keep the cosmological background unchanged predict a different rate of structure growth as the clustering of matter is driven by different effective gravity strength.

Galaxy spectroscopic surveys have become one of the most powerful probes of the cosmological model as they map the distribution of the large-scale structures over scales which contains information on how distances evolve in the universe and how those structures form in a given gravity scenario. A final advantage of spectroscopic surveys is that they give access to the radial distance from the observer through the measurement of redshift. This redshift and the celestial coordinates of objects are used to construct a three-dimension map of the cosmic structures using a fiducial cosmology. A particular challenge is that astrophysical objects have peculiar velocities arising from their infall towards overdense regions or from their binding to virialized systems. These velocities lead to an anisotropic clustering in redshift space which is known as Redshift Space Distortions (RSD), and was first described in~\citet{Kaiser87}. In practice, the observation of the RSD provides a measurement of the quantity $f(z)\sigma_8(z)$, where $\sigma_8(z)$ 
is the normalization of the linear power spectrum at redshift $z$ on scales of 8~$h^{-1}$Mpc. Anisotropy may also arise when the incorrect cosmology is used to transform redshifts and angular coordinates into comoving distances. These distortions appear in the radial and angular directions of the clustering signal and can be measured via the Alcock-Paczynski effect~\citep{AP}. The measurement of RSD thus allows for a test of the expansion history of the Universe and its structure growth independently of the assumed cosmology.

For these reasons, anisotropic clustering has received considerable attention from large-scale spectroscopic surveys. At low redshifts ($z<1$), where the precision on $f\sigma_8$ has reached $10\%$, there are measurements from 2dF~\citep{2df-rsd-04}, from 6dFGRS~\citep{6dFGS-rsd-2012}, from WiggleZ~\citep{wigglez-rsd-2011} and recently by SDSS-III BOSS~\citep{boss-dr12} and VIPERS~\citep{vipers+2017}. At redshifts greater than one, the data are sparse and only one recent exploratory measurement using Emission Line Galaxies (ELG) has been published~\citep[FastSound,][]{FastSound+16}. \\

Quasars are among the most luminous long-lived sources in the universe and they can be detected at redshifts $z>1$ at a number density high enough for cosmological measurements of BAO and RSD. Moreover, previous programs such as the 2dF QSO Redshift Survey~\citep[2QZ,][]{Cro09}, SDSS-I/II~\citep{Myers+07,Ross+09,Shen+09}, SDSS-III/BOSS~\citep{white12,Eftekharzadeh+15,Laurent+16} and a combination of quasar samples from the 2QZ and the 2dF-SDSS LRG and quasar Survey~\citep{2dQZ+08} have revealed that the observed correlation of quasars is the one expected for tracers of the underlying matter distribution, and that they can be used for clustering analysis. More recently,~\cite{RodriguezTorres+17} generated a first set of light-cones that reproduced the quasar clustering of the first year of SDSS-IV/eBOSS data and they compared their measurement on the mean mass of halos hosting quasars with previous analyses. It confirms the fact that quasars in the eBOSS redshift range reside in dark matter halos of mass ${\rm M} \sim 10^{12.5}{\rm M}_\odot$, although the halo properties are still studied to characterize the evolution the duty cycle with luminosity and redshift~\citep[for a recent study using eBOSS quasars see][]{Laurent+2017}. Other studies showed that the formation of such quasars could also be the result of a major merging of gas-rich galaxies~\citep{Sanders+88,Carlberg90, Hopkins+06}.
In addition to astrophysical motivations for studying the environment of quasars through their clustering, there is a strong interest to use them as direct tracers of the matter density field in the almost unexplored redshift range 1 $<$ z $<$ 2 in order to 
extend the test of GR. 
In particular, at the effective redshift of the eBOSS quasar sample, any deviation from GR predictions on the growth rate of structure would provide a promising discriminant between different modified gravity models.

In this work, we measure the redshift space 2-point correlation function of the spectroscopically-confirmed quasars of the SDSS-IV/eBOSS sample at effective redshift $z_{\rm eff} = 1.52$. We analyse the first three even Legendre multipoles and the three wedges of the anisotropic correlation function to constrain the angular diameter distance $D_{\rm A}(z)$, the Hubble parameter $H(z)$ and the linear growth rate of structure $f(z)\sigma_8(z)$. These measurements are presented in Figures~\ref{fig:bao} and ~\ref{fig:fs8} in this paper along with the result of the work presented here. \\

Our study complements the measurement of the BAO feature presented in~\citet{DR14-bao} and is accompanied by several companion papers that are all using the same DR14 quasar sample. \citet{Hector} analyses the clustering in Fourier space using multipoles and \citet{Hou+18} uses a different RSD model for the correlation function than the one we use in this work. Another type of RSD analaysis has been performed in~\citet{Ruggeri+18} and~\citet{Zhao+18} using a redshift weighting technique to probe the redshift evolution of the cosmological parameters across the redshift range. Further details on each analysis and a general comparison between all the methods is presented at the end of the paper.\\

The paper is structured as follows. We first present the data in Section~\ref{sec:data} and the mock catalogs we use for this analysis in Section~\ref{sec:mocks}. In Section~\ref{sec:model}, we focus on our adopted RSD model and the tests performed using mock catalogs to estimate the systematic uncertainties related to the modeling. Section~\ref{sec:analysis} details the potential sources of observational systematics, and we propose a new way of accounting for some of them which leads to observational systematic uncertainties that are much smaller than the statistical precision of the current sample. In Section~\ref{sec:results}, we compare the results obtained using 3-multipole and 3-wedge analyses, and then their consistency with the companion papers. We explore the cosmological implications of our measurements in Section~\ref{sec:cosmo} followed by a conclusion in Section~\ref{sec:conclusion}.

\section{The data}
\label{sec:data} 
The extended Baryon Oscillation Spectroscopic Survey~\citep[eBOSS,][]{ebossoverview} is one of the four programs of SDSS-IV~\citep{sdss4}. Observations were conducted at the 2.5m Sloan Foundation telescope~\citep{gunn06} at the Apache Point Observatory, and eBOSS uses the same two-arm optical fiber-fed spectrographs as BOSS~\citep{bossspectrometer}. The analysis presented in this paper uses the quasar catalogs DR14Q~\citep{DR14Q} from the Data Release 14 of SDSS~\citep{DR14}. The data and the construction of the catalogs are described in~\citet{DR14-bao}; we refer the reader to this article for further details. Quasar target selection~\citep{Myers+15} is based on the SDSS-I-II-III optical imaging data in the {\it ugriz}~\citep{Fukugita+1996} photometric pass band and on the Wide-field Infrared Survey Explorer~\citep[WISE,][]{Wright+10}. Selection is performed with the XDQSOz algorithm developed for BOSS~\citep{Bovy+2012} which is used to define a CORE homogenous sample suitable for cosmological clustering measurements. 

\begin{figure} 
\includegraphics[width=84mm]{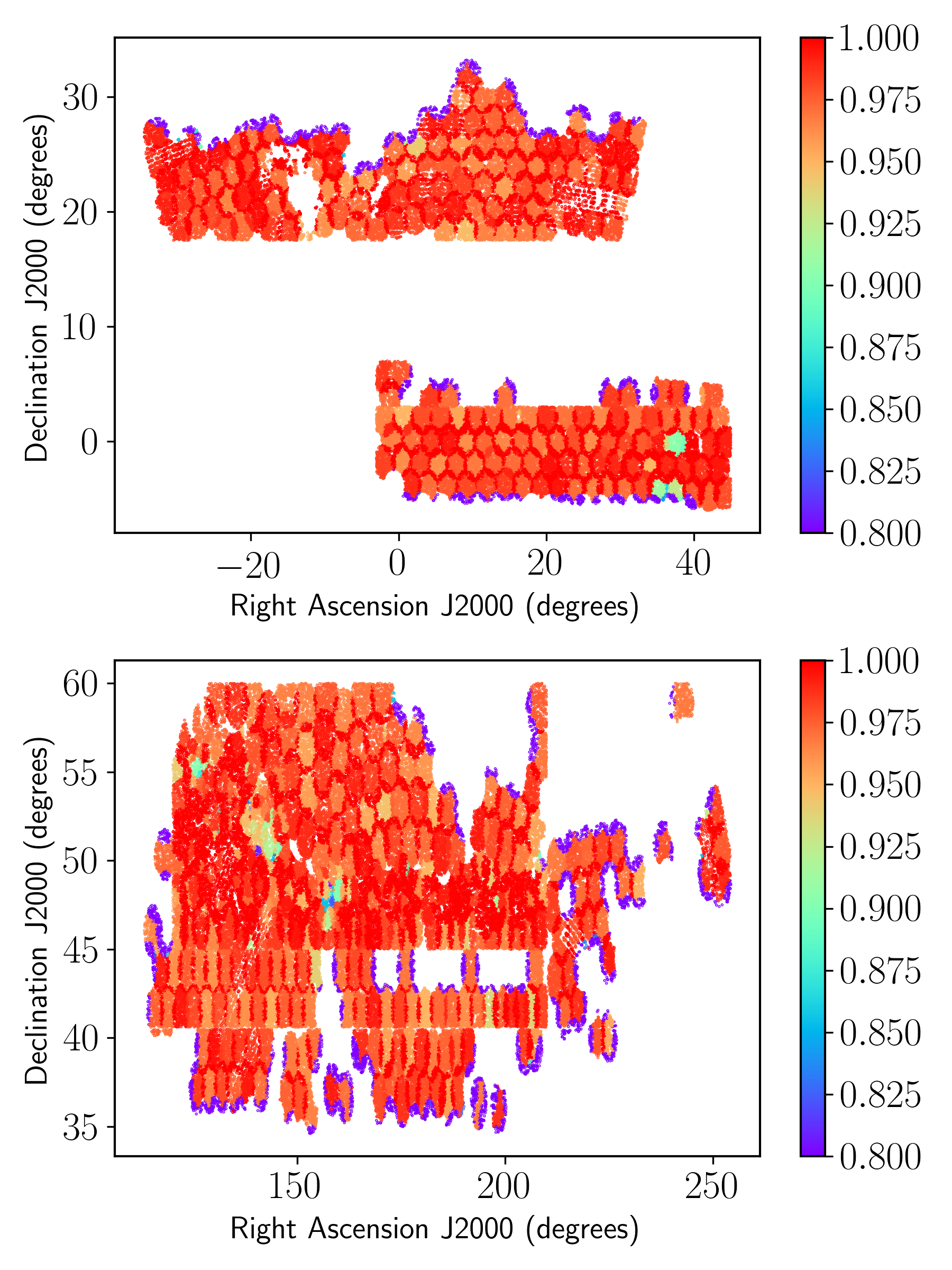}
\vskip -0.2cm
\caption{Footprint of the DR14Q catalog used for this analysis. The upper (lower) panel displays the South (Nothr) Galactic Cap resulting in a total effective area of 2112.9 deg$^2$. Each object is color-coded according to the completeness of the sector to which it belongs (object in purple have completeness between $0.5$ and $0.8$)}
\label{fig:footprint} 
\end{figure}

\begin{figure} 
\includegraphics[width=84mm]{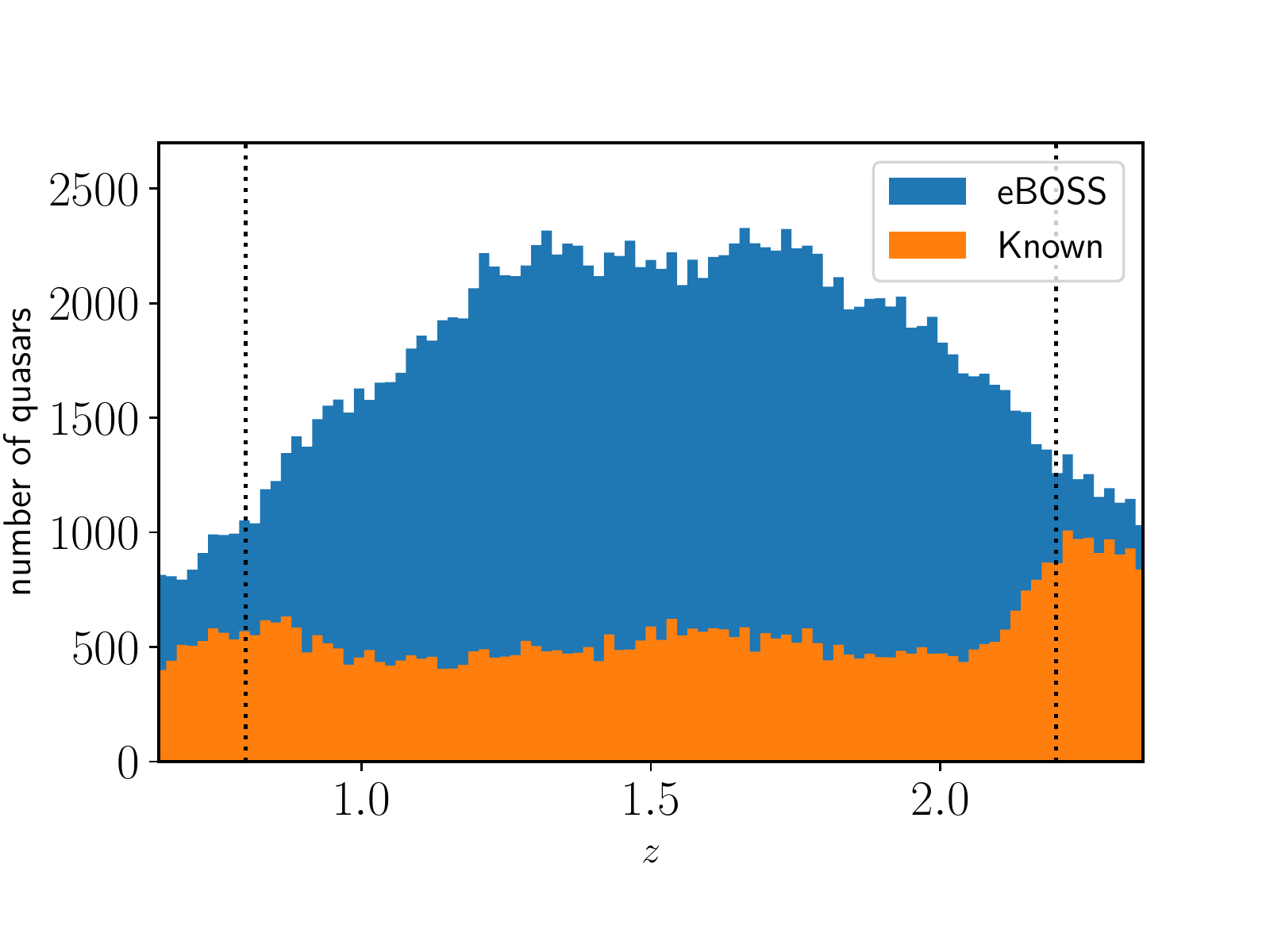} 
\vskip -0.5cm
\caption{Redshift distribution of the objects in the DR14 catalog corresponding to the CORE sample. The orange histogram shows the known quasars at the start of eBOSS data taking. Objects in $0.8\leq z \leq 2.2$ are kept for this analysis.}
\label{fig:zdistribution}
\end{figure}

The footprint of spectroscopically-observed objects is shown in Figure~\ref{fig:footprint} and the redshift distribution of the CORE quasars in the DR14 catalog is presented in Figure~\ref{fig:zdistribution}. The orange histogram corresponds to the distribution of the known quasars at the start of eBOSS data taking. Over 75\% of the new redshifts were obtained during the eBOSS program.
The number of objects and the effective area of the sample are detailed in Table~\ref{tab:nqso} and correspond to a maximum density of $2 \times 10^{-5} h^3{\rm Mpc}^{-3}$ and an effective redshift of $z_{\rm eff}=1.52$. It represents a sparse sample as the number density of quasars is relatively low compared to SDSS BOSS galaxies for instance but this drawback can be compensated by probing an important volume over a wide redshift range. While the BOSS galaxy sample can be considered as cosmic-variance limited, the eBOSS quasar sample is in the shot-noise dominated regime with $nP << 1$, where $n$ is the observed quasar density and $P$ is the amplitude of the power spectrum at the scale of interest.

In this section, we present some characteristics of the data that are important for our measurement.

\begin{table}
\caption{Number of quasars with $0.8\leq z \leq 2.2$ of the eBOSS CORE sample and effective area.}
\label{tab:nqso} 
\begin{tabular}{|c|c|c|c|}
                                      & NGC    & SGC    & Total\\
\hline
${\rm N}_{\rm quasar}$ $(0.8\leq z \leq 2.2)$ & 89233  & 59426  &  148659\\
Effective area (deg$^2$)              & 1214.6   & 898.3    &  2112.9
\end{tabular}
\end{table}

\subsection{Completeness} 
\begin{figure} 
\includegraphics[width=84mm]{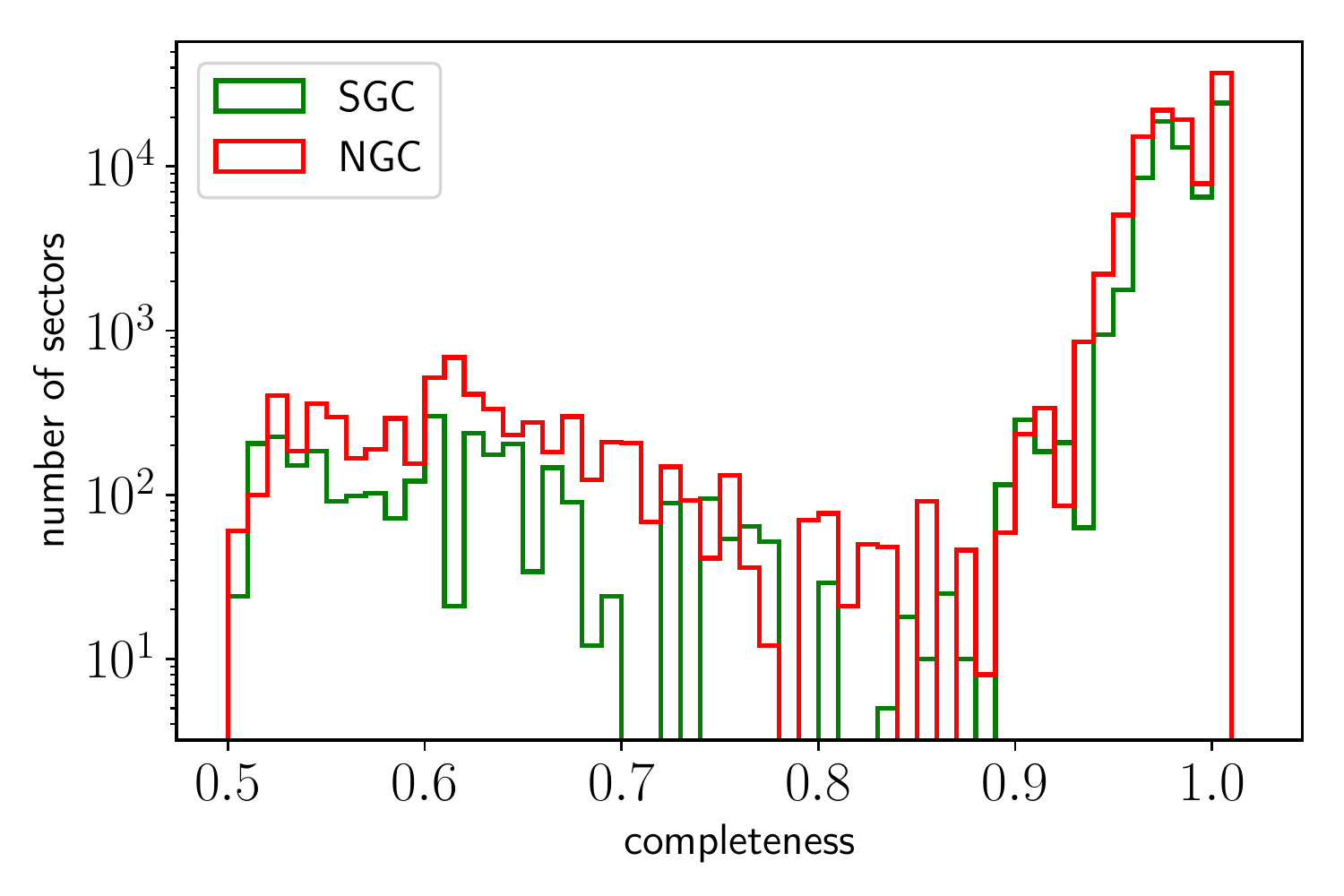} 
\vskip -0.2cm
\caption{Distribution of the completeness $C_{\rm eBOSS}$ per sector. Sectors with completeness smaller than 0.9 correspond to overlaping plates regions where only one plate has currently been measured. Objects with $C_{\rm eBOSS}>0.5$ are not considered for clustering analysis.} 
\label{fig:data-completeness}
\end{figure}

In general, the number of targets exceeds the number of available fibers, and a tiling procedure~\citep{Blanton+2003} is applied to maximize the completeness of the target sample, taking into account the constraints imposed by higher-priority targets and the physical size of the fibers.
The positions of the plates define sectors where multiple plates may overlap. For each sector, an observational completeness $C_{\rm eBOSS}$ is calculated from ${\rm N_{\rm target}}$, the number of imaging quasar targets selected, ${\rm N_{\rm fiber}}$, the number of targets that actually received a fiber after the tiling algorithm is applied, ${\rm N_{\rm cp}}$, the number of targets that were in collision within the 62$''$ exclusion radius around each target (set by physical size of the fiber) and thus did not receive a fiber, and ${\rm N_{\rm known}}$, the number of targets that are confirmed quasars measured by previous surveys at the time of tiling. The observational completeness is defined as : 
\begin{equation} 
C_{\rm eBOSS} = \frac{\rm N_{\rm fiber} + N_{\rm cp}  } {\rm N_{\rm
target} - N_{\rm known} }. 
\end{equation} 
With this definition of the completeness, targets in collision can be treated at the analysis level by upweighting by one unit the nearest identified quasar. The distribution of the completeness per sector for our survey is shown in Figure~\ref{fig:data-completeness}. It is high but it is not 100\% in all sectors because the CORE quasar targets do not get the highest priority and because of combinatorial requirements in the tiling algorithm. Low completeness sectors ($C<0.85$) are due to overlaping plates for which some plates have not yet been observed. Objects in sectors with $C_{\rm eBOSS}<0.5$ are removed from the catalogs. 

The completeness is used to create a random catalog which has the same angular selection function as the data. The redshifts of the objects in the random catalog are drawn from the redshift distribution of the data ensuring that the radial selection function is the same for the models as the data. The completeness is also used to downsample the legacy quasars in the same manner. 

\subsection{Redshif estimates}
\label{data-redshift}

The DR14Q quasar catalog~\citep{DR14Q} includes all SDSS-IV/eBOSS objects that were spectroscopically targeted as quasar candidates. It is based on a new automatic classification procedure which is detailed in Section 5.2 of~\citet{ebossoverview} and provides different redshift estimates for each quasar. Redshift determination proceeds from the analysis of the spectrum of the candidates. Quasar spectra contain broad emission lines due to the rotating gas located around the central black hole. These features are subject to matter outflows around the accretion disk which frequently give rise to systematic offsets when measuring redshifts.

The reported redshift estimates are based on the following methods :
\begin{itemize}
\item '$z_{\rm PL}$': the SDSS quasar pipeline redshifts. They are based on a Principal Component Analysis (PCA) using galaxy, star and quasar templates to fit a linear combination of four eigenspectra to each spectrum. Template-based redshifts are expected to be more stable since they use information from the full spectrum, but at $z \sim 1.5$, the CIV emission line (which is subject to significant shifts) enters the observed spectral range and drives the fit, which has an impact on the redshift acccuracy. 
\item '$z_{\rm MgII}$': For objects identified as quasars, the location of the maximum of the MgII emission line is fitted using a linear combination of five principal components. \citet{Hewett+2010} and \citet{Shen+2016} showed that the MgII feature is the quasar broad emission line that has the smallest velocity shifts ($\sim 200$~km~$s^{-1}$) because as it is a lower ionization species it presumably lies at a larger distance from the central black hole. Therefore, it provides the redshift estimate with the smallest systematic error, although its statistical precision is a bit degraded from the pipeline redshift, particularly for low signal-to-noise lines.
\item '$z_{\rm PCA}$': For objects identified as quasars, the redshift is measured using a dedicated PCA of the entire quasar spectrum, and the principal components are calibrated using the MgII emission as a reference. This approach allows a redshift determination for faint quasars at $z \simeq 2$ when the MgII line approaches the red limit of the SDSS-IV spectral coverage and is not clearly detected. 
\item '$z_{\rm VI}$': Redshift from visual inspection. For SDSS-III/BOSS, all quasar targets have been visually inspected; this is not the case for SDSS-IV eBOSS, where only the objects that the automated procedure considers as badly identified lead to a visual inspection~\citep[for more details, see Section 3.3 of][]{DR14Q}.
\end{itemize}
The DR14Q catalog also contains a redshift, 'z', which is equal to $z_{\rm PL}$ for the majority of the time, and for the $\sim$7\% visually-inspected quasars it can be either $z_{\rm PL}$ or $z_{\rm VI}$, depending on the robustness of each determination. This redshift estimate is available for all the DR14 quasars and is known to have the lowest rate of catastrophic failures ($<$1\%).  In what follows, this redshift measurement will be taken as the reference, and in Section~\ref{sec:analysis} we will compare the results obtained with this estimate to the result of the RSD analysis performed with special catalogs where the redshift is taken to be $z_{\rm MgII}$ (resp. $z_{\rm PCA}$) whenever it is available (i.e., 80\% of the time) and $z$ otherwise, such that these catalogs contain the same objects.

\begin{figure}
\includegraphics[width=70mm]{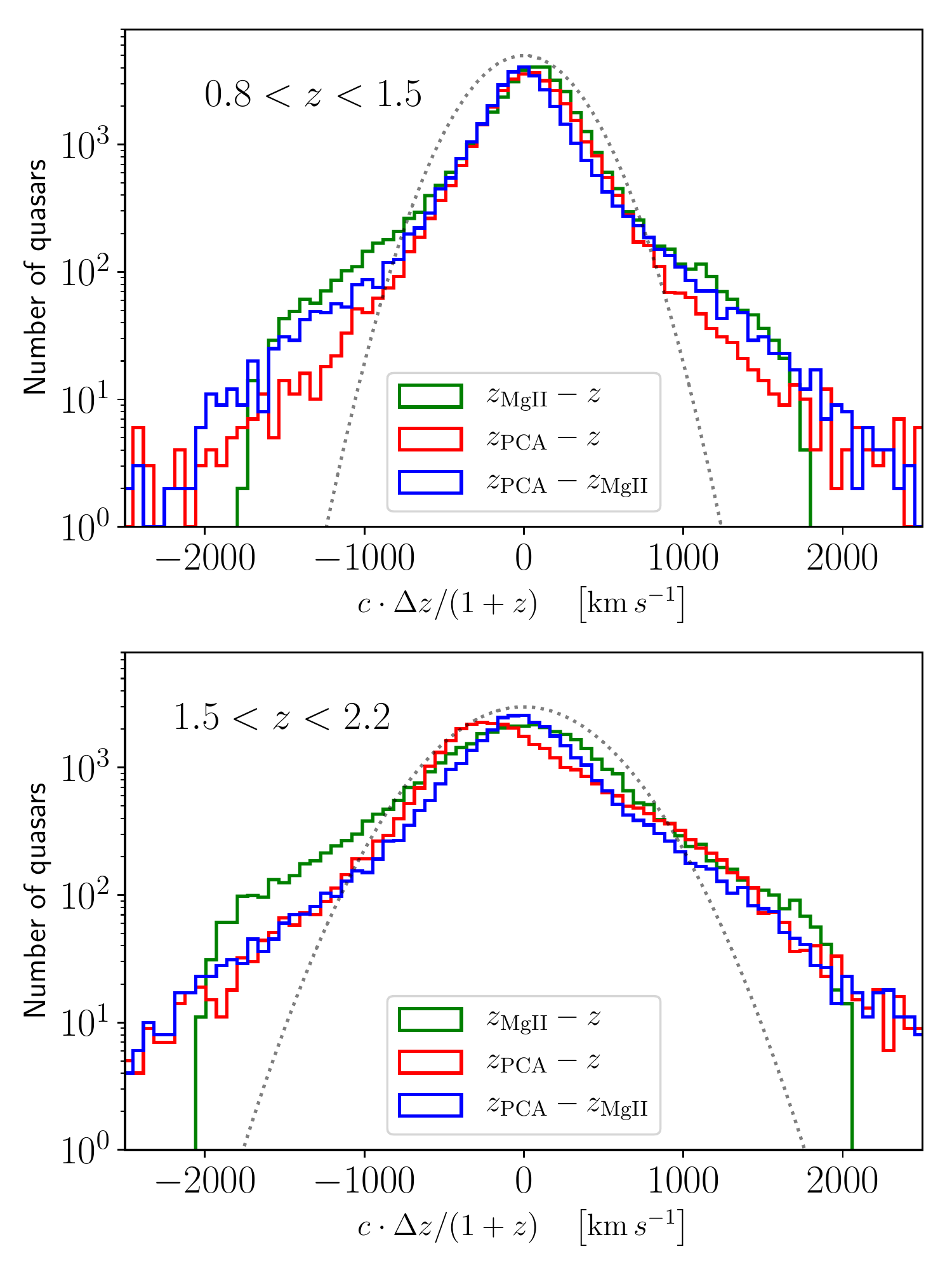}
\vskip -0.2cm
\caption{Physical distributions (solid lines) of $\Delta v = \Delta z \cdot c / (1 + z)$ between different redshift estimates for two redshift bins in our redshift range. The dotted line shows a gaussian distribution of width given by the survey requirements (see text). The most important feature is that the observed distributions present large non-Gaussian tails that extend to 3000~km~s$^{-1}$. At low redshifts (upper panel), the distributions are mostly symmetric although minor shifts can be observed. At high redshifts (lower panel), the distribution obtained for ${z}_{\rm PCA}-{z}$ (red) is asymmetric, and could yield systematic shifts in the separation of quasars}
\label{fig:data-dz}
\includegraphics[width=75mm]{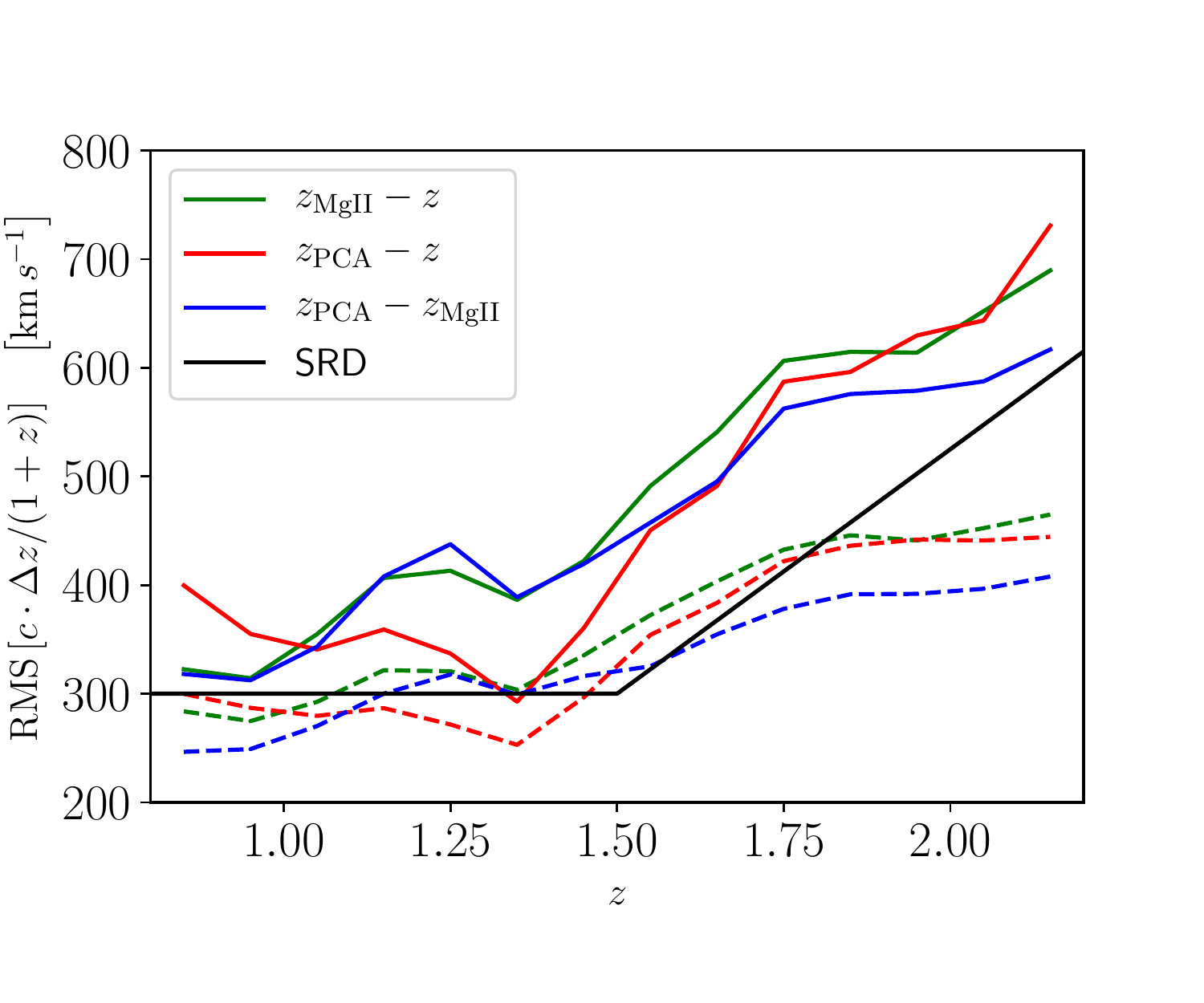}
\vskip -0.3cm
\caption{RMS of the scatter of $\Delta v = \Delta z \cdot c / (1 + z)$ for different redshift estimates as a function of redshift, compared to the survey requirements (black solid line). Solid lines (resp. dashed lines) are obtained requiring that $|\Delta v|< 3000{\rm km/s}$ (resp. $|\Delta v|< 1000{\rm km/s}$).}
\label{fig:data-dzbins}
\end{figure}

Figure~\ref{fig:data-dz} shows the distributions of $\Delta v = \Delta z \cdot c / (1 + z)$,  for the difference of redshift estimates: $\Delta {z}={z}_{\rm MgII}-{z}$, $\Delta {z}={z}_{\rm PCA}-{z}$ and $\Delta {z}={z}_{\rm PCA}-{z}_{\rm MgII}$ for the two redshift bins in our range of interest. We compare the discrepancies to a Gaussian distribution of width given by the survey requirements (SRD,~\cite{ebossoverview}) where the redshift resolution is expressed as:
\begin{align}
\sigma_v^{\rm SRD}(z) &= 300 \,\,{\rm km}\,s^{-1}  &z<1.5\\
\sigma_v^{\rm SRD}(z) &= 450 \times (z-1.5) +300 \,\,{\rm km}\,s^{-1} &z>1.5
\end{align}  
The most important feature is that the distributions present large tails extending to 3000~km~s$^{-1}$ so that they are clearly non-Gaussian. The distributions involving ${z}_{\rm MgII}-{z}$ (green) and ${z}_{\rm PCA}-{z}_{\rm MgII}$ (blue) are centered at zero offset (because of the calibration mentioned above) and are mostly symmetric. The distribution obtained for ${z}_{\rm PCA}-{z}$ (red) is asymmetric, suggesting that for the special catalogs which mix ${z}_{\rm PCA}$ and $z$, there could be systematic shifts in the separation of quasars.

Figure~\ref{fig:data-dzbins} presents the evolution of the standard deviations of the above distributions as a function of redshift and compare it to the SRD. When considering only quasars $|\Delta v|<$1000~km~s$^{-1}$ (dashed lines) in the calculation of the standard deviation, our result agrees with the SRD and with the results obtained in~\citet{ebossoverview}. When allowing larger values of $|\Delta v|<$3000~km~s$^{-1}$ (solid lines), the standard deviation increases as expected given the shape of the distributions. These results lead to a resolution which is slightly larger than the SRD.

In the following sections, we will demonsrtate that the redshift resolution has a large impact on the clustering signal, especially at scales below 40~$h^{-1}{\rm Mpc}$, and that the impact can be measured by fitting the data. Furthermore, we will investigate the impact of the redshift resolution on the Redshift Space Distortions modeling and on the ability to recover the cosmological parameters both in terms of shape and RMS of the redshift error distribution.

\subsection{Spectroscopic completeness}

\begin{figure}  
\includegraphics[width=75mm]{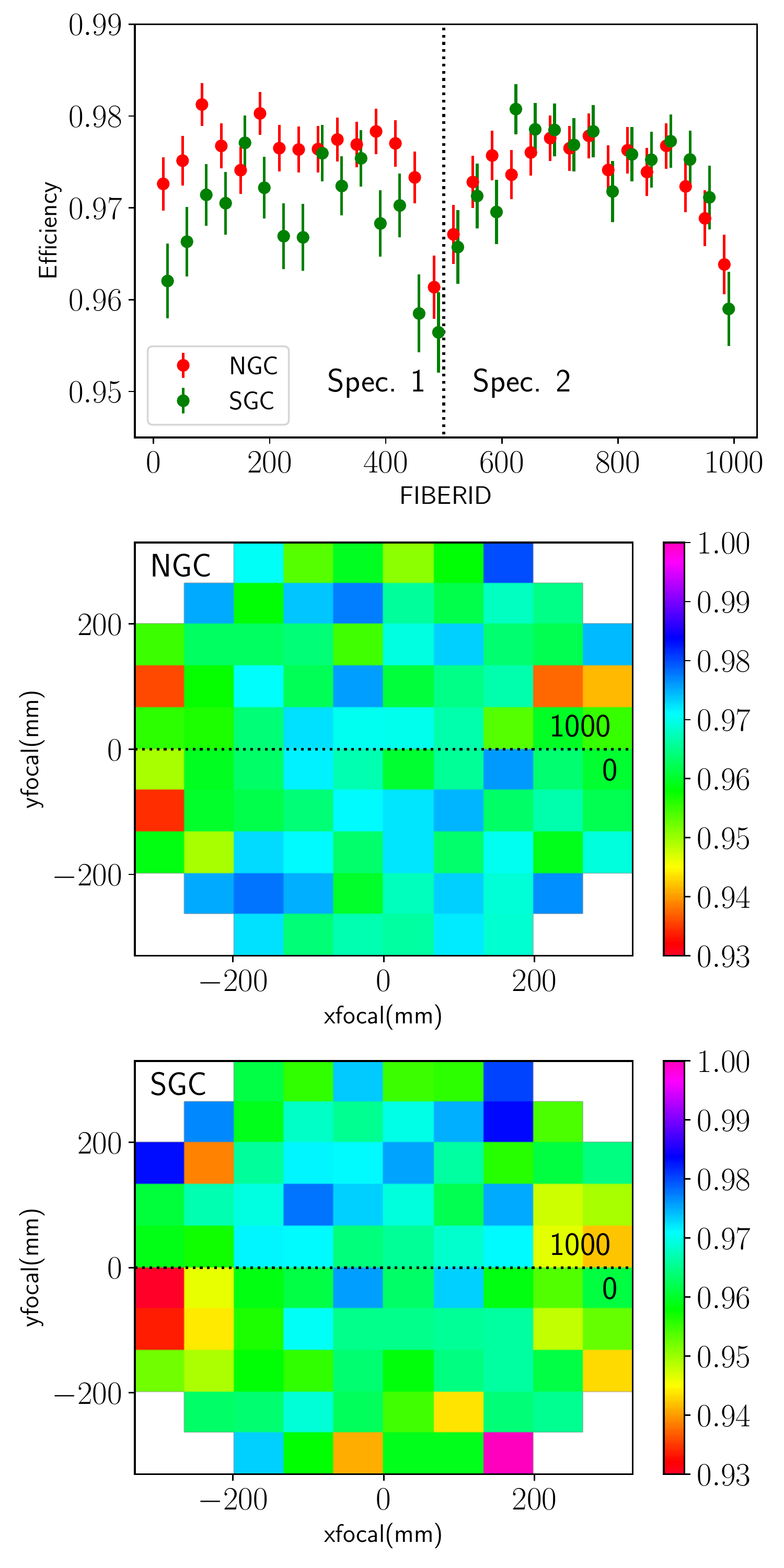} 
\caption{Top panel: Redshift efficiency as a function of the fiber number. The vertical dotted line shows the delimitation between the 2 spectrographs. Bottom panels: Redshift efficiency as a function of the focal plane coordinates for the NGC (middle panel) and SGC (lower panel). The fiber number goes clockwise from 0 to 1000.}
\label{fig:efficiency-fiberid}
\end{figure}

The probability of obtaining a reliable redshift from a spectrum depends on both observational and instrumentation parameters affecting the S/N. When the redshift from an identified quasar cannot be secured, the nearest quasar neighbour is marked such that we can track redshift efficiency and study the weighting scheme to take this into account. Here we extend the treatment applied in previous analyses and search for dependencies with the position in the focal plane. The redshift efficiency or spectroscopic completeness is defined as the ratio between the number of objects with a secured redshift to the number of quasars that received a fiber: 
\begin{equation} 
\epsilon =\frac{N_{\rm good}}{N(w_{zf}=1)+2\cdot N(w_{zf}=2)} 
\end{equation}
with $N_{\rm good}$ the number of quasars with robust redshift, and $N(w_{zf}=1,2$ the number of quasars without or with a neighbour with a redshift failure. This expression allows for the calculation of the redshift efficiency from the released catalog. 
The variation of the redshift efficiency for groups of fibers of the spectrographs is displayed in Figure~\ref{fig:efficiency-fiberid}. It confirms the findings of~\citet{Laurent+2017} that the quasar redshift efficiency is lower at the edges of the two spectrographs. Furthermore, the efficiency of the first spectrograph is found to be significantly lower for SGC observations. The variation of redshift efficiency across the focal plane is shown in the bottom panels of Figure~\ref{fig:efficiency-fiberid}. Regions with lower efficiency are at the left and right sides of the focal plane which correspond to edges of the spectrographs. 
Section~\ref{sec:analysis} examines the impact of redshift failures on the RSD measurement using mocks, and we propose a weighting scheme to mitigate their effect.

\subsection{Correlation function}
Measured redshift and angular coordinates are converted to comoving coordinates using the fiducial cosmology that was used for the BOSS DR12 analysis~\citep{boss-dr12} and for the eBOSS DR14Q BAO analysis~\citep{DR14-bao}: \begin{equation}
\begin{split}
H_{0} = 0.676, \,\, \Omega_{m} = 0.31, \,\,  \Omega_{\Lambda} = 0.69,\\
\Omega_{b}h^{2} = 0.022, \,\, \sigma_{8} = 0.80. \qquad
\end{split}
\label{eq:cosmo}
 \end{equation}
 The public code CUTE~\citep{CUTE} was used to calculate paircounts as a function of comoving separation $s$, and $\mu= \cos \theta$ where $\theta$ is the angle between the line of sight (LOS) and the orientation vector of the pair of tracers under consideration. The 2D correlation function $\xi(s,\mu)$ is calculated using the miminum variance estimator defined in~\citet{LandySzalay93}: 
\begin{equation}
\xi(s,\mu) = \frac{DD(s,\mu) - 2DR(s,\mu) + RR(s,\mu)}{RR(s,\mu)}\, , 
\end{equation} 
where $DD(s,\mu)$ is the number of pairs of quasars with separation $s$ in redshift space and orientation $\mu$, $DR(s,\mu)$ is the number of pairs between the quasar catalog and the random catalog, and $RR(s,\mu)$ is the number of pairs for the random catalog. The 2D correlation function is projected onto the Legendre polynomial basis through:
\begin{equation}
\xi_{l}(s)=\frac{2l+1}{2}\sum_{j}\xi(s,\mu_{j})P_l(\mu_{j})d\mu\, ,
\end{equation} 
where only $l=0,2,4$ are non-zero in linear theory. The analysis can also be performed by cutting the domain in $\mu$ into ``wedges":
\begin{equation}
\xi_{wi}(s)=\frac{1}{\mu_{i,{\rm max}}-\mu_{i,{\rm min}}}\int_{\mu_{i,{\rm min}}}^{\mu_{i,{\rm max}}}\xi(s,\mu)d\mu\,.
\end{equation} 

\subsection{Cosmological parameters}

The cosmological information is extracted through a fit of the measured correlation function with a model based on Convolution Lagrangian Perturbation Theory (CLPT) which is described in Section~\ref{sec:model}. In practice, the model establishes a prediction for the correlation function for a tracer of bias $b$ as a function of $f$ and uses a linear power spectrum $P_{\rm lin}$: 
\begin{equation}
\xi^{\rm CLPT}(\alpha_{\parallel}s_{\parallel},\alpha_{\perp}s_{\perp},b,f | P_{\rm lin}).
\end{equation}
where $P_{\rm lin}$ is fixed according to the fiducial cosmological parameters we use for the analysis. 
Here, we have introduced two additional parameters, $\alpha_{\parallel}$ and $\alpha_{\perp}$, to account for different dilation of scales for the directions along and perpendicular to the LOS. This approach allows the measured cosmology to differ from the fiducial cosmology from which distances are inferred using redshifts and angular coordinates. In linear theory, all terms in the correlation function (or power spectrum) including $f$ and $b$ are multiplied by $\sigma_8$ and the degeneracy cannot be broken~\citep{Percival+2009}. Therefore, results are reported in terms of $f\sigma_8$ and $b\sigma_8$. 

The parameters $\alpha_{\parallel}$ and $\alpha_{\perp}$ can be related to the expansion rate $H(z)$ and the angular diameter distance $D_{\rm A}$ through:
\begin{equation}
\alpha_{\parallel}=\frac{H^{\rm fid}(z)r_{s}^{\rm fid}}{H(z)r_{s}},\qquad \alpha_{\perp}=\frac{D_{\rm A}(z)r_{s}^{\rm fid}}{D_{\rm A}^{\rm fid}(z)r_{s}}
\end{equation}
where $r_{s}$ is the sound horizon at the end of the baryon drag epoch and quantities with the superscript~'fid' refer to quantities determined within the fiducial cosmology. 
From $\alpha_{\parallel}$, $\alpha_{\perp}$, and the fiducial cosmology, one can construct a volume averaged distance, $D_{\rm V}$:
\begin{equation}
D_{\rm V}=\left[ (1+z)^2cz\frac{D_{\rm A}^2}{H} \right]^\frac{1}{3}\,,
\end{equation} 
where $c$ is the speed of light. One can also define the Alcock-Paczynski parameter $F_{\rm AP}$, which is proportional to the ratio of scales along and perpendicular to the LOS: 
\begin{equation}
F_{\rm AP}= \frac{1+z}{c}D_{\rm A} H \,.
\end{equation} 

Alternatively, one can also use a combination of $\alpha_{\parallel}$ and $\alpha_{\perp}$ such that:
\begin{equation}
\alpha = \alpha_{\parallel}^{1/3}\alpha_{\perp}^{2/3},\qquad \epsilon = (\alpha_{\parallel}/\alpha_{\perp})^{1/3} - 1
\end{equation}
When using the monopole only, \citet{Ross+15} demonstrated that one can constrain the $\alpha$ variable; we often refer to this quantity as $\alpha_{\rm iso}$. It corresponds to an isotropic shift of the BAO feature and was measured in the eBOSS DR14Q analysis~\citep{DR14-bao} leading to a 3.8\% measurement of the spherically-averaged BAO distance $D_{\rm V}$ through:
\begin{equation}
\alpha_{\rm iso}=\frac{D_{\rm V}(z)r_{s}^{\rm fid}}{D_{\rm V}^{\rm fid}(z)r_{s}}\,.
\end{equation} 
For consistency, we also perform an analysis by assuming in the model that there is no anisotropic dilation of scales and fitting $\alpha_{\rm iso}$. We refer the reader to appendix A of~\citet{Hector} where it is shown that, given the statistical precision of the current sample, assuming $\alpha_{\rm iso} \simeq \alpha_{\parallel}^{1/3}\alpha_{\perp}^{2/3}$ is a valid approximation. In this framework, we can extract $f\sigma_8$ from:
\begin{equation}
\xi^{\rm CLPT}_{\rm iso}=\xi^{\rm CLPT}(\alpha_{\rm iso}s,b,f | P_{\rm lin})\, .
\end{equation} 
In the present work, we perform the anisotropic full-shape analysis for the multipoles up to the hexadecapole ($\xi_{l=0,2,4}$) and for three wedges in $\mu$ of constant size $\Delta \mu=1/3$ ($\xi_{w1,2,3}$). For a consistency check, we also present the results from the isotropic case when we evaluate the performance of the RSD modeling and when we summarize the tests on the final results.


\subsection{Parameter inference}
We extract the results of the fitting of either the three first Legendre multipoles or the three wedges by minimizing the $\chi^{2}$ defined by:
\begin{equation}
\chi^{2} = (\xi^{\rm Data} - \xi^{\rm Model}) C^{-1} (\xi^{\rm Data} - \xi^{\rm Model})^{t}
\end{equation}
where $\xi^{\rm Data}$ corresponds to the measurement, $\xi^{\rm Model}$ to the associated theoretical prediction, and $C^{-1}$ the inverse of the estimated covariance marix. The latter includes corrections due to number of mocks and number of bins in the analysis following the procedure described in~\citet{Hartlap+2007} and~\citet{Percival+2014} 
We find the $\chi^{2}$ minima using the MINUIT libraries\footnote{James, F. MINUIT Function Minimization and Error Analysis: Reference Manual Version 94.1. 1994.}. Error-bars are derived from the $\Delta \chi^{2}=1$ region of the marginalized $\chi^{2}$ profiles and are allowed to be asymmetric.

When comparing our results with the companion papers, we also run Markov-chains to compute the likelihood surface of the set of parameters. We use the $emcee$ package~\citep{emcee} which is a python implementation of the affine-invariant ensemble sampler for Markov chain Monte Carlo (MCMC); we check its convergence using the Gelman-Rubin convergence test $R - 1 < 10^{-2}$.

\section{The mock catalogs}
\label{sec:mocks}
In this work, we use the mocks (EZ and QPM mocks) that were produced for the analysis of the BAO feature in the eBOSS quasar sample presented in~\citet{DR14-bao} to determine the covariance matrix and to check for the impact of various observational systematic effects. We also developed a set of accurate mocks based on the OuterRim simulation in order to check systematics in the extraction of the cosmological parameters with the RSD model.

\subsection{Approximate mocks: EZ and QPM mocks}
\begin{figure}
\includegraphics[width=90mm]{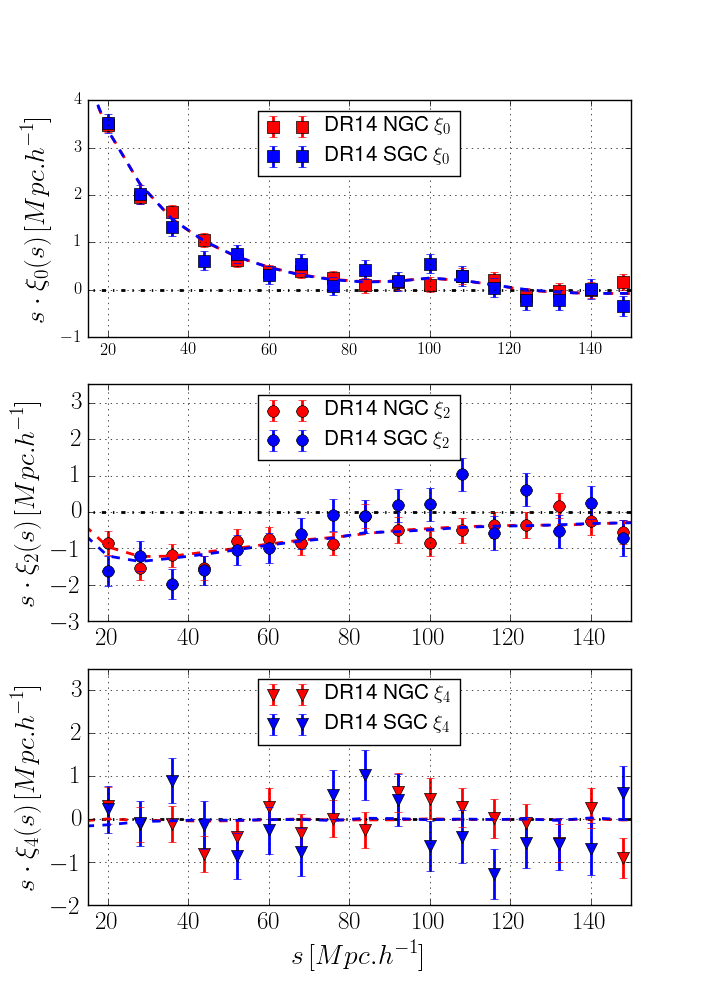}
\vskip -0.24cm
\caption{Top panel: Monopole of the eBOSS DR14 NGC (blue) and SGC (red) compared to the mean of the 1000 EZ mocks (dashed). Middle and bottom panels: Same for the quadrupole and the hexadecapole. The parameters of the EZ mocks are tuned on the observed clustering of the data for each Galactic cap separately.}
\label{fig:data-mock}
\end{figure}

EZ mocks are light-cone mock catalogs created with the Effective Zel'dovich approximation method~\citep{EZmocks} and based on seven redshift shells. We refer the reader to Section 5.1 of~\citet{DR14-bao} for more details on the generation of EZ mocks. The fiducial cosmology model is flat $\Lambda$CDM with $\Omega_{\rm m}=0.307115$, $h=0.6777$, $\sigma_8=0.8225$, $\Omega_{\rm b}=0.048206$ and $n_s=0.9611$. The simulation box size is 5~Gpc$^3$ and is large enough to include the DR14 volume. The relevant parameters were tuned on the data for the NGC and SGC separately to reproduce the difference in clustering as shown in Figure~\ref{fig:data-mock}. Given that the EZ mocks are adjusted on the data directly, they already contain the redshift resolution of the data that affect the clustering, especially the quadrupole on scales below $\sim$40$h^{-1}{\rm Mpc}$. We use 1000 independent realizations for each Galactic cap.

To study systematic effects arising from fiber collisions and spectroscopic completeness, we use a more realistic set of EZ mocks also described in Section 2.3.4 of ~\citet{Hector}. In these mocks, the plate geometry of the actual survey is applied to retrieve coordinates in the focal plane for each object. From these coordinates, one can determine whether the object belongs to a sector of overlaping plates and also estimate the redshift efficiency as measured in the data. This information provides the possibility to tag objects in collision (within 62$''$ of each other) and to downsample objects according to the redshift efficiency allowing for extensive tests of the weighting scheme. 

\begin{figure}
\hskip -0.6cm
\includegraphics[width=100mm]{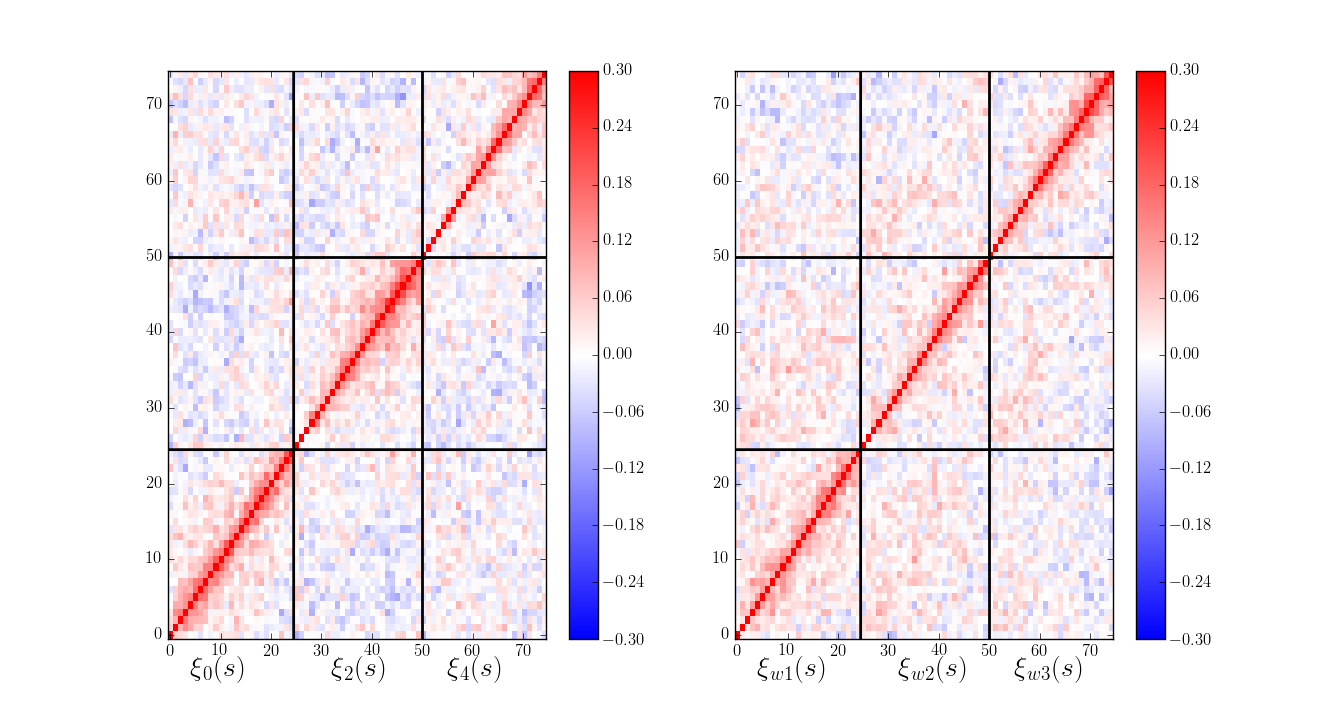} 
\vskip -0.1cm
\caption{Correlation matrices obtained from the 1,000 EZ mocks and used to fit the data for the 3-multipole (left) and 3-wedge (right) analyses. Values of the correlation above 0.3 (along the diagonal) are truncated to enhance the constrast in the lower correlation regions. Each individual square is $25\times 25$ bins of width 8~$h^{-1}{\rm Mpc}$ from 0 to 200~$h^{-1}{\rm Mpc}$.}
\label{fig:corrMat}
\end{figure}

The correlation matrices determined with these mocks are displayed in Figure~\ref{fig:corrMat} for the 3-multipole and 3-wedge analyses. The inverse of the covariance matrices used to fit the cosmological parameters include the correction procedure described in~\citet{Hartlap+2007} and~\citet{Percival+2014} due to finite number of mocks and number of bins in the analysis.

The eBOSS Quick Particle Mesh (QPM) mock catalogs were created from boxes of side ${\rm L}=5120$~$h^{-1}{\rm Mpc}$ with 2560$^3$ particles at $z$=1.55 from a low-resolution particle mesh solver procedure described in~\citet{White+14} which was used for BOSS galaxies. This procedure has been adapted to match the redshift range and to allow for lower halo masses for the eBOSS quasar analysis. A 5-parameter Halo Occupancy Distribution~\citep[HOD, e.g. ][]{Tinker+2012} model for biasing halos was tuned on the peak of the mean density of quasars as a function of redshift and on the projected quasar correlation function~\citep[see Figure 9 of ][]{DR14-bao}. In the approach taken for our data set, the number of satellites is independent of the presence of a central quasar; a sixth parameter, $\tau=1.2\%$, is used to model the duty cycle of the quasars. The expression of the HOD becomes :
\begin{eqnarray}
\langle{N}_{\rm cen}\rangle_\M &=& \tau \cdot \frac{1}{2} \left[ 1 + {\rm erf}\left( \frac{\log \M - \log \M_{\rm cen}} {\log \sigma_\M}\right)\right] \\
\langle{N}_{\rm sat}\rangle_\M &=& \left(\frac{\M}{\M_{\rm sat}}\right)^{\alpha_{\rm sat}} \cdot {\rm exp}\left( {-\frac{\M_{\rm cut}}{\M}}\right)
\end{eqnarray}
where $\langle{N}_{\rm cen}\rangle_\M$ is the probability for a halo of mass $\M$ to host a central quasar and $\langle{N}_{\rm sat}\rangle_\M$ is the number of sattelite in a halo of mass $\M$. The values of the parameters that reproduce the data are $\M_{\rm cen}=1.35\,10^{12}\M_{\odot}$, $\log \sigma_\M=0.2$, $\alpha_{\rm sat}=1$, $\M_{\rm cut}=10^8\M_{\odot}$ and $\M_{\rm sat}=1.93\,10^{15}\M_{\odot}$, which results in a population that consists of $\simeq$ 13\% satellites. Under these conditions the typical mass for dark matter halos hosting quasars is $\M_{\rm cen}=10^{12.5}\M_{\odot}$. We can apply the geometry of the DR14 survey more than once in the QPM cubic boxes which allows us to define four configurations with an overlap less than 1.5\% in order to produce 400 realizations per Galactic cap. Taking into account the fact that the NGC and SGC are produced using the same 100 original boxes, we combine them shifting the indices of the four realisations produced from each cubic box. We then apply the veto mask and survey geometry using the 'make survey' code used in~\citet{White+14}; these 400 QPM mocks were used to provide an alternative determination of the covariance matrix.

\subsection{Accurate mocks: OuterRim and MultiDark}
The purpose of accurate mocks is to check that the cosmological parameters extracted using the RSD model described in Section~\ref{sec:model} are in agreement with the input of the simulation. Indeed, we determined that approximate mocks like EZ and QPM mocks can produce discrepancy on the velocity statistics in real space up to 10\% on scales below $\simeq$40$h^{-1}{\rm Mpc}$ which will directly affect the anisotropic clustering in redshift space. Figure~\ref{fig:realspace} (Section~\ref{sec:real-model}) demonstrates that a full N-body simulation such as the OuterRim dataset is required to reproduce the clustering and velocity statistics at the scales of interest.

MultiDark simulations~\citep{MultiDark} were used at an earlier stage to study the performance of the model. They greatly assisted our understanding of the requirements in terms of box size and mass resolution. These simulations are at the limit for being used for the eBOSS quasar sample~\citep[see ][]{Comparat+2017}. We do not report results here although we found that infall velocities agreed well with the model in the quasi-linear regime up to scales of 60~$h^{-1}{\rm Mpc}$ for BigMDPL run of the MultiDark suite.

In the present work, we use the OuterRim N-body simulations~\citep{OuterRim} to evaluate the model of RSD. The volume of the OuterRim simulation is a cube of side $L=3000$~$h^{-1}{\rm Mpc}$ with 10240$^3$~dark matter particles and the force resolution is 6~$h^{-1}{\rm kpc}$. The mass resolution of a dark matter particle is ${\rm m}_p=1.82\,10^{9}\M_{\odot}h^{-1}$ and therefore halos hosting quasars of a typical mass of $\M=10^{12.5}\M_{\odot}$ are well resolved. The cosmological parameters are: $\Omega_{\rm cdm}h^2=0.1109$, $h=0.71$, $\sigma_8=0.8$, $\Omega_{\rm b}h^2=0.02258$, and $n_s=0.963$, and are consistent with the WMAP7 cosmology~\citep{WMAP7}. The initial conditions are calculated at $z$=200 using the Zel'dovich approximation.  

We build the mocks from a single snapshot at $z=1.433$ for which halos of more than 20 particles are available. In a first approach (dubbed ``mass bin"), we consider that only halos with mass $\M=10^{12.5\pm0.3}\M_{\odot}$ can host a quasar. In a refined approach, we apply a (5+1)-parameter Halo Occupancy Distribution using the parameters derived for the QPM mocks. For each halo, we determine the concentration from the halo mass using an ad-hoc parameterization of the data described in~\citet{Ludlow+2014}. The position of the satellites and their velocity are drawn from a profile according to the NFW prescription~\citep{NFW}. The fraction of satellites can be increased to $f_{\rm sat}=25$\% by setting $\M_{\rm sat}=10^{15}\M_{\odot}$ in the HOD model. 

We take advantage of the fact that the eBOSS quasar measurement is shot noise dominated, and that the duty cycle for quasars is low, to draw many realizations (up to 100) from the same parent box. We verified that increasing the duty cycle up to $\tau=10\%$ does not change the clustering signal.

Finally, the cartesian coordinates are transformed in right ascension, declination and redshift using OuterRim cosmology. Angular cuts are applied and an area of 1888~deg$^2$ with uniform redshift coverage can be selected. Finally, objects are downsampled in order to match the redshift distribution observed in the data. However, since the OuterRim mocks catalogs have been created from a single snapshot at $z=1.433$, and since we are just interested in evaluating the performance of the RSD model, we apply the redshift cut $0.8\leq z \leq 2.0$ to produce an effective redshift that matches the one of the single snapshot. It also allows us to compare, at the same effective redshift, the real space results using directly the output of the OuterRim simulation and the results in redshift space after applying the procedure we have just described.





\section{The CLPT-GS model}
 \label{sec:model}
 
 \subsection{Modeling the two-point correlation function}
 
 Redshift surveys provide a three-dimensional view of the large-scale structures of the universe whose statistical properties can be studied by modeling the two-point correlation function~\citep[see][for reviews on measuring large-scale structures]{Peebles1980,Bernardeau+2002}.
However, since the redshift we measure from spectroscopic surveys and from which we infer distance contains both a contribution from the Hubble expansion and the LOS velocity, galaxy redshift surveys actually measure a combination of the density and velocity fields in redshift space. Therefore, the two-point correlation in redshift space includes at least three types of non-linearities that are challenging to model theoretically: the non-linear evolution of density and velocity fields, the non-linear mapping from real to redshift space, and the non-linear relation between dark matter and tracers distribution. \\

The linear theory formalism was first derived by \citet{Kaiser87} ~\citep[see][for its development in configuration space]{Hamilton1998}; but its validity is limited to large scales (s $>$ 80~$h^{-1}{\rm Mpc}$) since it does not account for non-linear evolution as density fluctuations grow at smaller scales. It assumes a linear coupling between the density and the velocity fields, $\theta = -f \delta_{m}$, where the coupling factor $f$, the linear growth rate of structure, is scale-independent in the $\Lambda-$CDM+GR model. In the expression above, $\theta =  \nabla \cdot \vec{v}$, the divergence of the velocity field, is assumed to be irrotational and $\delta_{m}$ is the underlying matter density field. On large scales, where linear perturbation theory is valid, the background solution produces independent $\vec{k}$-modes evolution so that it is more natural to work in Fourier space. On smaller scales, and especially once non-linear couplings of the density and velocity fields become important, the choice of approach is not so clear.  Therefore, a variety of methods have been developed to model the galaxy clustering statistics on intermediate quasi-linear scales (20-80~$h^{-1}{\rm Mpc}$) using different perturbation theory (PT) models to reach beyond the linear theory~\citep[see][for a review and a comparison of different PT models for RSD]{Carlson+2009,White+2015}. 
The applicability of PT to interpret results from galaxy surveys is based on the fact that in the gravitational instability scenario, density fluctuations become small enough at intermediate scales (the "weakly non-linear regime") that a perturbative approach suffices to understand their evolution.

We choose to work in the Lagrangian framework where a perturbative expansion in the displacement field $\vec{\Psi} = \vec{\Psi}^{(1)} + \vec{\Psi}^{(2)} + \vec{\Psi}^{(3)} + \cdots$ is performed. In the Lagrangian picture, $\vec{\Psi}$ relates the Eulerian (final) coordinates $\vec{x}$ and Lagrangian (initial) coordinates $\vec{q}$ of a mass element or discrete tracer object:
\begin{equation}
\vec{x}(\vec{q}, t) = \vec{q} + \vec{\Psi}(\vec{q},t)\ .
\label{eq:lagriangian-disp}
\end{equation}
The first order solution to the perturbative expansion for $\vec{\Psi}$ is the well-known Zel'dovich approximation~\citep{Zeldovich1970},~\cite[see][for a recent use for RSD modeling]{White2015}. To move beyond the first order, we use the Convolution Lagrangian Perturbation Theory (CLPT) developed by \citet{Carlson+2013} which improves the work done by \citet{Matsubara2008a} by resumming more terms in the perturbative expansion. Despite the fact that CLPT dramatically improves the description of correlation function in real space, it remains inaccurate on quasi-linear scales for the quadrupole in redshift space. To overcome this deficiency, \citet{Wang+2014} extended the formalism to include the calculation of velocity moment statistics such as the pairwise infall velocity $v_{12}$ and pairwise velocity dispersion $\sigma_{12}$. These two ingredients with the correlation function in real space $\xi(r)$ are used as inputs in a Gaussian-Streaming (GS) model proposed by~\citet{RW2011}. The idea of a streaming model was first introduced by~\citet{Peebles1980} where the linear correlation function in real space is convolved along the LOS with a pairwise velocity distribution~\citep{Fisher1995,Peacock1994}). This work was extented by~\citet{Scoccimarro2004} and~\citet{RW2011}, leading to the following real to redshift space mapping encoded by the probability $\mathcal{P}$ that a pair of objects with real space LOS separation $r_{\parallel}$ will be observed with redshift space LOS separation $s_{\parallel}$ such that pair conservation implies:
\begin{equation}
  1+ \xi^{s}(s_{p}, s_{\parallel}) = \int_{-\infty}^{\infty} dy\
  [1+\xi(r)] \mathcal{P}( y = s_{\parallel} - r_{\parallel}, \vec{r} ).
\label{eq:GSmodel}
\end{equation}
This is the general expression of the Streaming model. In this work, $\mathcal{P}$ is assumed to be a Gaussian distribution centered at $y=\mu v_{12}(r)$ with dispersion $\sigma_{12}(r,\mu)$; we refer to this RSD model as the CLPT-GS model.

Given that all PT approaches are approximate methods to solve the dynamics of gravitational clustering, it is necessary to test the domain of validity of the theoretical predictions using numerical simulations. \citet{RW2011}~\citep[resp.][]{Carlson+2013,Wang+2014} tested the GS (resp. CLPT-GS) model using a set of N-body simulations presented in~\citet{White+2011} for halos of the appropriate mass range to host BOSS galaxies at redshift  $z \simeq 0.5$. 
In this work, we resort to the OuterRim simulation presented in Section~\ref{sec:mocks} from which we produce catalogs both in real and redshift spaces to investigate the performance of the CLPT-GS model for halos of masses of the order of $10^{12.5}{\rm M}_\odot$ which are hosting quasars at redshift $z \simeq 1.5$. 

Different sets of OuterRim catalogs have been produced in order to study the effect of satellite fraction and redshift smearing:
\begin{itemize}
 \item Satellite fraction: the presence of quasars hosted in satellite halos increases the amount of virialized objects within a halo; this increase modifies the small-scale clustering as it corresponds to a strong elongation of structures along the line of sight known as the Fingers-of-God~\citep[FoG,][]{Jackson72} effect. It also affects the amplitude of the clustering at all scales because of the dependence of the number of satellites with the mass of the halos. We study the case with $f_{\rm sat}=13\%$ satellite fraction as implemented in the QPM mocks but also the cases $f_{\rm sat}=0\%$ and $f_{\rm sat}=25\%$ for systematics checks.
 \item We apply two different redshift smearing models: a Gaussian redshift smearing according to the SRD and a redshift smearing according to the distribution $({z}_{\rm MgII}-{z})$ as seen in the data in Figure~\ref{fig:data-dz}. For the latter, we rescale the distribution so that the width matches the one of the SRD in order to focus the study on the effect of the exponential tails in the observed distributions.
\end{itemize}
In what follows, we first present the real space observables we can use to test quantitatively the CLPT predictions, then present the tests performed in redshift space to check the applicability of the CLPT-GS model for the RSD analysis of eBOSS DR14 quasar sample and estimate the systematic error related to the RSD modeling.

\begin{figure}
\includegraphics[width=90mm]{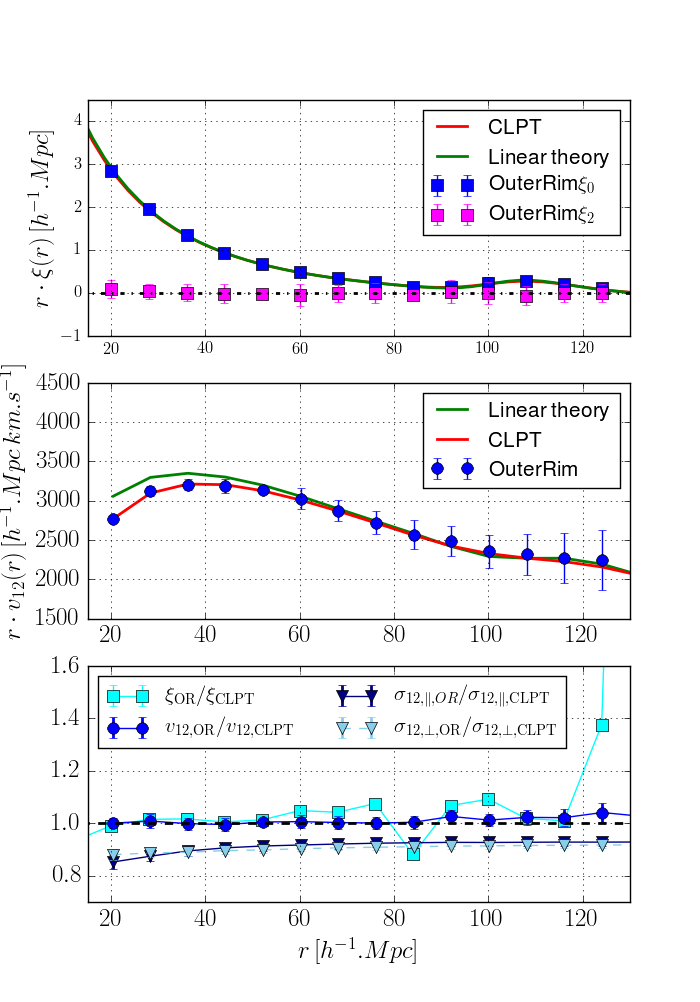}
\vskip -0.5cm
\caption{Real space observables for the case without satellite and redshift smearing. Blue points correspond to the results from the OuterRim simulation, the red curve is the CLPT prediction and the green one is the linear theory prediction. Top panel: correlation function in real space. Middle panel: pairwise infall velocity. Bottom panel: Ratio between OuterRim results and CLPT predictions for $\xi$ (square), $v_{12}$ (circle) and $\sigma_{12,\parallel,\perp}$ (triangles).}
\label{fig:realspace}
\end{figure}

\subsection{CLPT in real space}
 \label{sec:real-model}
 
CLPT gives predictions for tracers that are biased in a local Lagrangian formalism developed by \citet{Matsubara2008b} where the tracer density field, $\delta_{\rm tracer}$, is assumed to be a function, $F$, of a smoothed linear matter field $\delta_{m,R}$ at the same Lagrangian position $q$:
 \begin{equation}
  1 + \delta_{\rm tracer}(\vec{q}) = F[\delta_{m,R}(\vec{q})]\ ,
\end{equation}
The CLPT predictions for the clustering and velocity statistics depend on the first two Lagrangian parameters, $F'$ and $F''$, whose expressions can be found in \cite{Matsubara2008b}. On large scales the Eulerian bias $b$ is related to the first Lagrangian bias parameter by $b = 1 + F'$. Here, the local-Lagrangian bias scheme we use to connect the properties of tracers with the one of the underlying matter corresponds to a non-local bias in the Eulerian framework.


We inject the linear power spectrum corresponding to the OuterRim cosmology to the CLPT code \footnote{https://github.com/wll745881210/CLPT GSRSD.git} that calculates $\xi(r)$, $v_{12}$, $\sigma_{12}$ and compare the predictions with results from the OuterRim simulation.
Figure~\ref{fig:realspace} presents the agreement between the OuterRim results (blue points) for the catalog without satellites and the CLPT prediction (red) compared to the linear theory prediction (green) for the real-space correlation function, the mean infall pairwise velocity, and the velocity dispersion. 
The magenta curve in the top panel corresponds to the quadrupole of the correlation function in real space which is compatible with zero in the N-body simulation. It illustrates what we should measure if the redshift would accurately measure the radial distance from the observer in real space. However, the radial distances that are inferred from redshifts obtained in spectroscopic surveys also include components from peculiar velocities which give rise to the anisotropic clustering we observe, and thus to a non-zero quadrupole of the correlation function in redshift space.
The bottom panel displays the ratio between the results from the OuterRim simulation and the CLPT predictions. We confirm that at the mean redshift of eBOSS quasar sample, $z \simeq 1.5$, CLPT reproduces well the clustering and velocity statistics in real space for halos of masses of the order of $10^{12.5}{\rm M}_\odot$ on scales of interest (above $\simeq$20~$h^{-1}{\rm Mpc}$) which is in agreement with the \citet{Wang+2014} determination for ranges of halo masses that correspond to the BOSS LRG clustering at $z \simeq 0.5$.

Using the assumption of a linear coupling between the matter density field and the tracer velocity field , one can derive the linear theory prediction for the pairwise mean infall velocity:
\begin{equation}
v_{12}(r) = -r\frac{fb}{\pi^{2}} \int k P^{r}_{m}(k) j_{1}(kr) dk
\label{eq:v12}
\end{equation}
where $j_{1}(kr)$ is the first-order spherical Bessel function. We can see on the middle panel of Figure~\ref{fig:realspace} that the pairwise mean infall velocity measured on the N-body simulation deviates from the linear theory (green curve) on scales below $\sim$60~$h^{-1}{\rm Mpc}$. Given that it is directly proportional to the growth of cosmic structure, providing reliable cosmological constraint on this parameter requires precise modeling of the non-linear evolution of the matter and density fields.
A similar prediction can be derived for $\sigma_{12}$ \citep[both derivations can be found in][]{RW2011}. In this expression, the velocity field is assumed to be unbiased w.r.t. the matter density field. The effect of velocity bias was studied in~\citet{delaTorre+2012} and is expected to be of the order of a few percent for $f$. 

Equation~\ref{eq:v12} shows that the mean infall velocity is expected to be proportional to the bias and to the linear growth rate of structure on large scales. Since $\xi_{\rm tracer}=b^{2} \xi_{\rm m}$, another interesting test in real space is to check that the same bias value can reproduce both correlation function and infall velocity. Figure~\ref{fig:realspace} presents results for $F'=1.33$ which corresponds to $b\sigma_{8}=0.990$; this value is consistent with bias measurements in redshift space for the case without satellite and redshift smearing.
Here, the second bias parameter $F''$ is fixed under the peak-background split assumption~\citep{CK1989} using the Sheth-Tormen mass function~\citep[ST,][]{ST1999}. We show in Section~\ref{sec:redshift-model} the effect on the cosmological parameters of using the Press-Schetcher mass function (PS,~\citet{PS1974}) of setting $F''$ as a free parameter when fitting on observables in redshift space. 

The bottom panel of Figure~\ref{fig:realspace} also shows that the velocity dispersion terms parallel (dark blue triangles) and perpendicular (light blue triangles) to the separation of the pair are not well reproduced by CLPT. This issue has previously been discussed in~\citet{RW2011} and~\citet{Wang+2014}. We will see in Section~\ref{sec:redshift-model} that adding a constant shift to the CLPT predictions to match the velocity dispersion observed in OuterRim does not affect the cosmological parameters when fitting on observables in redshift space.

\subsection{CLPT-GS model in redshift space}
 \label{sec:redshift-model}

In this section, we investigate the response of model by fitting the redshift space correlation function of the mocks created from the OuterRim simulations and comparing the cosmological parameters to the expected values ($f\sigma_{8}=0.382$ and $\alpha_{\parallel}=\alpha_{\perp}=1$). When not specified, the reference uses a covariance matrix from the NGC EZ mocks without adding close-pairs or redshift failures treatment and is rescaled to match the statistics of the OuterRim catalogs. $F''$ is fixed according to the peak-background split assumption using the ST mass function, and the fit uses data from 16~$h^{-1}{\rm Mpc}$ to 138~$h^{-1}{\rm Mpc}$ with bin width of 8~$h^{-1}{\rm Mpc}$. The results of the fits are presented in Table~\ref{tab:OR}.

\subsubsection{Bias models and redshift smearing}

\begin{figure}
\includegraphics[width=85mm]{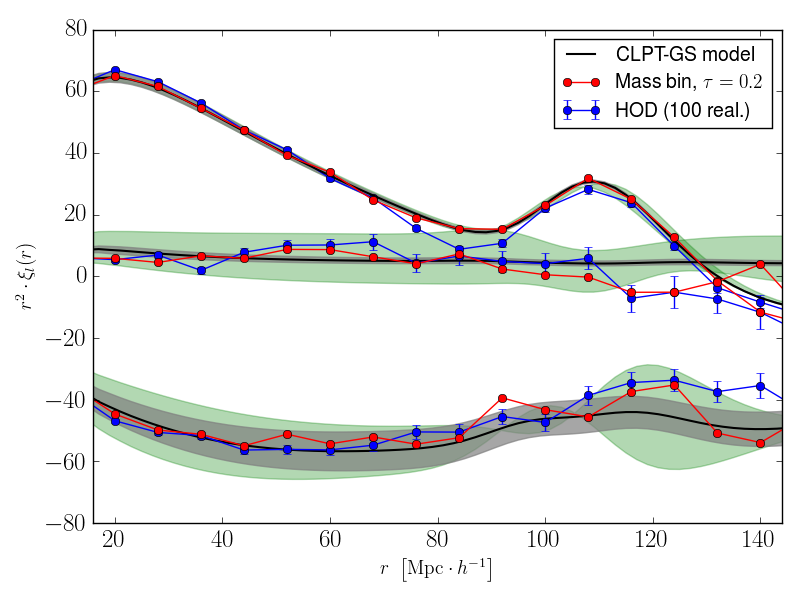}
\includegraphics[width=85mm]{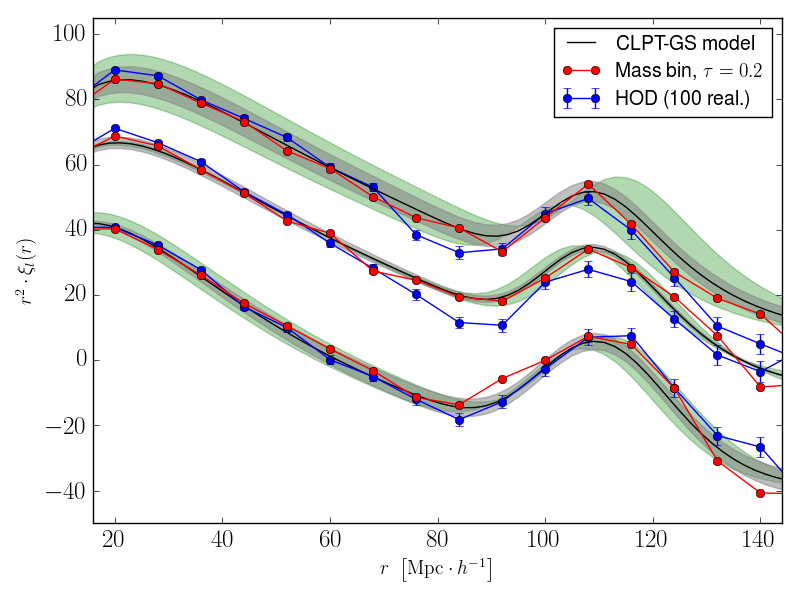}
\vskip -0.5cm
\caption{Top panel: Monopole of the correlation functions for the two bias models considered : ``mass bin" (red) and HOD (blue). For the HOD the data points are obtained from the average of 100 realisations. The CLPT model has been adjusted on the ``mass bin" points (solid line). The green band shows the effect of a $\pm$10\% variation of the parameter $F_{\rm AP}(\propto \alpha_\perp/\alpha_\parallel)$ and the grey band shows the effect of a $\pm$10\% variation of  $f\sigma_8$. Bottom panel: Same for the three wedges.}
\label{fig:OR-bias-models}
\end{figure} 

Figure~\ref{fig:OR-bias-models} compares the multipoles (top panel) and the wedges (bottom panel) of the correlation function for the ``mass bin" and the HOD biasing scenarios. For the latter, the data points are obtained from the average of 100 realisations. 
At the largest scales shown, the results from OuterRim tend to deviate from the predictions of the model in a region where these predictions do not differ from linear theory. This disappears due to the simulation box size effects, but we do not use scales larger than 138~$h^{-1}{\rm Mpc}$ in our fit range and the deviation is much smaller than the statistical precision of the data. For the monopole, it is clear that the ``mass bin" scenario is better reproduced by the model at all scales and that, in the region of the BAO feature, the HOD presents an unexpected behaviour. Therefore, with the present version of these mocks we may anticipate differences in the extracted geometrical parameters $\alpha_\parallel$ and $\alpha_\perp$. Furthermore, Figure~\ref{fig:OR-bias-models} reveals the impact of a $\pm$10\% variation of the parameter $F_{\rm AP}$ (green band) and a $\pm$10\% variation of  $f\sigma_8$ (grey band), showing that the quadrupole is equally sensitive to variations of $F_{\rm AP}(\propto \alpha_\perp/\alpha_\parallel)$ and $f\sigma_8$. But it also demonstrates that the hexadecapole is mostly sensitive to the variations of the geometrical parameters and hence will contribute to break this degeneracy. As expected, for the wedges, since the sum of the three wedges corresponds to the monopole, the effect is more degenerate among the three wedges and the wedge in the middle is the least affected as it probes pairs with intermediate angles between parallel and perpendicular to the LOS.

\begin{figure}
\includegraphics[width=90mm]{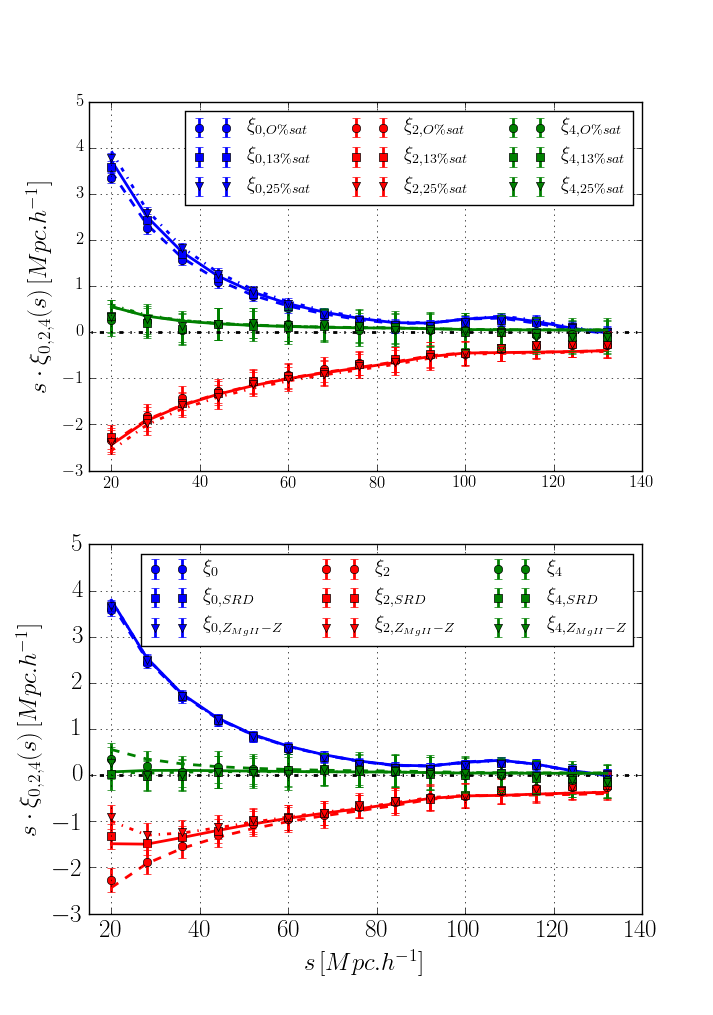}
\vskip -0.5cm
\caption{Top panel: Monopole (blue), quadrupole (red) and hexadecapole (green) for 3 satellite fractions without redshift smearing with the model set to the best fitting parameters for 0\% satellite (dashed line), 13\% satellite (solid line) and 25\% satellite (dashdot line). Bottom panel: Monopole (blue), quadrupole (red) and hexadecapole (green) for 3 redshift smearing and 13\% satellite with the model set to the best fitting parameters for no smearing (dashed line), SRD smearing (solid line) and $(z_{\rm MgII}-z)$ smearing (dashdot line).}
\label{fig:OR-configurations}
\end{figure}

For the HOD case, we varied the satellite fraction and present the measured monopole (blue), quadrupole (red) and hexadecapole (green) obtained in the top panel of Figure~\ref{fig:OR-configurations}. Increasing the satellite fraction mildly enhances the amplitude of the clustering, and the quadrupole and hexadecapole are almost unaffected. While no large difference between satellite fractions is seen in the mocks, previous analyses of the data tend to favour a satellite fraction around 0.15. This behaviour is shown in Figure 9 of~\cite{DR14-bao} which compares the projected quasar correlation function measurements to the HOD model we use in QPM mocks and in the OuterRim for the case $f_{sat}=0.15$. The exact satellite fraction for the halos hosting quasars however, is not known precisely, and is degenerate with the duty cycle of quasars that probably varies with luminosity and redshift. We therefore report the average value obtained for the three satellite fractions (0\%,13\% and 25\%) in Table~\ref{tab:OR} when estimating the systematic error related to the modeling. Further studies to constrain the dark matter halo mass and duty cycle using the final eBOSS quasar clustering measurements would provide a superior statiscal power to investigate these effects, following the approach developed in~\citet{Eftekharzadeh+15} and~\citet{Laurent+2017}. \\

In Table~\ref{tab:OR}, we report the results for the 3-multipole and 3-wedge analyses where small systematic shifts between the two methods can be observed at the level of $\Delta f\sigma_8=0.006$ , $\Delta \alpha_\parallel=0.005$ and $\Delta \alpha_\perp=0.006$.

We first investigate the response of the model when no smearing due to redshift error is applied. For all the cases considered, we observe a systematic shift of $f\sigma_8$ towards lower values and the maximum offset w.r.t the input cosmology is $\Delta f\sigma_8=-0.014$. For $\alpha_\parallel$, the maximum offsets for the HOD ($\Delta \alpha_\parallel=0.038$) is much larger than for the ``mass bin" ($\Delta \alpha_\parallel=0.016$); this situation probably arises from the difference observed on the monopole and demonstrates the need for a better understanding of the impact of the astrophysics conditions leading to the formation of quasars. For $\alpha_\perp$, the results are consistent with an offset smaller than $\Delta \alpha_\perp=0.006$. All these estimates receive contributions from both the biasing scenarios and from the modeling of the correlation function, but presently they should be viewed as global intrinsic systematic errors in our measurement.

The impact of redshift resolution is studied either by drawing the redshift from a Gaussian distribution according to eBOSS SRD (solid lines) or by drawing the redshift from the ``physical" distribution of (${z}_{\rm MgII}-{z}$) as shown in Figure~\ref{fig:data-dz}. For this comparison, the ``physical" distribution is rescaled such that the standard deviation is the same as for the Gaussian case, which allows for the estimation of the contribution of the tails in the redshift distribution. The bottom panel of Figure~\ref{fig:OR-configurations} reveals that, for the two types of smearing, the quadrupole and the hexadecapole are affected at scales below $\sim$50~$h^{-1}{\rm Mpc}$ and that the monopole is unaffected. It also shows that applying a more physical smearing has a larger effect on the quadrupole.

To account for redshift smearing in the RSD modeling, we add a constant dispersion velocity term to the width of the Gaussian distribution used for $\mathcal{P}$ in equation~\ref{eq:GSmodel} following the approach in~\citet{Reid+2012}:
\begin{equation}
\sigma^{2}_{12}(r,\mu) = \sigma^{2}_{12,\rm CLPT}(r,\mu) + \sigma^{2}_{\rm tot}\, .
\end{equation}
This additional term can be decomposed as $\sigma^{2}_{\rm tot}=\sigma^{2}_{\rm FoG} + \sigma^{2}_{z}$ where $\sigma_{\rm FoG}$ is produced by the Finger-of-God effect due to virialized motions of the quasars within their host halo and $\sigma_{z}$ arises from the smearing due to redshift resolution. However, the two parameters are degenerate, and in the model a single total nuisance parameter is used to represent this effect. 

A sizeable effect of the redshift smearing is observed for the parameter $f\sigma_8$ extracted from the fits as presented in Table~\ref{tab:OR}. For the cases considered, an average systematic shift of $\Delta f\sigma_8\simeq-0.010$  exists for the SRD smearing and an effect of $\Delta f\sigma_8\simeq-0.021$ for the physical redshift smearing. This systematic shift could, in principle, be reduced by using the actual shape of the redshift error distribution in a future modified streaming model. For $\alpha_\parallel$, there is a small compensation of the large effect seen for the HOD when applying the SRD smearing which is slightly reduced when using the physical smearing. For the ``mass bin", a similar behaviour is observed but remains smaller that for the HOD. No effect is seen on $\alpha_\perp$.

In summary, the overall systematic shifts due to the modeling of the full-shape anisotropic correlation function with our CLPT-GS model are $\Delta f\sigma_8=0.033$ , $\Delta \alpha_\parallel=0.038$, and $\Delta \alpha_\perp=0.006$ where we use the maximum deviation observed for the two biasing scenarios and the two redshift smearing options.

Table~\ref{tab:OR2} presents the results for more restrictive hypotheses on the cosmology where the cosmology is either fixed to the input cosmology of OuterRim or where we allow for an isotropic variation of the geometrical parameters, namely $\alpha_\parallel=\alpha_\perp=\alpha_{\rm iso}$. This test is performed using the physical redshift smearing in the case of the multipole analysis; for the HOD, we list the results for the set with $f_{\rm satt}=13\%$ that is favoured by the data. In these conditions, the systematic shift on $f\sigma_8$ w.r.t. the input cosmology is reduced to $\Delta f\sigma_8=0.017$; for the parameter $\alpha_{\rm iso}$, the maximum variation observed among all mocks that were produced is $\Delta \alpha_{\rm iso}=0.024$.


\begin{table*}
\caption{Impact on measured cosmological parameters for the different halo populating approaches and redshift smearing options. For the input cosmology  $f\sigma_{8}=0.382$ and $\alpha_{\parallel}=\alpha_{\perp}=1$.}
\label{tab:OR}
\begin{center}
\begin{tabular}{|c|c|c|c|c|c|c|c|c|}
 \hline
 \hline
 config                  &  smearing    & $b\sigma_{8}$         & $f\sigma_{8}({\rm OR}=0.382) $       & $\alpha_{\parallel}({\rm OR}=1.0)$  & $\alpha_{\perp}({\rm OR}=1.0)$& $\sigma_{\rm tot}$  \\
 \hline
 3-multipole           & & & & & & & \\
 \hline
HOD & no                             & $1.024\pm0.001$ & $0.377\pm0.002$ & $1.031\pm0.002$ & $1.001\pm0.001$ & $1.026\pm0.1$  \\
HOD & SRD                            & $1.024\pm0.001$ & $0.363\pm0.002$ & $1.021\pm0.002$ & $1.005\pm0.001$ & $5.48\pm0.03$  \\
HOD & $({\rm z}_{\rm MgII}-{\rm z})$ & $1.028\pm0.002$ & $0.355\pm0.003$ & $1.028\pm0.003$ & $0.998\pm0.001$ & $6.73\pm0.03$  \\
 \hline
mass bin & no     & $0.966\pm0.005$ & $0.377\pm0.006$ & $1.014\pm0.006$ & $1.002\pm0.005$ & $1.26\pm0.130$ \\
mass bin & SRD    & $0.971\pm0.005$ & $0.368\pm0.006$ & $1.011\pm0.007$ & $1.002\pm0.005$ & $5.60\pm0.040$  \\
mass bin & $({\rm z}_{\rm MgII}-{\rm z})$ & $0.976\pm0.005$ & $0.355\pm0.007$ & $1.025\pm0.008$ & $0.994\pm0.005$ & $6.84\pm0.036$ \\
\hline
 \hline
 3-wedge & &  & & & & &\\
 \hline
HOD & no                             & $1.025\pm0.001$ & $0.368\pm0.003$ & $1.038\pm0.002$ & $0.995\pm0.002$ & $1.58\pm0.1$ \\
HOD & SRD                            & $1.029\pm0.001$ & $0.360\pm0.003$ & $1.025\pm0.003$ & $1.003\pm0.002$ & $5.42\pm0.03$ \\
HOD & $({\rm z}_{\rm MgII}-{\rm z})$ & $1.031\pm0.001$ & $0.353\pm0.003$ & $1.025\pm0.003$ & $1.002\pm0.002$ & $6.55\pm0.03$  \\
 \hline
mass bin &no    & $0.968\pm0.005$ & $0.372\pm0.007$ & $1.016\pm0.007$ & $1.001\pm0.005$ & $1.34\pm0.130$\\
mass bin &SRD   & $0.974\pm0.005$ & $0.362\pm0.008$ & $1.012\pm0.008$ & $1.002\pm0.006$ & $5.58\pm0.043$  \\
mass bin &$({\rm z}_{\rm MgII}-{\rm z})$& $0.979\pm0.005$ & $0.349\pm0.008$ & $1.022\pm0.008$ & $0.996\pm0.006$ & $6.57\pm0.042$ \\
 \hline
\hline
 \end{tabular}
\end{center}

\caption{Comparison between different hypotheses on the cosmology : cosmology is fixed to the input of OuterRim ($ \alpha_{\parallel}=\alpha_{\perp}=1.$), isotropic case ($ \alpha_{\parallel}=\alpha_{\perp}=\alpha_{\rm iso}$), and anisotropic case ($\alpha_{\parallel}$ and $\alpha_{\perp}$). Results are given for physical redshift smearing and for the 3-multipole analysis}
\label{tab:OR2}
\begin{center}
\begin{tabular}{|c|c|c|c|c|c|c|c|}
config & cosmology & $b\sigma_8$ & $f\sigma_{8}$ & $\alpha_{\parallel}$ & $\alpha_{\perp}$ & $\sigma_{\rm tot}$ \\
\hline
\hline
mass bin                  & OuterRim    & $0.961\pm0.005$ & $0.370\pm0.005$ & fixed & fixed & $6.40^{+0.26}_{-0.27}$  \\
mass bin                  & isotropic   & $0.966\pm0.005$ & $0.371\pm0.006$ & $\alpha_{\rm iso}=1.005\pm0.005$ & -- & $6.44^{+0.28}_{-0.29}$ \\
mass bin                  & anisotropic & $0.976\pm0.005$ & $0.355\pm0.007$ & $1.025\pm0.008$ & $0.994\pm0.005$ & $6.84^{+0.36}_{-0.36}$ \\
\hline
HOD $f_{\rm sat}=13\%$   & OuterRim    & $1.015\pm0.001$ & $0.368\pm0.002$ & fixed & fixed & $6.50^{+0.82}_{-1.07}$  \\
HOD $f_{\rm sat}=13\%$   & isotropic   & $1.022\pm0.002$ & $0.366\pm0.002$ & $\alpha_{\rm iso}=1.005\pm0.001$ & - & $6.37^{+0.89}_{-1.07}$   \\
HOD $f_{\rm sat}=13\%$   & anisotropic & $1.027\pm0.002$ & $0.351\pm0.003$ & $1.027\pm0.002$ & $0.994\pm0.002$ & $6.73^{+1.38}_{-1.44}$ \\
\hline
HOD $f_{\rm sat}=0\%$    & OuterRim    & $0.980\pm0.001$ & $0.363\pm0.002$ & fixed & fixed & $6.05^{+1.07}_{-1.40}$ \\
HOD $f_{\rm sat}=0\%$    & isotropic   & $0.999\pm0.002$ & $0.367\pm0.002$ & $\alpha_{\rm iso}=1.017\pm0.001$ & - & $6.08^{+1.10}_{-1.38}$  \\
HOD $f_{\rm sat}=0\%$    & anisotropic & $1.006\pm0.002$ & $0.355\pm0.003$ & $1.035\pm0.003$ & $1.010\pm0.002$ & $6.35^{+1.63}_{-1.73}$ \\
\hline
HOD $f_{\rm sat}=25\%$   & OuterRim    & $1.046\pm0.001$ & $0.376\pm0.002$ & fixed & fixed & $6.48^{+0.72}_{-0.90}$ \\
HOD $f_{\rm sat}=25\%$   & isotropic   & $1.050\pm0.002$ & $0.374\pm0.002$ & $\alpha_{\rm iso}=1.002\pm0.001$ & - & $6.39^{+0.77}_{-0.88}$  \\
HOD $f_{\rm sat}=25\%$   & anisotropic & $1.052\pm0.002$ & $0.360\pm0.002$ & $1.020\pm0.002$ & $0.990\pm0.002$ & $6.72^{+1.04}_{-1.14}$\\
\hline

\hline
\end{tabular}
\end{center}

\caption{Additional tests performed when varying hypotheses on the second order bias parameter $F''$, on the total velocity dispersion $\sigma_{\rm tot}$, and on the lower bound of the fit range $r_{\rm min}$. These tests are performed for the ``mass bin" case with gaussian redshift smearing and for the multipole analysis.}
\label{tab:OR3}
\begin{center}
\begin{tabular}{|c|c|c|c|c|c|c|c|c|c|c|}
 config & hypothesis & $b\sigma_8$ & $f\sigma_{8}$ & $\alpha_{\parallel}$ & $\alpha_{\perp}$ & $\sigma_{\rm tot}$ \\
 \hline
 mass bin SRD & ref: uses $F''=F''_{\rm ST}$                      & 
 $0.971\pm0.005$ & $0.368\pm0.006$ & $1.011\pm0.007$ & $1.002\pm0.005$ & $5.60^{+0.38}_{-0.40}$ \\
 mass bin SRD & $F''=F''_{\rm PS}$                          &
 $0.969\pm0.005$ & $0.369\pm0.007$ & $1.011\pm0.007$ & $1.003\pm0.005$ & $5.49^{+0.40}_{-0.42}$ \\
 mass bin SRD & $F''_{\rm free}=-3.461^{+1.803}_{-1.239}$   & 
 $0.934\pm0.009$ & $0.376\pm0.007$ & $1.001\pm0.007$ & $1.000\pm0.005$ & $4.04^{+1.05}_{-1.12}$  \\
 mass bin SRD & $\sigma_{\rm tot}=5.7$~$h^{-1}{\rm Mpc}$ &
 $0.969\pm0.005$ & $0.369\pm0.007$ & $1.019\pm0.005$ & $1.000\pm0.005$ & fixed \\
 mass bin SRD & $\sigma_{\rm shift}$                     & $0.967\pm0.005$ & $0.370\pm0.006$ & $1.010\pm0.006$ & $1.003\pm0.005$ & $5.71^{+0.38}_{-0.40}$ \\
 mass bin SRD & $r_{\rm min}= 24$~$h^{-1}{\rm Mpc}$      & 
 $0.948\pm0.006$ & $0.369\pm0.007$ & $0.998\pm0.008$ & $0.997\pm0.005$ & $5.18^{+0.93}_{-1.09}$\\
 \hline
 
 \end{tabular}
 \end{center}
\end{table*}

%
%
%

\subsubsection{Additional tests}
Finally, we perform a series of tests for the mocks on the ``mass bin" case with SRD redshift smearing to study the impact of the ingredients of the model on the cosmological parameters. In particular, we examine the following effects whose results are summarized in Table~\ref{tab:OR3}:
\begin{itemize}
 \item $F''$: in the reference, $F''$ is fixed under the peak-background split assumption using the ST mass function and the result of the fit using the PS mass function is the same. When $F''$ is set free in the fit, small changes in the cosmological fit parameters are observed and are compatible with the variations of the statistical errors. The nuisance parameter $\sigma_{\rm tot}$ and its error are affected, suggesting a probable degeneracy with  $F''$. There is also an effect on the linear bias parameter with a shift $\Delta b\sigma_8=0.037$. We therefore do not report any bias measurement in the final results for this sample; further investigations on the bias models and prescriptions are needed for the final sample if we want to constrain the astrophysical properties of quasars using bias measurement from full-shape analysis.
\item $\sigma_{\rm tot}$: When fixing $\sigma_{\rm tot}=5.7 h^{-1}{\rm Mpc}$ (i.e., the average value of the SRD resolution used to create the mocks), the cosmological parameters of the simulation are recovered and the precision on $\alpha_\parallel$ is improved by 30\%. This result is achieved because, when fixing $\sigma_{\rm tot}=5.7 h^{-1}{\rm Mpc}$, the small scale statistical power is available for constraining $\alpha_\parallel$. Although this result should be viewed as a consistency check only, it demonstrates that a better knowledge of the redshift precision is important for the analysis of the full eBOSS quasar sample. 
 \item $r_{\rm min}$: Setting the lower bound of the fit range to $r_{\rm min}= 24$~$h^{-1}{\rm Mpc}$ instead of 16~$h^{-1}{\rm Mpc}$ produces almost no variation of $f\sigma_{8}$ and an effect on $\alpha_\parallel$ which is within the statistical precision. 
\item $\sigma^{2}_{12}(r,\mu)$ : Adding a constant shift to the CLPT predictions for the velocity dispersion to match the one observed in the OuterRim simulation in real space produces no effect on the measured cosmological parameters. The amplitude of the quadrupole depends mostly on the infall velocity $v_{12}$, and velocity dispersion variations are a second order effect. As expected, slightly changing the CLPT prediction on $\sigma_{12}$ has a negligible impact.
\end{itemize}

In light of this study, we conclude that the CLPT-GS model can be used for the clustering analysis of the eBOSS quasar sample at  $0.8\leq z \leq 2.2$ with overall systematic errors due to the modeling of the full-shape anisotropic correlation function of: 
\begin{equation}
\Delta f\sigma_8=0.033 \quad \Delta \alpha_\parallel=0.038  \quad \Delta \alpha_\perp=0.006
\end{equation}
The systematic errors between the 3-multipole and 3-wedge methods are similar and the errors reported are always the largest of the two possibilities.
For the analysis of the final eBOSS quasar sample, further work on improving the fidelity of the OuterRim-based mocks and understanding the difference in the bias models is needed. In particular, \citet{Vlah+2016} extended the CLPT-GS formalism to take into account contributions from Effective Field Theory (EFT) and additional bias terms. They showed that the effects of the biasing scheme are as important as higher-order corrections to the theoretical predictions. Therefore it would be interesting to see how this model performs for the analysis of the final eBOSS sample. Improvements in the model to account for the shape of the redshift error distribution would also be valuable.

\section{Analysis} 
\label{sec:analysis}

This section reviews the weighting scheme applied to the data to treat the potential systematic effects. We denote the total weighting scheme by ${\cal W_{\rm X}}$, where the subscript $\rm X$ specifies the different methods to compute the total weight.
With this notation, the total weighting scheme used for the DR14 quasar BAO analysis~\citep{DR14-bao} is:
\begin{equation}
{\cal W}_{\rm noz}=w_{\rm FKP} \cdot w_{\rm photo}\cdot  ( w_{\rm cp} + w_{\rm noz} - 1 ) 
\end{equation}
The first term, $w_{\rm FKP}=1/(1+n(z)P_0)$, is the FKP weight~\citep{FKP} that corrects for the variations of the observed quasar density $n(z)$ across the redshift range and also depends on the amplitude of the power spectrum at the scale at which the FKP weights optimize the measurements (here we choose~$k=0.14 h{\rm Mpc}$ which is the typical scale at which the BAO signal is optimally detected which gives $P_0=6 \times 10^{3} h^{-3}{\rm Mpc}^{3}$). The second term,$w_{\rm photo}$, is a photometry weight which corrects for the variation of the depth across the survey;  $w_{\rm cp}$, is a weight that accounts for the quasar targets that could not be measured due to fiber collision; and $w_{\rm noz}$ is a weight that adresses for the number of confirmed quasars for which a secure redshift could not be determined.
We will show that the use of the redshift efficiency, $\epsilon(x,y)$, across the focal plane as a weight, $w_{\rm focal}=1/\epsilon(x,y)$ is more appropriate than $w_{\rm noz}$. We adopt the following definition of the total weight ${\cal W}_{\rm focal}$: 
\begin{equation}
{\cal W}_{\rm focal}=w_{\rm FKP} \cdot w_{\rm photo} \cdot  w_{\rm cp} \cdot  w_{\rm focal} 
\label{eq:w_ref}
\end{equation}
Each quasar in the DR14 catalog is thus weighted by ${\cal W}_{\rm focal}$ to correct for any spurious variation of the quasar densities and to provide a more isotropic selection. For the random catalog, we apply the FKP weight alone as it corresponds to a Poisson sampling which should not be affected by inhomogeneities in the selection.
In tests defined throughout the following subsections, using the ability to test against unbiased samples, we will demonstrate that this new weight reduces systematic effects on the quadrupole by a factor of three. Furthermore, the region close to $\mu=1$ is responsible for the remaining systematic shift, and we propose a method to take this into account.

The fits are performed with the CLPT-GS model. The model prediction uses a linear power spectrum based upon the fiducial cosmology given in equation~\ref{eq:cosmo}, the second order bias parameter $F''$ is calculated according to the peak-background split using the Sheth-Tormen mass function. The fit is performed under the same conditions for the data and the EZ mocks from 16~$h^{-1}{\rm Mpc}$ to 136~$h^{-1}{\rm Mpc}$ using binwidth of 8~$h^{-1}{\rm Mpc}$. The covariance matrices are determined from the EZ mocks with a correction to equalize small differences in area. The priors on the fit parameters are :

\begin{center}
\label{tab:prior}
\begin{tabular}{|l||l|}
parameter & prior\\
\hline
$F'$               & flat prior, range (0.1, 2.8) \\
$f$                & flat prior, range (0,5) \\
$\alpha_\parallel$ & flat prior, range (0,2) \\
$\alpha_\perp$     & flat prior, range (0,2) \\
$\sigma_{\rm tot}$ & flat prior, range (0,20) \\
\end{tabular}
\end{center}

\subsection{Impact of photometric weights} 
\begin{figure} 
\includegraphics[width=84mm]{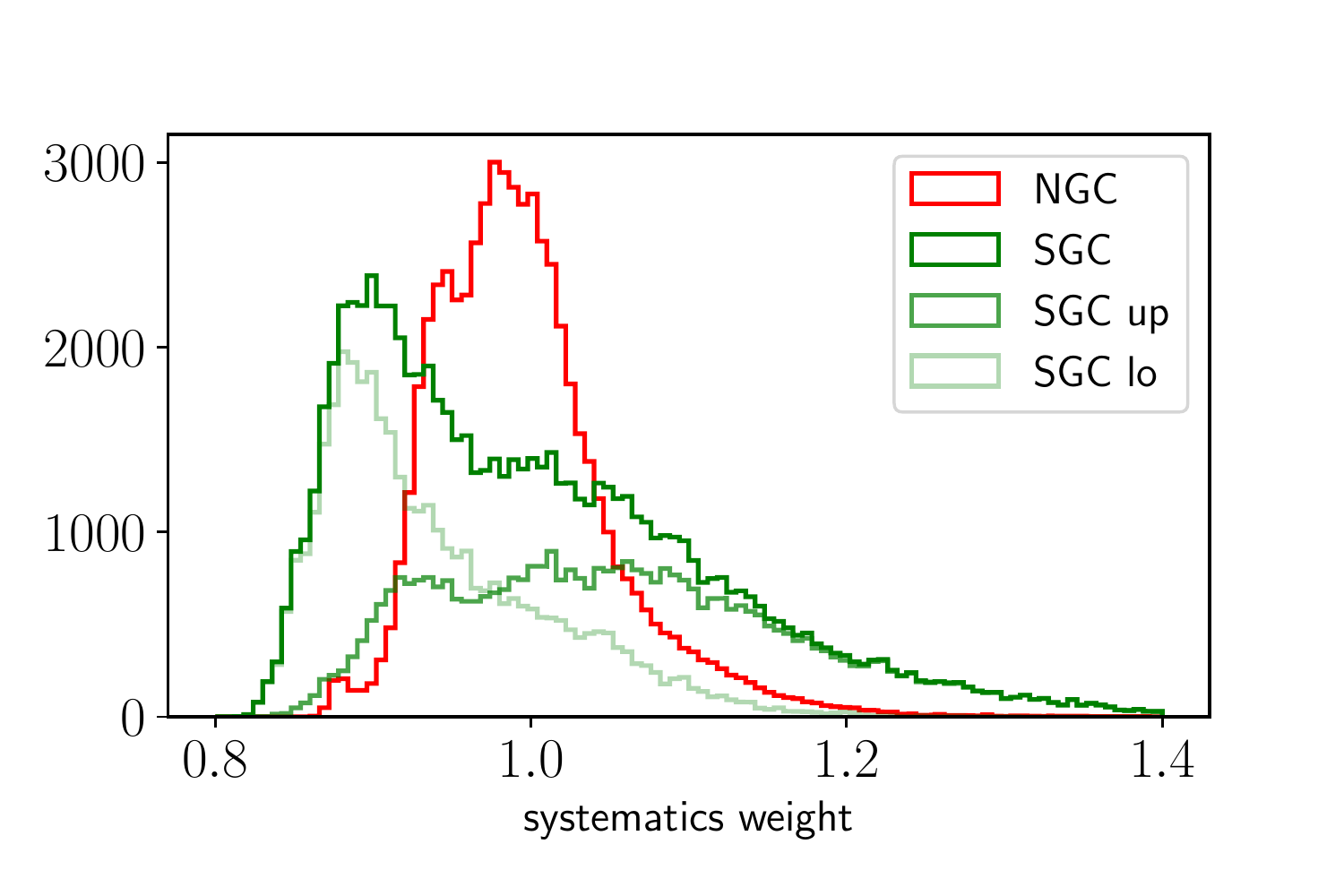} 
\includegraphics[width=84mm]{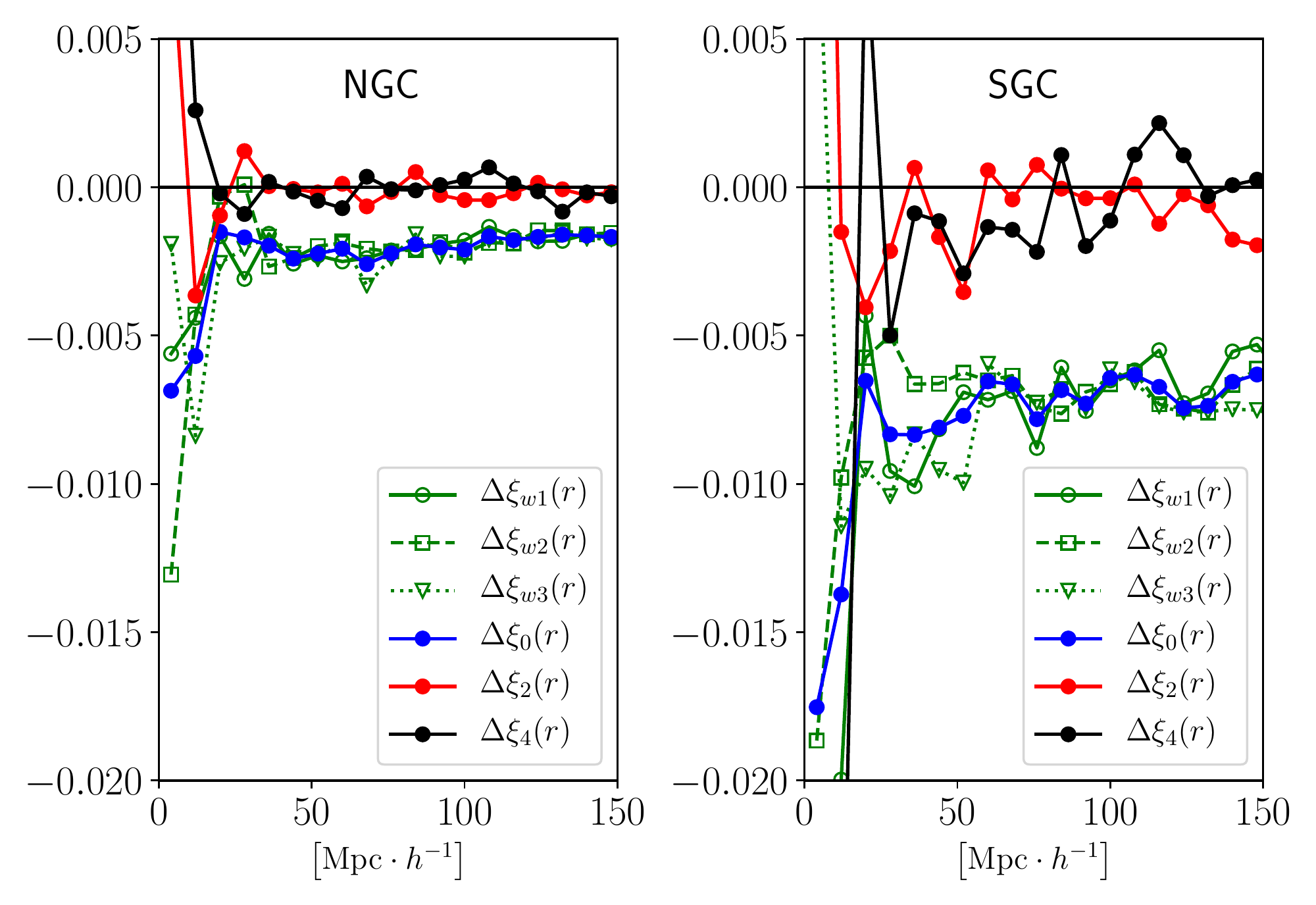} 
\hskip -0.5cm
\caption{Top panel: Distribution of the systematic weights applied to the data to correct for inhomogeneity in the depth of the photometric sample used at the targeting stage. Bottom plots: effect of photometric weights on the monopole, quadrupole and wedges of the correlation function for the NGC (left) and SGC (right). Note that the correlation function is not multiplied by $s$ on these plots.} 
\label{fig:sysweights} 
\end{figure}

The impact of the inhomogeneity of the quasar target selection on the observed quasar density was first studied on~\citet{Laurent+2017} using the early eBOSS quasar sample. Following the approach of~\citet{RossA+11,RossA+12,RossA+17} for BOSS analyses, they introduced a photometric weight according to the 5-$\sigma$ detection in magnitude for a point source, also called depth. By studying the variation of the observed quasar density as a function of depth which contains the dependence on airmass, seeing and Galactic extinction, one can compute photometric weights based on linear fits according to the dependency with the depth. These weights actually mitigate the systematic errors in the evaluation of the correlation function induced by the variation of the depth across the footprint.
We use the same weights as those presented in Section 3.4 of~\citet{DR14-bao} which have been computed for the DR14 sample with separate correction for each Galactic cap. 
The impact of the photometric weights on the clustering statistics is shown in Figure~\ref{fig:sysweights}. The top panel represents the distribution of photometric systematic weights computed for both Galactic caps, and for the two regions of the SGC separately; the spread of weights is much larger for the SGC. By computing the photometric weights in each cap separetely, we can correct for the variation in targeting efficiency due to differences in imaging properties. As explained in~\citet{Myers+15}, we expect the target selection to be more efficient in the NGC.
The bottom panels of Figure~\ref{fig:sysweights} show the impact of the photometric weights on the correlation function for the NGC (left panel) and SGC (right panel). The effect of weights on the correlation function is almost constant across the range of separation considered for this analysis. The effect on the correlation function for wedges in $\mu$ is similar for all wedges; as a consequence, the effects on the quadrupole and on the hexadecapole are small although some spread is observed in the correction for the SGC. As observed in~\cite{Laurent+2017}, the effect on the monopole is much larger for the SGC than for the NGC, but the corrected correlation function shows no remaining systematics within the current precision (e.g., the top panel of Figure~\ref{fig:results-correlation-function} in the next section).

Additional tests were conducted on the WISE photometry which also enters the target selection algorithm. We used the method developed in~\cite{Prakash+2016} to estimate the weights from the linear regression of the target density w.r.t. the photometric parameters including WISE, and no significant effect was observed. Moreover, the regions where there is some contamination from the moon (mostly in the SGC) were removed; this deletion produced no impact on our results.

\subsection{Spectroscopic completeness} 
\begin{figure*}
\includegraphics[width=180mm]{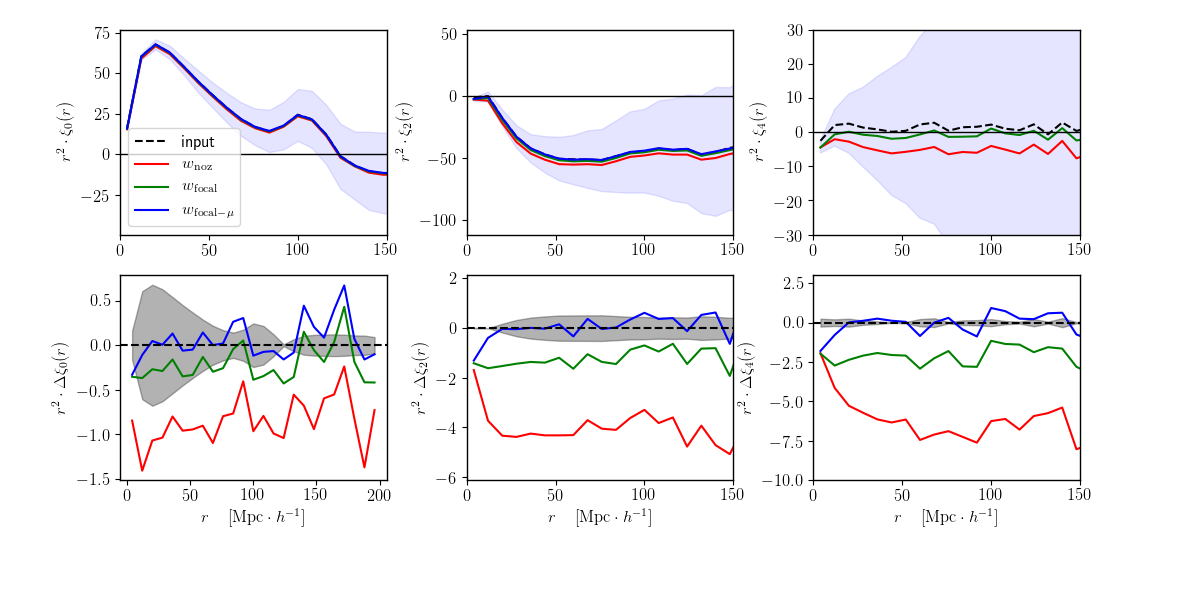} 
\vskip -0.8cm
\caption{Effect of the different weighting schemes on the correlation function multipole (left: monopole, middle: quadrupole, right: hexadecapole). ${\cal W}_{\rm noz}$ (red curves): upweighting of the nearest neighbour for redshift failures. ${\cal W}_{\rm focal}$ (green curves): weight according to the inverse of the spectroscopic efficiency. ${\cal W}_{{\rm focal-}\mu}$ (blue curves): same as ${\cal W}_{\rm focal}$ but the $\mu >1-1/480$ region is removed as described in the text. The light-blue shaded bands on the top plots represent the dispersion of the mocks. Bottom plots : difference between each weighting scheme and the input; the shaded bands represent a $\pm1\%$ effect for the monopole and the quadrupole and a $\pm10\%$ for the hexadecapole.} 
\label{fig:cut-mu} 
\end{figure*}

To study the impact of the spectroscopic completeness we use the special set of EZ mocks~(see Section~\ref{sec:mocks}) that  includes the redshift failures. Figure~\ref{fig:cut-mu} shows the difference between the measured correlation function to the correlation function without redshift failures and fiber collisions (both estimated with the EZ mocks). For the quadrupole, using the upweighting of the nearest neighbor (${\cal W}_{\rm noz}$, red curves) yields a systematic shift of 8\% at large scales. An effect is also observed on the monopole but, at first order, it only affects the bias determination. The hexadecapole displays a large effect, although the offset is well within the statistical precision. Results on the fit parameters for the 1,000 EZ mocks are summarized in Table~\ref{tab:weigthing} and exhibit a large shift (e.g. $\Delta f\sigma_8=0.105$) for the 3-multipole case which exceeds even the statistical precision of our measurement. For the 3-wedge analysis the shifts are smaller but still large w.r.t. our precision, especially on $f\sigma_8$.

In the proposed modified weighting scheme the observed quasars are weighted by the inverse of the efficiency calculated from the coordinates of the object in the focal plane, $w_{\rm focal}$. The results are preesnted (green curves) in Figure~\ref{fig:cut-mu}, which reveals a reduction of a factor three of the effect on the quadrupole. As a consequence the average shift estimated from the mocks is decreased to $\Delta f\sigma_8=0.033$ (resp. $\Delta f\sigma_8=0.013$) for the 3-multipole (resp. 3-wedge) analysis. The parameters $\alpha_{\parallel}$ and $\alpha_{\perp}$ are also shifted by $0.02$ in the case of the 3-multipole analysis, probably as a consequence of the sensitivity of the hexadecapole to these parameters.

\subsection{Weighting of close pairs} 

In previous analyses, unmeasured targets due to fiber collision are corrected by increasing by one unit the weight of the identified quasar in the collision group. This approach means that any target within 62" of a measured quasar will be displaced along the LOS and brought to the position of the measured quasar. This action inevitably creates a lack of objects at all scales and at $\mu\simeq1$ and hence will affect the correlation function evaluation. In their measurement of $f\sigma_8$ at small scales where the effect of collisions is large, \citet{Reid+2014} redefined the correlation function multipoles by excluding the region above a given threshold $\mu_s(s)$ depending on the separation $s$ and defined by the minimum angular distance between two objects (62$''$). Here we adopt a similar approach, but for simplicity we recalculate the value of the correlation function for the last $\mu$-bin in the same manner for all bins in separation.

At $z \simeq 1.5$, a 62'' radius exclusion corresponds to $\sim$0.4 Mpc; when considering scales larger than 20~$h^{-1}{\rm Mpc}$, we observed that pairs for which $1-\frac{1}{480}<\mu<1$ are affected by the upweighting due to close pairs. Since we use 30 $\mu$ bins in our analysis, the affected orientation correspond to 1/16 of the last $\mu$ bin. To mitigate this effect, we discard the paircounts in this region and rescale the counts of the last $\mu$ bin by 16/15. 

The results obtained after this correction was applied to the EZ~mocks are shown as the blue curves of Figure~\ref{fig:cut-mu}. With this method, for scales larger than 15~$h^{-1}{\rm Mpc}$, the true quadrupole is recovered to an accuracy better than 1\%, and no systematic behaviour is found on the monopole. The result on the cosmological parameters extracted from the fit of the 1000 EZ mocks with our model are in agreement with the reference with $\Delta f\sigma_8 < 0.001$, $\Delta \alpha_{\parallel} < 0.003$ and $\Delta \alpha_{\perp} < 0.001$. This method allows for a mitigation of the effect of fiber collisions and redshift efficiency variations across the focal plane at the level where it will not be a limitation even when the full eBOSS quasar sample will be available. 

\begin{table}
\caption{Effect on the EZ mocks of the different weighting schemes to mitigate systematic effects arising from spectroscopic completeness and fiber collisions. The values and the errors are obtained from 1000 realisations. The reference is given by the same set of mocks but where neither fiber collisions nor spectroscopic completeness are considered.}
\label{tab:weigthing}
\begin{tabular}{|c|c|c|c|}
3M AP  & $f\sigma_8$ ($\Delta f\sigma_8$)& $\alpha_{\parallel}$ ($\Delta \alpha_{\parallel}$) & $\alpha_{\perp}$ ($\Delta \alpha_{\perp}$)\\
\hline
reference                     & 0.3733$\pm$0.0022 & 0.9950$\pm$0.0023 & 0.9926$\pm$0.0020 \\

${\cal W}_{\rm noz}$          & +0.1050           & -0.0522           & 0.0559 \\
${\cal W}_{\rm focal}$        & +0.0338           & -0.0169           & +0.0184 \\
${\cal W}_{{\rm focal-}\mu}$  & -0.0003           & +0.0009           & -0.0007 \\
\hline
3W AP  & $f\sigma_8$ ($\Delta f\sigma_8$)& $\alpha_{\parallel}$ ($\Delta \alpha_{\parallel}$) & $\alpha_{\perp}$ ($\Delta \alpha_{\perp}$)\\
\hline
reference                    & 0.3784$\pm$0.0031 & 0.9966$\pm$0.0028 & 0.9963$\pm$ 0.0025 \\
${\cal W}_{\rm noz}$         & 0.0424            &-0.0086            & 0.0158 \\
${\cal W}_{\rm focal}$       & 0.0130            &-0.0007            & 0.0050 \\
${\cal W}_{{\rm focal-}\mu}$ &-0.0004            & 0.0029            &-0.0003 \\
\hline
\end{tabular}

\caption{eBOSS DR14 quasar sample : Effect on the data of the different weighting schemes  obtained from the 3-multipole and 3-wedge analyses. Differences are calculated w.r.t. the ${\cal W}_{{\rm focal-}\mu}$ case and a given in parentheses.}
\label{tab:tests-data2}
\begin{tabular}{|c|c|c|c|}
\hline
\hline
 3-multipole            & $f\sigma_{8} $ & $\alpha_{\parallel}$ & $\alpha_{\perp}$    \\
\hline
${\cal W}_{\rm noz}$       & $0.436^{+0.071}_{-0.072}$ & $0.999^{+0.078}_{-0.070}$ & $1.031^{+0.050}_{-0.048}$ \\
   & ($0.024$) & ($-0.015$) & ($0.006$) \\
\hline
 ${\cal W}_{\rm focal}$       & $0.426^{+0.070}_{-0.070}$ & $1.014^{+0.070}_{-0.063}$ & $1.030^{+0.050}_{-0.048}$ \\
  & ($0.014$) & ($0.000$) & ($0.005$) \\
\hline
${\cal W}_{{\rm focal-}\mu}$ & $0.412^{+0.069}_{-0.070}$ & $1.014^{+0.070}_{-0.062}$ & $1.025^{+0.049}_{-0.048}$ \\
  & $-$ & $-$ & $-$ \\
\hline
\hline
 3-wedge            & $f\sigma_{8} $ & $\alpha_{\parallel}$ & $\alpha_{\perp}$   \\
\hline
${\cal W}_{\rm noz}$     & $0.343^{+0.084}_{-0.088}$ & $1.089^{+0.141}_{-0.097}$ & $1.008^{+0.053}_{-0.053}$ \\
  & ($-0.021$) & ($0.035$) & ($-0.006$) \\
\hline
 ${\cal W}_{\rm focal}$     & $0.365^{+0.082}_{-0.083}$ & $1.064^{+0.107}_{-0.081}$ & $1.015^{+0.052}_{-0.052}$ \\
  & ($0.001$) & ($0.010$) & ($0.001$) \\
\hline
${\cal W}_{{\rm focal-}\mu}$ & $0.364^{+0.081}_{-0.081}$ & $1.054^{+0.101}_{-0.078}$ & $1.014^{+0.052}_{-0.052}$ \\
  & $-$ & $-$ & $-$ \\
\hline
\end{tabular}
\end{table}

\subsection{Tests on the data}

\subsubsection{Effect of weighting schemes}
The different weighting schemes are also applied on the data and the fits results are given in Table~\ref{tab:tests-data2}. The differences in the fits between the weighting schemes are found to be smaller than in the mocks. The largest differences between favoured schemes ${\cal W}_{{\rm focal-}\mu}$ and ${\cal W}_{\rm focal}$ are $\Delta f\sigma_8 = 0.014$ and $\Delta \alpha_{\parallel}=0.010$; these differences represent only ~20\% of the statistical precision. From the distribution of differences in the mocks what is observed in the data is not unusual, although an alternative explanation is that the effect of close pairs and redshift failures is somehow magnified in our improved set of EZ mocks. In the following we use ${\cal W}_{{\rm focal-}\mu}$ weighting scheme as our reference. For consistency with other analyses which do not employ this weighting scheme we will also present the results for the case ${\cal W}_{\rm focal}$ .

\subsubsection{Effect of redshift estimates}
As explained in Section~\ref{data-redshift}, we can use three redshift estimates to measure the clustering of the DR14 quasar sample. We adopt the redshift '${z}$' as the reference throughout this analysis and compare its results with catalogs where the redshift is taken to be ${z}_{\rm MgII}$ (resp. ${z}_{\rm PCA}$) whenever it is available (i.e. 80\% of the time) and '$z$' otherwise such that these catalogues have the same objects. The results in Table~\ref{tab:tests-data-anisotropic} are consistent within 1$\sigma$, although the results from ${z}_{\rm PCA}$ exhibit a stronger deviation than ${z}_{\rm MgII}$. This behaviour could be an argument in favor of the astrophysical motivations to use MgII-based redshift, since it is supposed to be the more systematics-free redshift estimate, but further investigation on the reliability of the MgII line across our entire redshift range is required before stating firm conclusions. In addition, these measurements that use redshift estimates should not be considered as independent, and we lack equivalent different redshift estimates for mocks we cannot simply combine the redshifts. 

\begin{figure}
\hskip -0.5cm
\includegraphics[width=100mm]{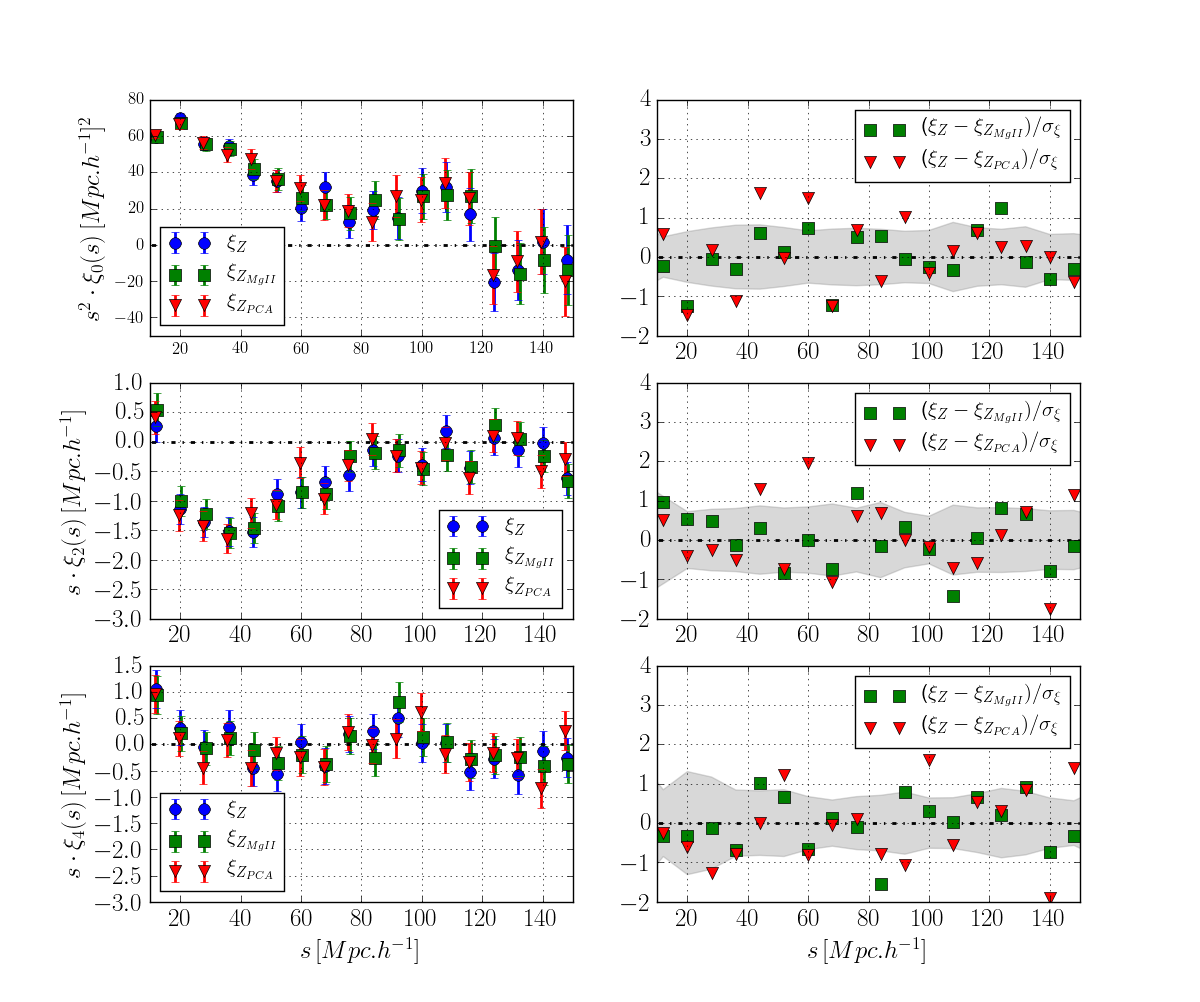}
\vskip -0.4cm
\caption{Top panels: Monopole (left) and quadrupole (right) for $z$ (blue), $z_{\rm MgII}$ (green) and $z_{\rm PCA}$ (red). Bottom panels: Difference $\xi_{z_{\rm MgII}} - \xi_{z}$ and $\xi_{z_{\rm PCA}} - \xi_{z}$ divided by the error using EZ mocks, compared to the dispersion of 30 realizations for the same mock with SRD redshift smearing. The differences in clustering are consistent with the expected dispersion from statistically independent redshift smearing.}
\label{fig:z-dispersion}
\end{figure}

Differences in clustering between ${\rm z}_{\rm MgII}$ (resp. ${\rm z}_{\rm PCA}$) w.r.t 'z' can also be compared to the dispersion due to different realizations of the same mock for a given redshift smearing. For a specific OuterRim mock catalog, we can draw several realizations of a given redshift smearing on the same mock. We investgate the case with 13\% satellite fraction and with SRD redshift smearing since it is the closest configuration to the data and we draw 30 realizations. We then examine the dispersion on the monopole and quadrupole, which corresponds to the grey envelope in Figure~\ref{fig:z-dispersion}. 
The differences in clustering using different redshift estimates lie within the dispersion expected from statistically independent redshift smearing and that they do not show any systematic trend. We conclude that difference between the results of the fit with the different redshift estimates are due to statistics and we do not quote an additional systematic error. 

\begin{table*}
\caption{Results of the fit of the data when changing cosmological assumptions, redshifts estimates, covariance matrix determination, and second order bias. The upper part (resp. lower part) of the table presents results for the 3-multipole (resp. 3-wedge) analysis. Results for each individual Galactic cap are given at the end of each table.}
\label{tab:tests-data-anisotropic}
\begin{tabular}{|c|l|c|c|c|c|c|c|}
config                        & $b\sigma_{8}$         & $f\sigma_{8} $       & $\alpha_{\parallel}$ & $\alpha_{\perp}$ & $\sigma_{\rm tot}$    & $\chi^{2}/ dof$ \\
\hline
3-multipole & & & & & & \\
NGC+SGC ref             & $1.042^{+0.059}_{-0.056}$ & $0.412^{+0.069}_{-0.070}$ & $1.014^{+0.070}_{-0.062}$ & $1.025^{+0.049}_{-0.048}$ & $6.00^{+1.19}_{-1.41}$ & $40.5 (45-5) $ \\
NGC+SGC isotropic & $1.044^{+0.056}_{-0.053}$ & $0.406^{+0.054}_{-0.053}$ & $\alpha_{\rm iso}=1.021^{+0.039}_{-0.037}$ & -- & $6.10^{+0.89}_{-1.12}$ & $40.5 (45-4) $ \\
NGC+SGC fiducial  & $1.019^{+0.030}_{-0.030}$ & $0.398^{+0.050}_{-0.051}$ & fixed & fixed & $5.91^{+0.81}_{-1.09}$ & $40.9 (45-3)$ \\
NGC+SGC covQPM     & $1.054^{+0.068}_{-0.060}$ & $0.386^{+0.069}_{-0.071}$ & $1.055^{+0.083}_{-0.068}$ & $1.022^{+0.051}_{-0.049}$ & $6.60^{+1.21}_{-1.35}$ & $39.9$ \\
NGC+SGC '${\rm z}_{\rm PCA}$'  & $0.997^{+0.066}_{-0.065}$ & $0.387^{+0.072}_{-0.073}$ & $0.988^{+0.085}_{-0.080}$ & $1.005^{+0.048}_{-0.048}$ & $5.37^{+1.36}_{-1.74}$ & $39.3$ \\
NGC+SGC '${\rm z}_{\rm MgII}$'  & $0.966^{+0.074}_{-0.066}$ & $0.424^{+0.070}_{-0.073}$ & $0.972^{+0.095}_{-0.079}$ & $0.994^{+0.052}_{-0.049}$ & $6.27^{+1.33}_{-1.38}$ & $31.8$ \\ 
NGC+SGC  $\sigma_{\rm tot}$ fixed & $1.042^{+0.057}_{-0.055}$ & $0.412^{+0.069}_{-0.070}$ & $1.014^{+0.051}_{-0.045}$ & $1.025^{+0.049}_{-0.047}$ & $6.00$ (fixed) & $40.5$ \\
NGC+SGC $F''_{PS}$ & $1.042^{+0.059}_{-0.056}$ & $0.412^{+0.069}_{-0.069}$ & $1.015^{+0.070}_{-0.062}$ & $1.025^{+0.049}_{-0.047}$ & $6.06^{+1.18}_{-1.39}$ & $40.5$ \\
NGC AP & $0.960^{+0.083}_{-0.076}$ & $0.440^{+0.083}_{-0.084}$ & $0.950^{+0.102}_{-0.083}$ & $0.992^{+0.058}_{-0.054}$ & $6.30^{+1.50}_{-1.54}$ & $28.8$ \\
SGC AP & $1.142^{+0.085}_{-0.078}$ & $0.383^{+0.097}_{-0.096}$ & $1.048^{+0.100}_{-0.077}$ & $1.086^{+0.071}_{-0.072}$ & $3.68^{+2.61}_{-3.68}$ & $55.0$ \\
\hline

\hline
\hline
3-wedge & & & & & & \\
NGC+SGC ref              & $1.069^{+0.066}_{-0.059}$ & $0.364^{+0.081}_{-0.081}$ & $1.054^{+0.101}_{-0.078}$ & $1.014^{+0.052}_{-0.052}$ & $6.08^{+1.57}_{-1.74}$ & $37.8 (45-5)$ \\
NGC+SGC isotropic & $1.060^{+0.055}_{-0.054}$ & $0.385^{+0.057}_{-0.056}$ & $\alpha_{\rm iso}=1.027^{+0.039}_{-0.037}$ & -- & $5.70^{+1.07}_{-1.42}$ & $38.0 (45-4)$ \\
NGC+SGC fiducial & $1.028^{+0.031}_{-0.031}$ & $0.374^{+0.053}_{-0.054}$ & fixed & fixed & $5.42^{+1.00}_{-1.42}$ & $38.5 (45-3)$ \\
NGC+SGC covQPM     & $1.042^{+0.080}_{-0.063}$ & $0.383^{+0.078}_{-0.078}$ & $1.029^{+0.119}_{-0.078}$ & $1.027^{+0.056}_{-0.055}$ & $6.01^{+1.75}_{-1.84}$ & $42.4$ \\                      
NGC+SGC '${\rm z}_{\rm PCA}$'  & $1.064^{+0.067}_{-0.075}$ & $0.277^{+0.097}_{-0.092}$ & $1.130^{+0.113}_{-0.123}$ & $0.963^{+0.054}_{-0.055}$ & $6.85^{+1.65}_{-1.96}$ & $44.1$ \\
NGC+SGC '${\rm z}_{\rm MgII}$'  & $1.027^{+0.097}_{-0.115}$ & $0.362^{+0.124}_{-0.113}$ & $1.094^{+0.168}_{-0.181}$ & $0.975^{+0.059}_{-0.055}$ & $7.71^{+2.07}_{-2.50}$ & $38.6$ \\ 
NGC+SGC  $\sigma_{\rm tot}$ fixed & $1.069^{+0.062}_{-0.058}$ & $0.364^{+0.081}_{-0.080}$ & $1.054^{+0.066}_{-0.056}$ & $1.014^{+0.050}_{-0.050}$ & $6.09$ (fixed) & $37.8$ \\
NGC+SGC $F''_{PS}$ & $1.070^{+0.066}_{-0.059}$ & $0.364^{+0.081}_{-0.081}$ & $1.055^{+0.101}_{-0.078}$ & $1.014^{+0.052}_{-0.052}$ & $6.13^{+1.56}_{-1.71}$ & $37.8$ \\
NGC AP & $1.001^{+0.146}_{-0.090}$ & $0.373^{+0.108}_{-0.133}$ & $1.007^{+0.249}_{-0.122}$ & $0.973^{+0.061}_{-0.059}$ & $6.63^{+2.99}_{-2.12}$ & $28.9$ \\
SGC AP & $1.143^{+0.087}_{-0.081}$ & $0.387^{+0.120}_{-0.110}$ & $1.032^{+0.109}_{-0.081}$ & $1.094^{+0.085}_{-0.081}$ & $3.14^{+3.08}_{-3.14}$ & $46.8$ \\
\hline
\end{tabular}
\end{table*}

\subsubsection{Additional tests}

Table~\ref{tab:tests-data-anisotropic} summarizes the different tests we perform on data to compare to the reference and to study the robustness of our measurements. In particular, we review at the following effects:
\begin{itemize}
\item Isotropic analysis: As a consistency check, using $\alpha_{\rm iso} \simeq \alpha_{\parallel}^{1/3} \alpha_{\perp}^{2/3}$ with the reference values from the anisotropic fitting of the three multipoles for instance for $\alpha_{\parallel}$ and $\alpha_{\perp}$, we compute $\alpha_{\rm iso}=1.021 \pm 0.057$, which matches well the result from the isotropic fit. The effect on $f\sigma_{8}$ is also consistent, and no significant shift is reported.
\item Fixing the fiducial cosmology produces consistent results with the anisotropic and isotropic cases, and as expected given the degeneracy between the AP parameters and $f\sigma_{8}$, this approach provides a better constraint on $f\sigma_{8}$. However, if one wishes to constrain modified gravity models based on different assumptions than the one of $\Lambda$CDM-GR for structure formation, one must use the results obtained by the full anisotropic clustering using AP parameters.
\item Effect of covariance matrix: The QPM mocks are used to compute another covariance matrix, and there is no significant effect on the cosmological parameters $f\sigma_{8}$, $\alpha_{\parallel}$ and $\alpha_{\perp}$.
\item Effect of redshift resolution: When fixing $\sigma_{\rm tot}$ to the best-fitting values, the precision on $\alpha_\parallel$ is improved by 30\% as seen in tests on the OuterRim catalogs. This results provides clear motivation to improve our knowledge of the redshift uncertainty for future quasar samples.
\item Effect of $F''$ prescription: as shown in the model, there is no significant difference on the fitted cosmological parameters when using PS mass function instead of ST. We do not report any result when letting $F''$ free because, as explained in Section~\ref{sec:redshift-model}, we are not sufficiently sensitive to this parameter to derive useful constraints. In addition, since $F''$ accounts for non-linearities in the bias model at small scales, it may be degenerated with $\sigma_{tot}$.
\end{itemize}

\citet{Hector} also investigated the redshift evolution of the parameters across three redshift bins and reported no significant redshift-dependence on $f\sigma_{8}$ given the current statistical precision. Alternatively, one can use a different parametrization for the cosmological parameters such as proposed by \citet{Zhu+15} and more recently adapted for RSD in~\citet{Ruggeri+17a} and validated on mocks in~\citet{Ruggeri+17b}. We expect this optimal redshift weighting technique to provide tighter constraints on the final eBOSS sample, where it will be possible to compare results from that technique to results from multipoles and wedges decomposition by sub-dividing the full redshift range.

\subsection{Consistency between NGC and SGC}

\begin{figure*} 
\includegraphics[width=70mm]{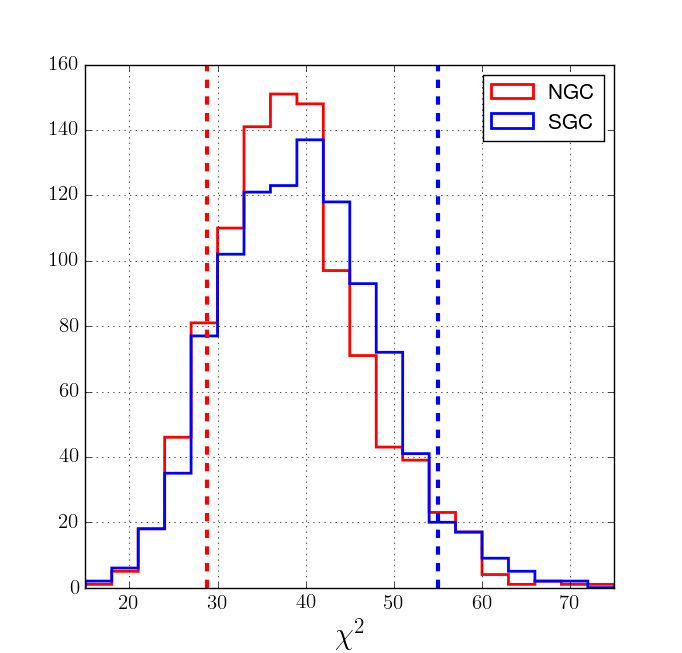}
\includegraphics[width=70mm]{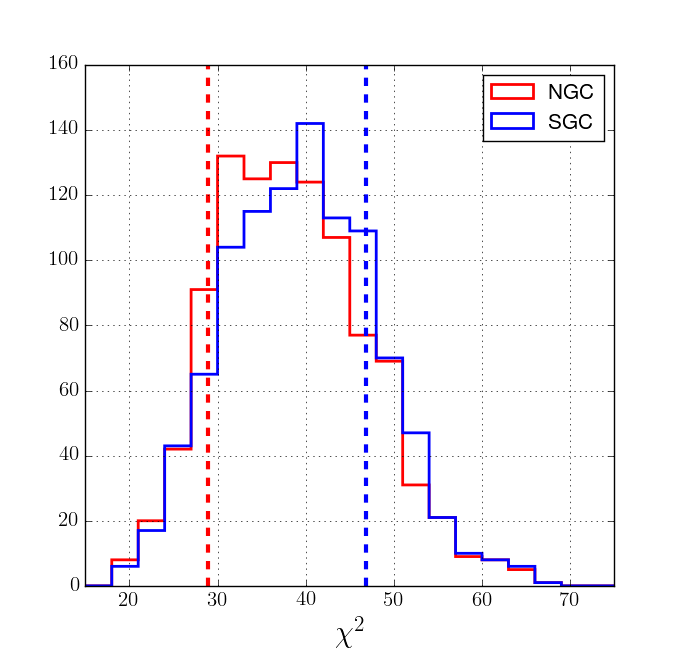}
\caption{Distribrution of the $\chi^2({\rm d.o.f}=40)$ of the fits of the 1000 EZ mocks per Galactic cap (solid line) and comparison with the $\chi^2$ obtained from the data (dashed line). Left: for 3-multipole. Right: for 3-wedge. The $\chi^2$ on the data are found to be within the distribution of the EZ mocks, even for the $\chi^2$ in the SGC (blue dashed line) which is larger for the 3-multipole analysis. Results on each cap thus represent a statistical realization of the EZ mocks.} 
\label{fig:chi2}  
\end{figure*}

The results of the fits performed on the two Galactic caps separately are given in Table~\ref{tab:tests-data-anisotropic} for the 3-multipole and 3-wedge fits. The fit parameters are in agreement, although the $\chi^2$ of the fit of the SGC using 3-multipole reaches $\chi^2=55.0({\rm d.o.f}=40)$. We conducted extensive tests in order to isolate a potential source for this effect. This increase in $\chi^2$ has been located in the $\delta>10°$ area of the SGC which is the region where the spread of the photometric weights distribution is the greatest. After removing regions of extreme values of the systematic weights, with moon contamination in WISE photometry, or regions of high Galactic extinction, no obvious source could be identified. \\
Figure~\ref{fig:chi2} compares the $\chi^2$ on the data in each cap (dashed line) with the $\chi^2$ distribution obtained for the results of the 1,000 EZ mocks (solid) by cap (NGC in red and SGC in blue) for the 3-multipole (top panel) and 3-wedge (bottom panel) fits. As described in Section~\ref{sec:mocks}, the NGC and SGC EZ mocks are created from separate simulations whose bias parameters have been adjusted on the observed DR14 eBOSS quasar clustering on each cap directly. As is clearly visible in Figure~\ref{fig:chi2}, the $\chi^2$ in the SGC (blue dashed) for the 3-multipole analysis is large but is not unusual compared to the EZ mocks distribution. \\

\section{Results}
\label{sec:results}
 
We now present the main results of this work. We compare the results between the 3-multipole and 3-wedge analyses for the data and for the EZ mocks, and examine the consistency between this work and the companion analyses as an excellent evidence of the robustness of the clustering measurements using the eBOSS DR14 quasar sample.


\subsection{Consistency between 3-multipole and 3-wedge analyses}

\begin{figure}
\includegraphics[width=90mm]{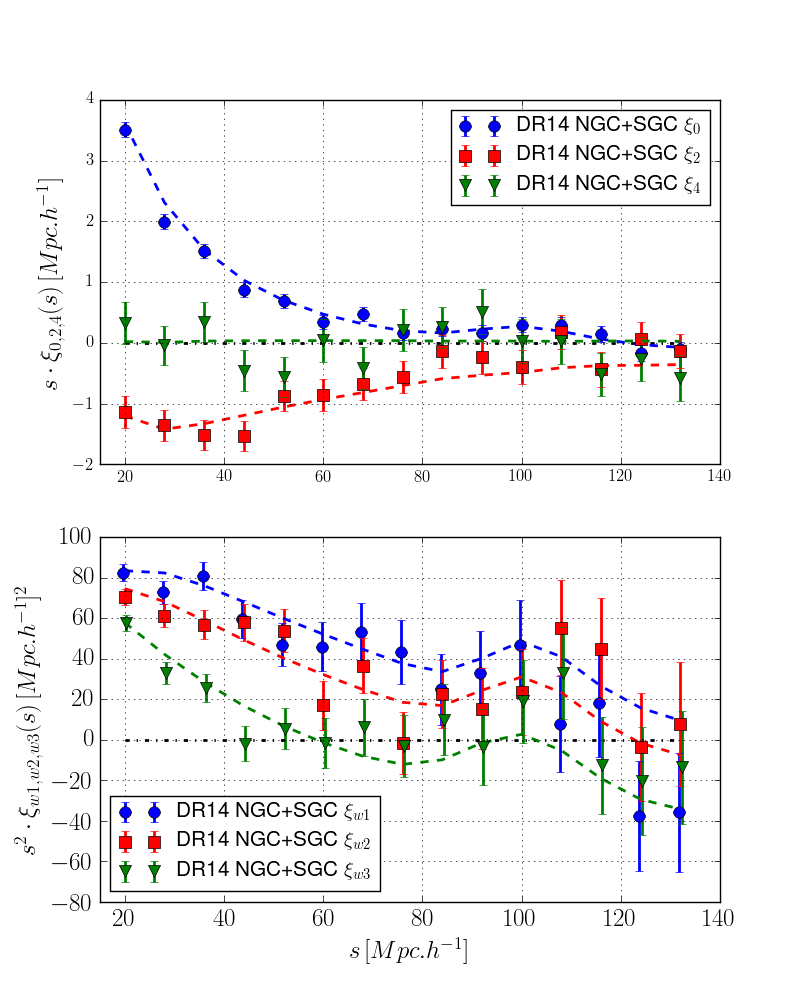}
\vskip -0.5cm
\caption{Top panel: Monopole (blue) and quadrupole (red) and hexadecapole (green) of correlation function of the NGC+SGC eBOSS DR14 quasar sample fitted using the CLPT-GS model (dashed line) set to the best-fit parameters. Bottom panel: Same for the three wedges: 0$<\mu<$1/3 (blue), 1/3$<\mu<$2/3 (red) and 2/3$<\mu<$1 (green).The fit is performed from 16~$h^{-1}{\rm Mpc}$ to 136~$h^{-1}{\rm Mpc}$ using binwidth of 8~$h^{-1}{\rm Mpc}$. The covariance matrices are determined from the EZ mocks with a correction to equalize small differences in area.}
\label{fig:results-correlation-function}
\end{figure}

\begin{table*}
\caption{Results for the anisotropic full-shape analysis}
\label{tab:results-AP}
\begin{tabular}{|c|l|c|c|c|c|c|c|}
type              & config                        & $b\sigma_{8}$         & $f\sigma_{8} $       & $\alpha_{\rm par}$  & $\alpha_{\rm perp}$& $\sigma_{\rm tot}$    & $\chi^{2}/ dof$ \\
\hline
3-multipole & NGC+SGC & $1.038^{+0.060}_{-0.057}$ & $0.426^{+0.070}_{-0.070}$ & $1.012^{+0.071}_{-0.064}$ & $1.031^{+0.050}_{-0.048}$ & $5.94^{+1.19}_{-1.40}$ & $42.9 / (45-5) $ \\
\hline
3-wedges     & NGC+SGC  & $1.068^{+0.066}_{-0.062}$ & $0.363^{+0.082}_{-0.081}$ & $1.054^{+0.102}_{-0.078}$ & $1.013^{+0.052}_{-0.052}$ & $6.10^{+1.57}_{-1.73}$ & $37.5 / (45 - 5)$ \\
\hline
\end{tabular}
\end{table*}

\begin{figure*}
\includegraphics[width=120mm]{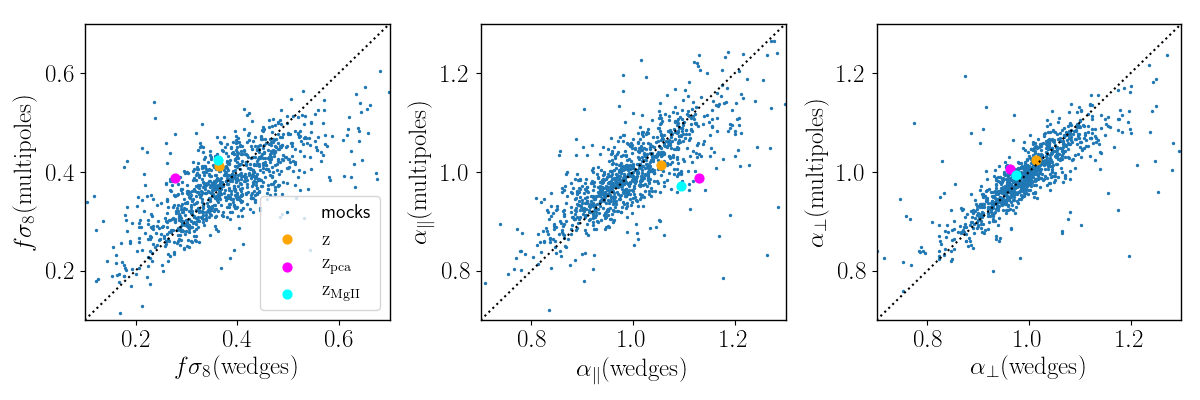}
\includegraphics[width=120mm]{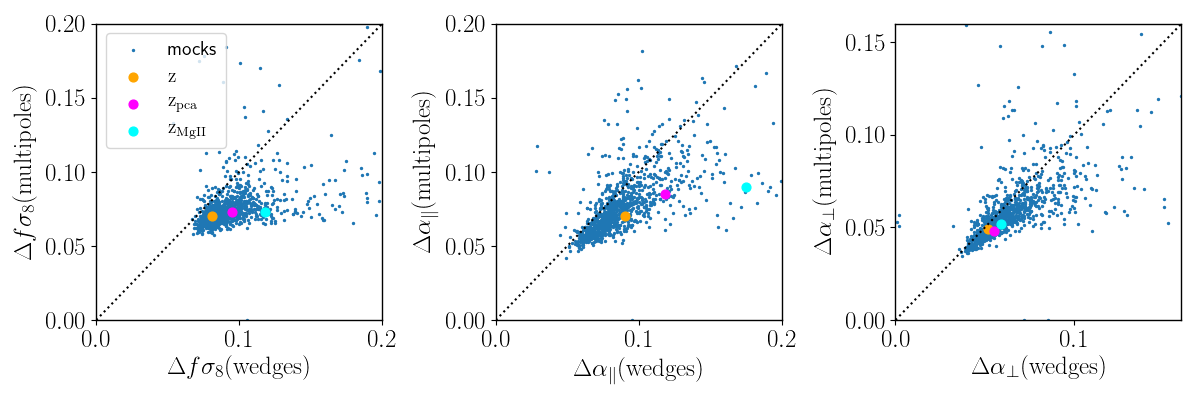}
\vskip -0.25cm
\caption{Upper row : comparison between the 3-multipole and 3-wedge results on the cosmological parameters for the 3 resdshift estimates and for the 1,000 EZ mocks. Bottom row : comparison of the errors obtained for the two methods.
and the results on data for the different redshift estimates for 3-multipole. Bottom panel: Same for 3-wedge.}
\label{fig:scatter-mock}
\end{figure*}

The correlation function multipoles and wedges of the eBOSS DR14 quasar sample with the weighting scheme described in the previous section and the CLPT-GS model with parameters set to the best-fitting values is presented in Figure~\ref{fig:results-correlation-function}. As mentioned previously, the error bars shown in the figure are estimated from the covariance matrix of the EZ mocks. The reference results for the two analyses are displayed in Table~\ref{tab:results-AP}, and in appendix~\ref{appA}, we show the corresponding likelihood contours for a selection of pairs of parameters.
The differences observed between the two methods are within one standard deviation. The performance of the two methods can be compared using the EZ mocks; the results are shown in Figure~\ref{fig:scatter-mock} along with the measurments obtained from the data for the three redshift estimates ('$z$','$z_{\rm PCA}$','$z_{\rm MgII}$'). For the redshift estimate '$z$', the results obtained from the data are similar w.r.t the distribution of the 1000 mocks. For the other redshift estimates, the results from the data are deviate further from the spread of the mocks but one should bear in mind that the statistical errors for these measurement can be much larger in the case of the 3-wedge analysis (see Table~\ref{tab:tests-data-anisotropic}). This behaviour is confirmed using the measurement of the errors for the EZ mocks displayed  in bottom row of Figure~\ref{fig:scatter-mock}. The error on $\alpha_{\rm par}$, when considering the redshift estimate '$z$', is already shifted from the highest density region obtained from the mocks; this shift also explains why there is a large gain in precision on $\alpha_{\rm par}$ for the 3-multipole analysis.\\

Finally, we conclude that the differences observed between the two methods are consistently explained by the expected statistics, and we consider the 3-multipole analysis as the results of this work. In Table~\ref{tab:param3M}, we summarize the results of this work and the correlation between the five parameters obtained from the 3-multipole analysis.

\begin{table}
\caption{Results of the best fit parameters, and the statistical and systematical uncertainties for the 3-multipole analysis. The lower table shows the correlation coefficient between the 5 parameters in the RSD modeling.}
\label{tab:param3M}
\begin{center}
\begin{tabular}{|l|c|c|c|}
parameter & best fit & stat. error  & syst. error \\
\hline
 $b\sigma_{8}$       & $1.038$ & $^{+0.060}_{-0.057}$ & \\
 $f\sigma_{8}$       & $0.426$ & $^{+0.070}_{-0.070}$ & $0.033$\\ 
 $\alpha_{\rm par}$  & $1.012$ & $^{+0.071}_{-0.064}$ & $0.038$\\  
 $\alpha_{\rm perp}$ & $1.031$ & $^{+0.050}_{-0.048}$ & $0.006$\\ 
 $\sigma_{\rm tot}$  & $5.94$  & $^{+1.19}_{-1.40}$   &  \\
\end{tabular}
\vskip 0.5cm
\begin{tabular}{|c|c|c|c|c|c|}
                 & $\alpha_{\parallel}$ & $\alpha_{\perp}$ & $b\sigma_{8}$ & $f\sigma_{8}$ & $\sigma_{\rm tot}$ \\
\hline
$\alpha_{\parallel}$ & 1 & -0.05 & 0.70 & -0.38 & 0.68 \\
$\alpha_{\perp}$     &    & 1        & 0.42 & 0.58  & -0.15 \\ 
$b\sigma_{8}$        &    &           & 1      & -0.33 & 0.18 \\    
$f\sigma_{8}$         &    &           &         & 1        & 0.06 \\
$\sigma_{\rm tot}$  &    &           &         &           & 1 \\
\hline
\end{tabular}
\end{center}
\end{table}

The degeneracy of the parameters are indicated by the likelihood contours presented in appendix~\ref{appA}. As previously mentioned when performing tests on the bias prescription using the OuterRim catalogs and on the data, there is an important correlation between $\alpha_{\parallel}$ and $b\sigma_{8}$. We also confirm a correlation between $\alpha_{\parallel}$ and $\sigma_{\rm tot}$ which is consistent with the fit on the data when fixing the redshift resolution improves the precision on  $\alpha_{\parallel}$. The degeneracy between $f\sigma_{8}$ and the AP parameters demonstrates the importance of fitting them jointly in order to provide a measurement of the growth rate of structure independent of the fiducial cosmology. \\

Our measurement of the isotropic shift of the BAO feature is: $\alpha_{\rm iso}= 1.021^{+0.046}_{-0.044}$. The errors include both the statistical precision and the systematic error related to the RSD modeling. In section~\ref{sec:cosmo}, we compare our measurement with the one obtained using BAO-only analysis described in~\citet{DR14-bao}.

The measured dilation of scales using the eBOSS DR14 quasar sample, $\alpha_\parallel$ and $\alpha_\perp$, can be converted into cosmological parameters according to the equations given at the end of Section~\ref{sec:data}. We measured the expansion rate $H(z)$ and the angular diameter distance $D_A(z)$:
\begin{eqnarray}
H(z_{\rm eff})\cdot r_s(z_d) &=& 23.5^{+1.7}_{-1.9}\,10^3\,{\rm km.s}^{-1} \\ 
D_A(z_{\rm eff})/r_s(z_d)  &=& 12.58^{+0.61}_{-0.78}
\end{eqnarray}
where that $r_s(z_d)$ is the comoving sound horizon at the end of the baryon drag epoch. 
In the case of the isotropic analysis, $\alpha_{\rm iso}$, can be converted into the spherically averaged distance $D_V$ :
\begin{eqnarray}
D_V (z_{\rm eff}=1.52) /r_s (z_d) &=& 26.8 \pm1.1
\end{eqnarray}
where all the quoted uncertainties include systematic and statistical contributions which are added in quadrature.
In the next section, we compare our result on $f\sigma_{8}$, $H(z_{\rm eff})\cdot r_s(z_d)$ and $D_A(z_{\rm eff})/r_s(z_d)$ with four companion papers performing complementary RSD analyses using the same sample.

\subsection{Consensus results}
\label{sec:consensus}

The clustering analysis presented in this paper is based on the eBOSS DR14 quasar sample in the redshift range $0.8\leq z \leq 2.2$, using Legendre multipoles with $\ell=0,2,4$ and three wedges of the correlation function on the $s$-range from 16~$h^{-1}{\rm Mpc}$ to 138~$h^{-1}{\rm Mpc}$. 
We use the Convolution Lagrangian Perturbation Theory (CLPT) with a Gaussian Streaming (GS) model and demonstrate its applicability for dark matter halos of masses of the order of $10^{12.5}{\rm M}_\odot$ hosting eBOSS quasar tracers at mean redshift $z\simeq1.5$ using the OuterRim simulation. We find consistent results between the two methods although in our case the Legendre multipoles basis decomposition provides the cosmological measurements with the best statistical precision. So we use the constraints on the cosmological parameters obtained using the 3-multipole fit as our reference results. 

Four companion papers also present complementary RSD analyses using the same sample and the identical fiducial cosmology. All the companion analyses used the weighting scheme based on ${\cal W}_{\rm focal}$ with a weight according to the inverse of the spectroscopic efficiency. In this work we also discard the paircounts in the region $\mu >(1-1/480)$ to account for the effect of upweighting due to close pairs (${\cal W}_{{\rm focal-}\mu}$). 
We briefly describe the companion papers below and outline the differences:
\begin{itemize}

\item The analysis reported in~\citet{Hector} uses the power spectrum monopole, quadrupole and hexadecapole measurements on the $k$-range, $0.02\leq k\,[h{\rm Mpc}^{-1}]\leq 0.30$, shifting the centres of $k$-bins by fractions of $1/4$ of the bin size and averaging the four derived likelihoods. Applying the TNS model along with the 2-loop resumed perturbation theory, they are able to effectively constrain the cosmological parameters $f\sigma_8(z_{\rm eff})$, $H(z_{\rm eff})r_s(z_d)$ and $D_A(z_{\rm eff})/r_s(z_d)$, along with the remaining `nuisance' parameters, $b_1\sigma_8(z_{\rm eff})$, $b_2\sigma_8(z_{\rm eff})$, $A_{\rm noise}(z_{\rm eff})$, and $\sigma_P(z_{\rm eff})$, in all cases with wide flat priors. 

\item \citet{Hou+18} performs an analysis using Legendre polynomial with order $\ell=0,2,4$ and clustering wedges. They use the "gRPT" to model the non-linear matter clustering and a streaming model extended to one-loop contribution developed by~\citet{Scoccimarro+04} and~\citet{TNS10} along with a nonlinear corrected FoG term. The bias is modelled as described in~\citet{Chan+12}, which includes both local and nonlocal contribution. Additionally, they include the modelling for spectroscopic redshift error. Finally, they provide constraints on $f\sigma_8(z_{\rm eff})$, $D_{\rm V}(z_{\rm eff})/r_{\rm d}$, $F_{\rm AP}(z_{\rm eff})$.

\item \citet{Ruggeri+18} perform a Fourier space RSD analysis using a redshift-dependent weighting scheme that has been developed for RSD analysis~\citep{Ruggeri+17a} to measure cosmological parameters. Such a technique avoids binning in redshift and accounts for the redshift evolution of the geometry and structure growth parameters across the sample. The comparison presented in this section uses the results from the traditional analysis where only FKP weights are taken into account as they correspond to the limit when there is no redshift dependence of the cosmological parameters. Moreover, the results come from the fitting of the first two even multipoles of the power spectrum.

\item \citet{Zhao+18} develop an alternative approach to extract the information in redshift and perform a joint BAO and RSD analysis. It is also based on a power spectrum analysis using the monopole and the quadrupole only (in the $k$-range of $0.02\leq k\,[h{\rm Mpc}^{-1}]\leq 0.30$). They construct an optimally redshift-weighted sample and compare to a power spectrum template based on the regularised perturbation theory up to second order. Using four redshift-weighted power spectra, they constrain $\alpha_{\bot}, \alpha_{\|}$ and $f\sigma_8$ at four effective redshifts (0.98, 1.23, 1.53 and 1.94). The comparison presented in this section uses the traditional weighting scheme,${\cal W}_{\rm focal}$, presented in this work without the additional redshift weight.
\end{itemize}

\begin{figure}
\centering
\includegraphics[width=85mm]{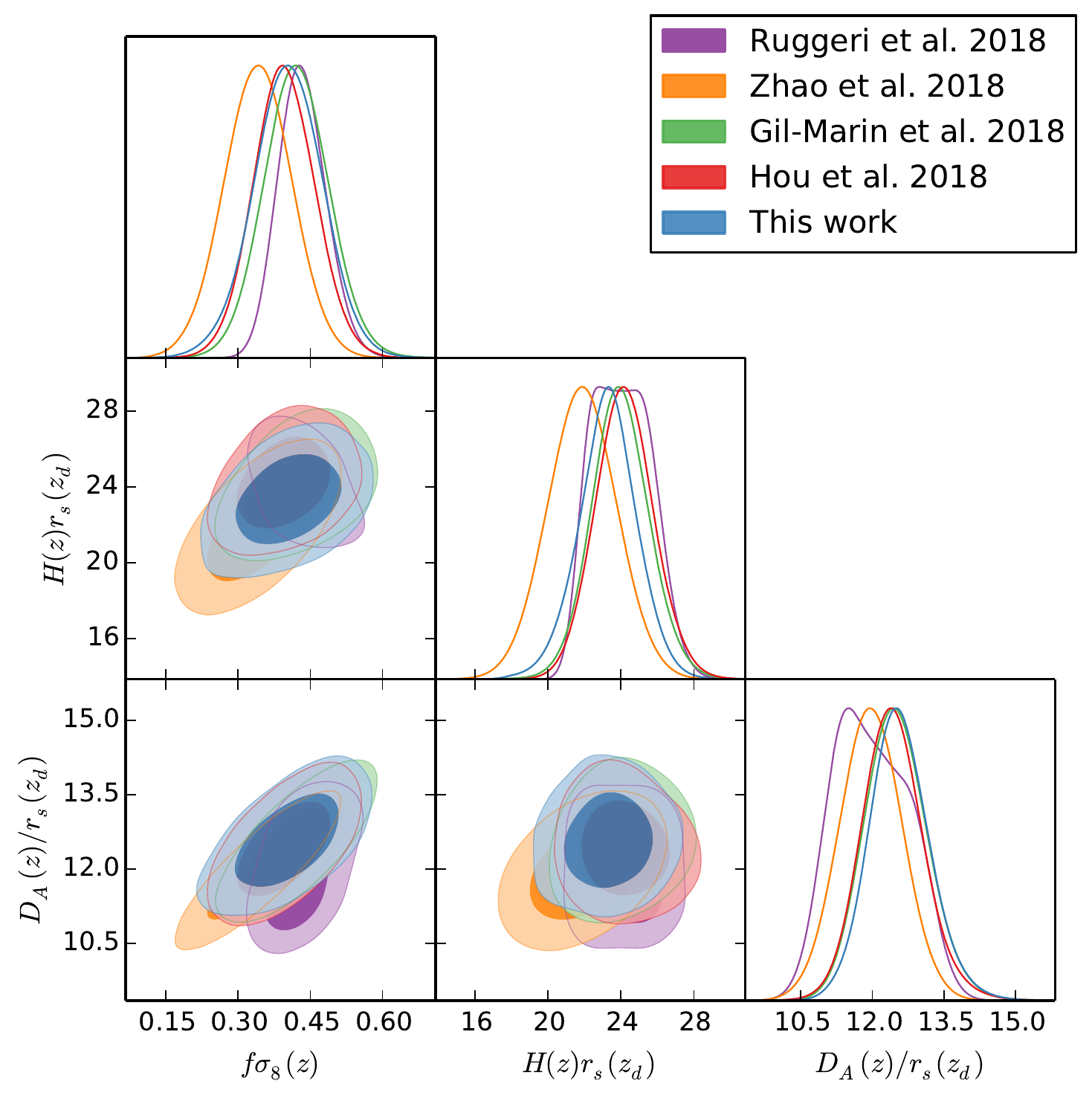}
\caption{Parameter contours for $f\sigma_8$, $D_A$ and $H$ for the predictions by the 5 companion papers using the same DR14Q dataset for traditional RSD analyses. Blue contours show the results presented in this work in configuration space, and red contours show the predictions by \citet{Hou+18} in configuration space too using a second RSD modeling. The Fourier Space based analyses are shown in green contours for the results by \citet{Hector} using a third RSD modeling, in magenta contours for the results by \citet{Ruggeri+18} and in orange contours for \citet{Zhao+18}, both using redshift weighting techniques but with a different model.}
\label{plot:consensus}
\end{figure}

The likelihood contour constraints for the cosmological parameters $f\sigma_8$, $H(z)r_s$, and $D_A(z)/r_s$ at z$_{\rm eff}=1.52$ for the five analyses described above are shown in Figure~\ref{plot:consensus}. Each analysis uses a different model for the 2-point statistics, three are in Fourier space and two in configuration space. Despite those differences, there is good agreement between all analyses. These contours only show the statistical precision which is also similar. The one-dimensional likelihood for each parameter better displays the consistency between the measurements. For the three traditional analyses \citep{Hector,Hou+18 and this work}, the agreement is excellent. The systematic errors, which are not included in these contours, are estimated by the different groups and found to be up to 40\% of the statistical precision.

The likelihood distribution for the two different redshift-weighting techniques~\citep{Ruggeri+18,Zhao+18} when using no redshift-dependent weights are slightly wider but remain consistent with the others. In fact, the results from the analyses using redshift weights are obtained by fitting the monopole and quadrupole only. Adding the hexadecapole provides additional information that increases the sensitivity of the clustering observables to the cosmological parameters. We report no results using the first two even multipoles but we found that adding the hexadecapole could improve the statistical precision by few percents which is consistent to what is reported on table 9 of~\citet{Hector} in Fourier space. We refer the reader to Section 5 of each paper for additional information on the different approaches and on the comparison between the redshift-dependent weights and the traditional analysis at a singe effective redshift on the data.

We do not show any consensus plot on the other parameters such as $b\sigma_{8}$ and $\sigma_{tot}$ as each model uses a different modeling that biases the comparison. Regarding the linear bias, we found a $\sim$1$\sigma$ discrepancy between the Fourier space~\citep{Hector} and the configuration space (this work) that can be explained in our case by different bias model assumptions for the non-linear bias $F''$ as reported in Section~\ref{sec:redshift-model}.

Two additional BAO analyses, presented in~\citep{Wang+18,Zhu+18}, are released along with this paper and complement the measurement of the spherically-averaged distance presented in~\citet{DR14-bao}. These analyses use redshift weights according to the method presented in~\citet{Zhu+15} to compute optimal estimators for $H(z)$ and $D_{A}$ parameters at different redshifts accross the full sample. Consistency between the two methods and their comparison to~\citet{DR14-bao} can be found in each paper.




\section{Cosmological implications}
\label{sec:cosmo}

\subsection{Cosmological distances measurements}
\begin{figure}
\hskip -0.5cm
\includegraphics[width=95mm]{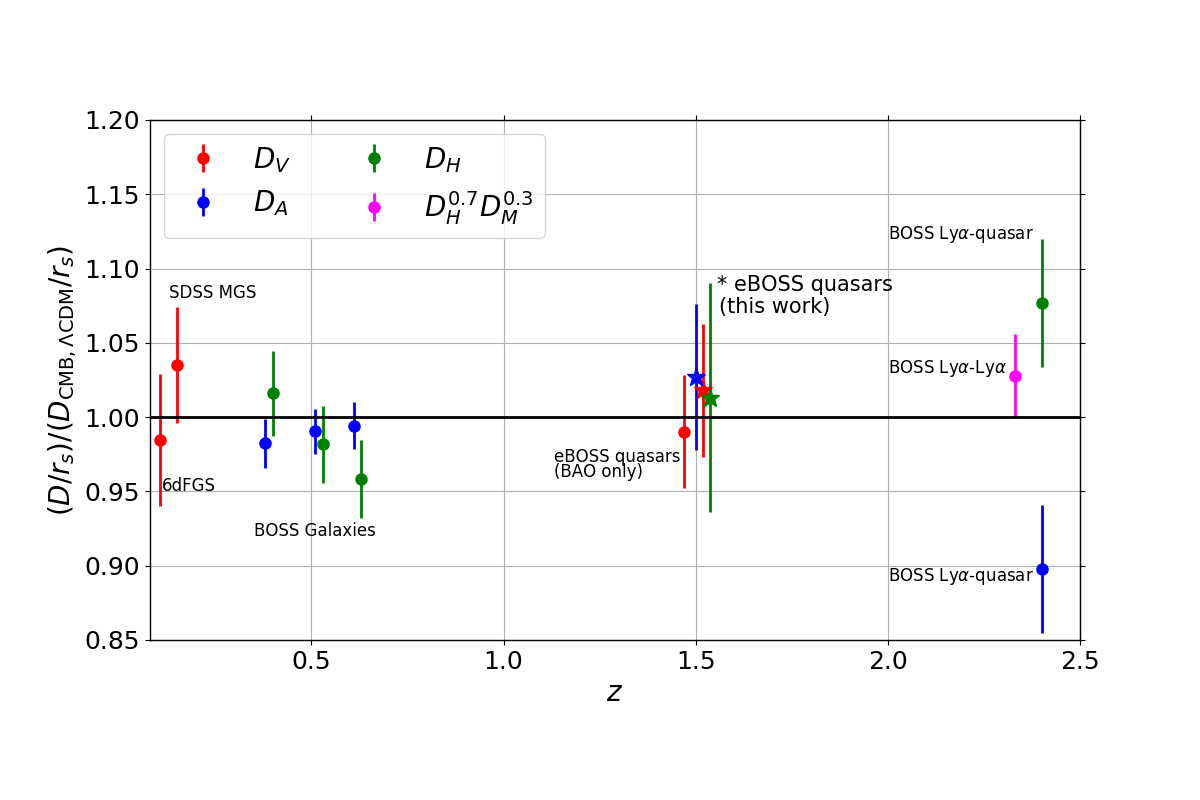}
\vskip -0.25cm
\caption{Evolution of the BAO distances with redshift compared to the prediction from the flat $\Lambda$-CDM model with Planck parameters. The Hubble distance $D_{\rm H}$ is related to the Hubble parameter $H$ by $D_{\rm H}=c/H$ and $D_{\rm M}=(1 + z)D_{\rm A}$ where $D_{\rm M}$ is the comoving angular diameter distance. The BAO results from this work using the eBOSS DR14 quasars are represented by the * marker and are compared to previous analyses using galaxies and Ly-$\alpha$ forests to probe different epochs.}
\label{fig:bao}
\end{figure}

Figure~\ref{fig:bao} presents our measurements of cosmological distances estimates compared with the prediction of $\Lambda$-CDM using Planck results~\citep{planck15}. Also shown are the results of previous measurements: 6dFGS from~\citet{Beutler+11}, SDSS MGS from~\citet{Ross+15}, BOSS DR12 from~\citet{boss-dr12}, WiggleZ from~\citet{Kazin+14}, and BOSS Ly$\alpha$ from the combination of the DR12 Ly$\alpha$ auto-correlation from~\citet{Bautista+17}  and the measurement from~\citet{dMdB+17} using the cross-correlation of the Ly$\alpha$ forest and quasars. Our measurements are consistent with previous analyses and all measurements agree with the expansion history predicted by the $\Lambda$-CDM+GR concordance model using Planck measurements of the cosmological parameters.

We also compare the measurement of the spherically-averaged BAO distance between full-shape analysis (this work) and BAO-only~\citep{DR14-bao}. The two measurements are in agreement and that they provide similar constraints on this parameter (3.8\% precision using BAO-only and 4.1\% using full-shape correlation function).


\begin{figure}
\hskip -0.5cm
\includegraphics[width=45mm]{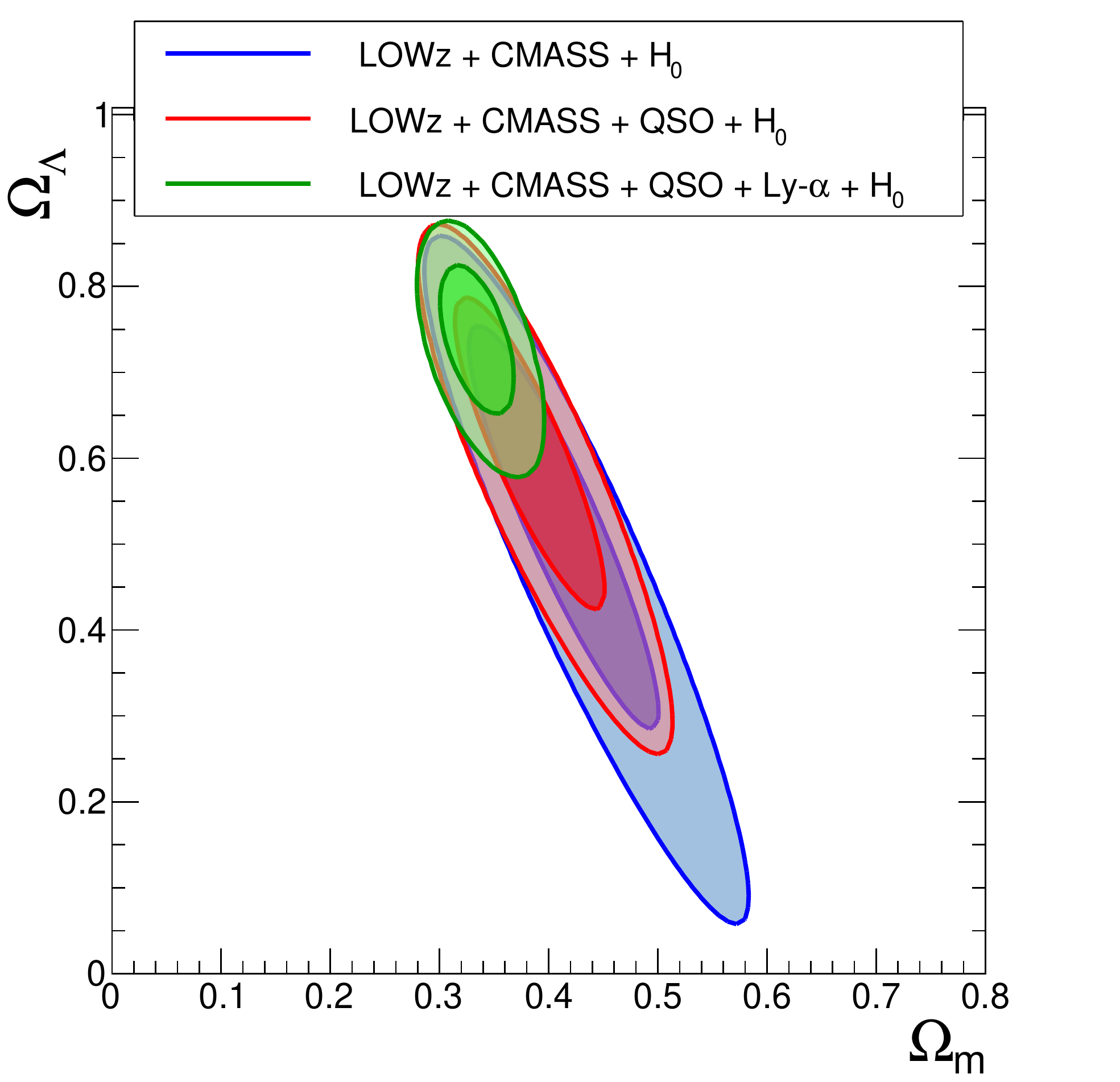}
\includegraphics[width=45mm]{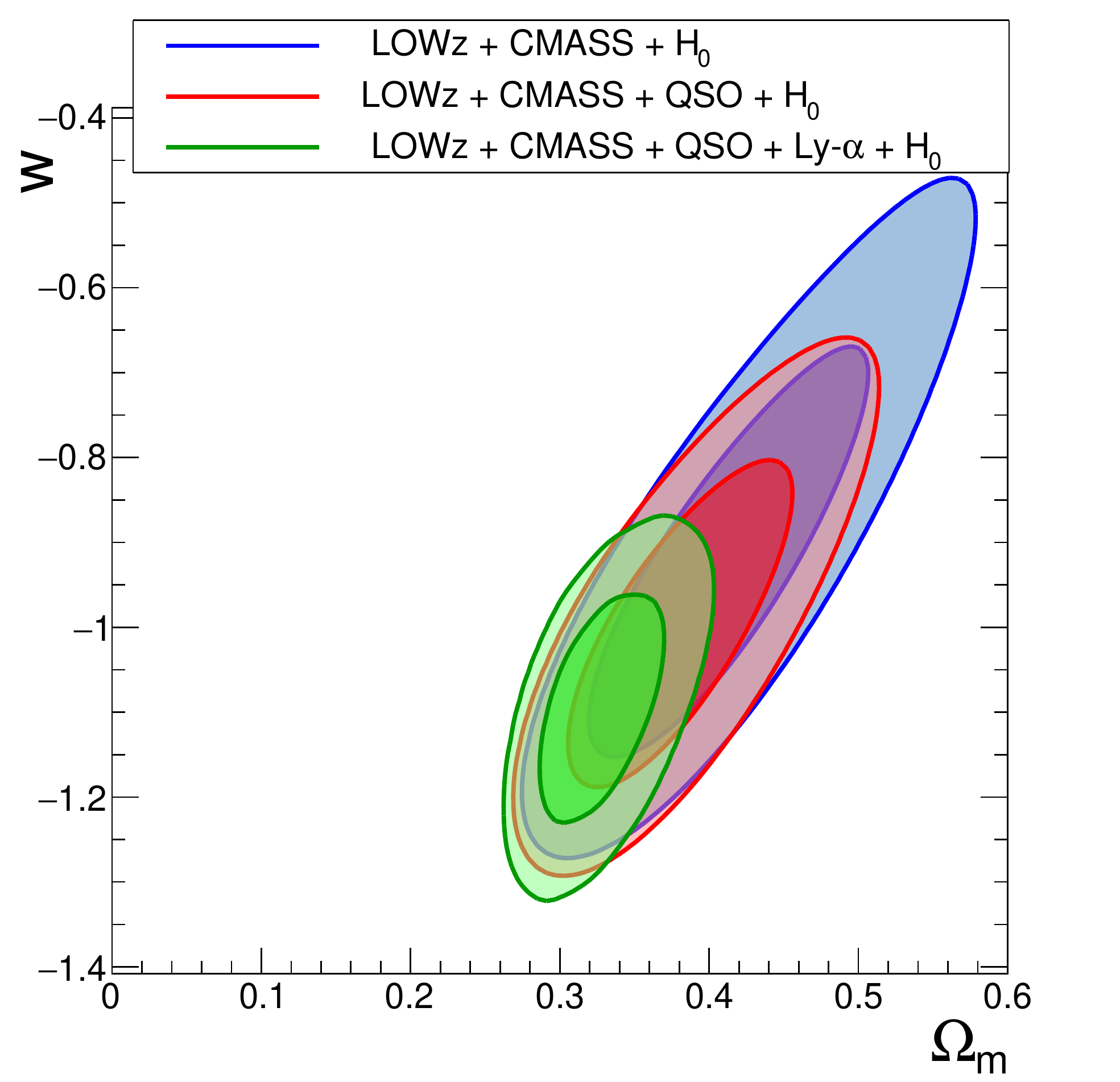}
\vskip -0.25cm
\caption{Left : Cosmological constraints in the $\Omega_\Lambda$ vs $\Omega_m$ plane. Right:  Cosmological constraints in the $w$ vs $\Omega_m$ plane. The inner and outer contours show the 68 and 95\% confidence-level two-dimensional marginalised constraints.
All contours are showed assuming a flat $\Lambda$CDM-model. The blue contour represents the cosmological constraints using BOSS DR12 galaxies, the red contour shows the gain when adding the eBOSS quasar sample and the green contour also includes the results from Ly-$\alpha$ measurements. All results are consistent with a $\Lambda$CDM Universe.}
\label{fig:oCdm}
\end{figure}

Similarly to the study of~\citet{DR14-bao}, we evaluate the impact of our distance measurements on extensions of $\Lambda$CDM. The left panel (resp. right panel) of Figure~\ref{fig:oCdm} shows the contour in the $\Omega_\Lambda$ vs $\Omega_m$ plane (resp.~$w$ vs $\Omega_m$) to test predictions of oCDM (resp.~wCDM). We see that, when using $H_0$ from Planck, adding the current eBOSS quasar RSD measurement (red contour) to the BOSS CMASS sample~\citep[blue contour,][]{boss-dr12} substantially improves the constraints on the extensions of $\Lambda$CDM, and we expect a factor $\sim$2 improvement on the BAO distances for the final eBOSS quasar sample. Finally, the Ly-$\alpha$ BAO measurements (green contour) provide an additional strong constraint in full agreement with a flat universe and pure cosmological constant universe.

\subsection{Growth rate measurements}
\begin{figure}
\hskip -0.5cm
\includegraphics[width=90mm]{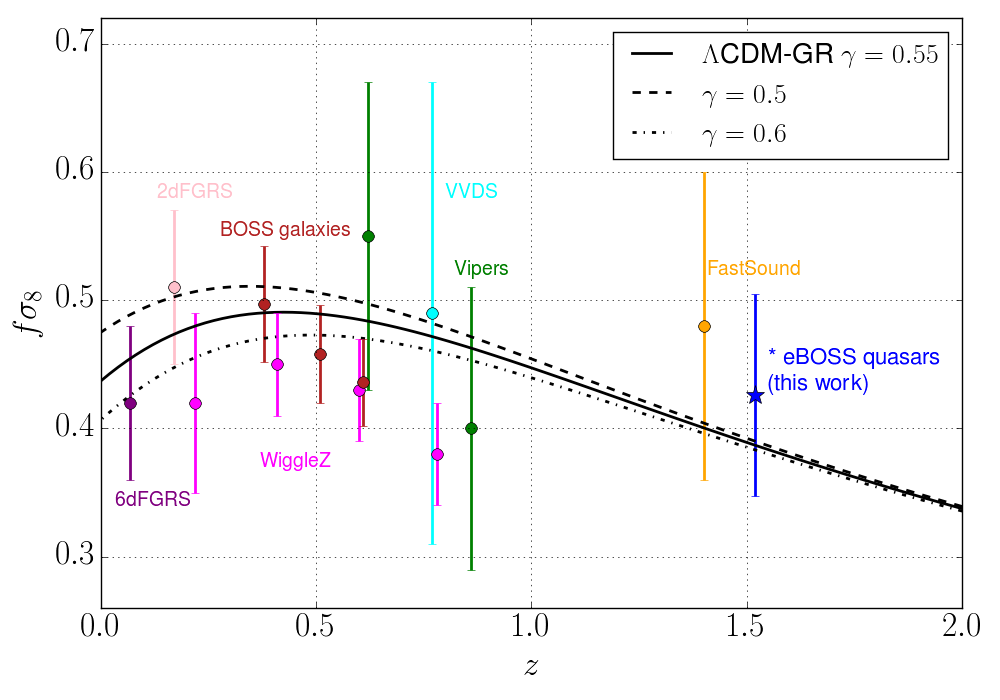}
\vskip -0.25cm
\caption{Measurements of $f\sigma_8(z)$ with redshift compared to the prediction from the flat $\Lambda$-CDM+GR model with Planck parameters. The $f\sigma_8(z)$ result presented in this work for the quasar sample is represented by the * marker and is obtained using 3-multipole fit. The error bar represents the total systematic error that includes the statistical precision and the systematic error related to the RSD modeling used in this analysis.}
\label{fig:fs8}
\end{figure}

The measurement of the anisotropic clustering of the DR14 eBOSS quasar sample produces the constraint on $f\sigma_{8}(z_{eff}=1.52)=0.426\pm0.079$ that is presented in Figure~\ref{fig:fs8}. The result is obtained from a fit of the $l=0,2,4$ Legendre multipoles of the correlation function, and the uncertainty includes systematic errors due to the modelling of the RSD and statistical contributions added in quadrature. The results obtained from the present work are compared with previous measurements from the 2dfGRS~\citep{2df-rsd-04} and 6dFGSN~\citep{6dFGS-rsd-2012}, WiggleZ~\citep{wigglez-rsd-2011}, VVDS~\citep{Guzzo+08}, VIPERS~\citep{vipers+2017} and FastSound~\citep{FastSound+16} surveys, as well as the BOSS DR12 completed sample~\citep{boss-dr12}.




As originally highlighted in~\citet{Guzzo+08}, the measurement of the growth rate of structure can be a direct test of GR, our fundamental theory of gravitation. Our results confirm the validity of GR in the intermediate redshift range ($1 < z < 2$) probed by eBOSS quasars and there is consistency between our result and the measurement done by previous surveys. 
Not all these measurements perform the anisotropic clustering fit using the AP parameters to extract $f\sigma_{8}$, e.g., in~\citet{FastSound+16} they analysed the clustering of Emission Line Galaxies (ELG) sample and provided a single $f\sigma_{8}$ measurement with ~25\% precision without marginalizing over $D_A$ and $H$. 
Since we must assume a fiducial cosmology to infer distances from redshift, an approach of providing a measurement of $f\sigma_{8}$ that is valid in other background cosmologies is to perform a full AP fit. 
We therefore provide a $\sim$18\% measurement of $f\sigma_{8}$ when marginalizing over $D_A$ and $H$, and for comparison, when fixing to their fiducial values, we reach a $\sim$12\% precision.



The GR prediction that $\gamma=0.55$ can not be accurately tested given the statistical precision of the eBOSS quasar sample only. Combining our data to the measurement of $\Omega_m$ from Planck produces $\gamma=-0.2\pm1.2$. The lack of precision arises because in the eBOSS quasar redshift range, $\Omega_m$ is close to 1 and the sensitivity to $\gamma$ is therefore reduced as can be seen from the black curves in Figure~\ref{fig:fs8}, which shows theoretical predictions on $f\sigma_{8}$ for different values of $\gamma$. \\

As for the cosmological distances, the growth rate measurement uncertainty should be reduced by a factor $\sim$2 once the final eBOSS sample will be complete. However, the clustering measurements using the current eBOSS quasar sample represent the most precise $f\sigma_{8}$ measurements to date in the almost unexplored redshift range $1 < z < 2$. 

\section{Conclusion}
\label{sec:conclusion}

We analyse the anisotropic clustering of the eBOSS DR14 quasar sample that includes 148,659 quasars spread over the redshift range $0.8\leq z \leq 2.2$ and spanning 2112.9 square degrees. This sample represents two years of data from eBOSS, and we present the first clustering measurements using the full-shape correlation function that we decompose both into three multipoles and three wedges. We use the Convolution Lagrangian Perturbation Theory (CLPT) with a Gaussian Streaming (GS) model and demonstrate its applicability for dark matter halos of masses of the order of $10^{12.5}{\rm M}_\odot$ hosting eBOSS quasar tracers at mean redshift $z\simeq1.5$. 

We check that the multipoles and wedges approaches yield consistent results. The decomposition into Legendre multipoles provides the cosmological measurements with the best statistical precision. At the effective redshift $z_{\rm eff} = 1.52$, the growth rate of structures $f\sigma_{8}(z_{\rm eff})= 0.426 \pm 0.079$, the expansion rate $H(z_{\rm eff})= 159^{+12}_{-13}(r_{s}^{\rm fid}/r_s){\rm km.s}^{-1}.{\rm Mpc}^{-1}$, and the angular diameter distance $D_{A}(z_{\rm eff})=1850^{+90}_{-115}\,(r_s/r_{s}^{\rm fid}){\rm Mpc}$ where  $r_{s}$ is the sound horizon at the end of the baryon drag epoch and $r_{s}^{\rm fid}$ is its value in the fiducial cosmology. 

The quoted uncertainties include both systematic and statistical contributions. In order to estimate the systematic errors related to the RSD modeling, we use the N-body OuterRim simulation to test the pregdictions of CLPT in real space and then evaluate the performance of the model in redshift space using a hundred mock catalogs created for that purpose. 
We investigate both the effect of the bias model and the spectroscopic resolution in the RSD modeling. Given the statistical precision of the current quasar sample, the reported systematic error is not dominant in our analysis, but further investigations including a full blind mock challenge similar to that undertaken for BOSS is in progress and will be available in time for the analysis of the final eBOSS sample. The eBOSS quasar sample suffers from an important systematic uncertainty related to spectroscopic redshift precision: we study its effect of by modeling a Gaussian redshift resolution and a more physical resolution using the comparison between different redshift estimates ${\rm z}$ and ${\rm z}_{\rm MgII}$.
We demonstrate that accounting for the non-Gaussian tails of the physical distributions has a sizeable impact on the response of the model. In fact, about half of the quoted uncertainty on $\Delta f\sigma_8$ arises from redshift resolution effects. In this analysis, we propose a way of investigating spectroscopic redshift resolution using mock catalogs. In parallel, further improvements in the model to take into account the shape of the redshift error distribution are also considered to reduce the systematic error budget.

We also propose to move beyond the traditional weighting scheme that was used for BOSS galaxies and the BAO measurement with the DR14 sample to account for redshift failures and close-pairs. We validate the procedure on a thousand EZ mock catalogs. This approach allows the observational systematics to be much smaller than the current statistical precision and should be sufficient for the final eBOSS quasar sample. 

The results presented here are compared to the other companion papers using the same data sample but analysed with different techniques; all are found to be in excellent agreement, demonstrating the complementary and the robustness of each method.

The results on the evolution of distances are consistent with the predictions of $\Lambda$CDM with Planck parameters assuming the existence of a cosmological constant to explain the late-time acceleration of the expansion of the Universe. The measurement of $f\sigma_{8}$ is consistent with General relativity (GR) in the almost unexplored redshift range probed by the eBOSS quasar sample. This measurement of the growth rate of structure can also be used to extend the tests of modified gravity models at higher redshift ($z>1$).

This study is a first use of eBOSS quasars for full-shape analysis and will be included for the final eBOSS sample. We expect a reduction on the statistical error of a factor $\sim$2 by the end of the experiment. The improvement of statistics would allow different methods to be combined and thus to provide even tighter constraints on cosmological parameters. Together with BOSS, eBOSS is of particular interest since it paves the way for future programs such as the ground-based Dark Energy Spectroscopic Instrument~\citep[DESI,][]{desi2016a,desi2016b} and the ESA space-based mission, Euclid~\citep{euclid-2013}. Both programs will extensively probe the intermediate redshift range $1 < z < 2$ with millions of spectra, pushing an order of magnitude beyond current measurements. In addition to the quasar sample, eBOSS is also acquiring data for Luminous Red Galaxy and Emission Line Galaxy samples. A companion paper on BAO-only measurement with the LRG sample \citep{Bautista+18} was recently released and the first clustering analyses using the ELG sample are already ongoing.

\section*{acknowledgements}
PZ and EB would like to thank Martin White for helpful comments on the RSD modeling and on requirements for the use of N-body simulations to the analysis of the quasar sample. \\
PZ and EB acknowledge support from the P2IO LabEx (reference ANR-10-LABX-0038). HGM acknowledges support from the Labex ILP (reference ANR-10-LABX-63) part of the Idex SUPER, and received financial state aid managed by the Agence Nationalede la Recherche, as part of the programme Investissements d'avenir under the reference ANR-11-IDEX-0004-02. AJR is grateful for support from the Ohio State University Center for Cosmology and ParticlePhysics. SH's and KH's work was supported under the U.S. Department of Energy contract DE-AC02-06CH11357.GBZ is supported by NSFC Grant No. 11673025, and by a Royal Society Newton Advanced Fellowship. GR acknowledges support from the National Research Foundation of Korea (NRF) through Grant No. 2017077508 funded by the Korean Ministry of Education, Science and Technology (MoEST), and from the faculty research fund of Sejong University in 2018. 
Funding for SDSS-III and SDSS-IV has been provided by
the Alfred P. Sloan Foundation and Participating Institutions.
Additional funding for SDSS-III comes from the
National Science Foundation and the U.S. Department of Energy Office of Science. Further information about
both projects is available at \url{www.sdss.org}.
SDSS is managed by the Astrophysical Research Consortium
for the Participating Institutions in both collaborations.
In SDSS-III these include the University of
Arizona, the Brazilian Participation Group, Brookhaven
National Laboratory, Carnegie Mellon University, University
of Florida, the French Participation Group, the German Participation Group, Harvard University,
the Instituto de Astrofisica de Canarias, the Michigan
State / Notre Dame / JINA Participation Group, Johns
Hopkins University, Lawrence Berkeley National Laboratory,
Max Planck Institute for Astrophysics, Max Planck
Institute for Extraterrestrial Physics, New Mexico State
University, New York University, Ohio State University,
Pennsylvania State University, University of Portsmouth,
Princeton University, the Spanish Participation Group,
University of Tokyo, University of Utah, Vanderbilt University,
University of Virginia, University of Washington,
and Yale University.
The Participating Institutions in SDSS-IV are
Carnegie Mellon University, Colorado University, Boulder,
Harvard-Smithsonian Center for Astrophysics Participation
Group, Johns Hopkins University, Kavli Institute
for the Physics and Mathematics of the Universe
Max-Planck-Institut fuer Astrophysik (MPA Garching),
Max-Planck-Institut fuer Extraterrestrische Physik
(MPE), Max-Planck-Institut fuer Astronomie (MPIA
Heidelberg), National Astronomical Observatories of
China, New Mexico State University, New York University,
The Ohio State University, Penn State University,
Shanghai Astronomical Observatory, United Kingdom
Participation Group, University of Portsmouth, University
of Utah, University of Wisconsin, and Yale University.
This research used resources of the Argonne Leadership Computing Facility, which is a DOE Office of Science User Facility supported under contract DE-AC02-06CH11357.
This work made use of the facilities and staff of the UK Sciama High Performance Computing cluster supported by the ICG, SEPNet and the University of Portsmouth.
This research used resources of the National Energy Research
Scientific Computing Center, a DOE Office of Science User Facility 
supported by the Office of Science of the U.S. Department of Energy 
under Contract No. DE-AC02-05CH11231.

\bibliographystyle{mnras}
\bibliography{biblio}




\appendix

\section{Likelihood contours of the 3-multipole and 3-wedge analyses}
\label{appA}

\begin{figure}
\includegraphics[width=90mm]{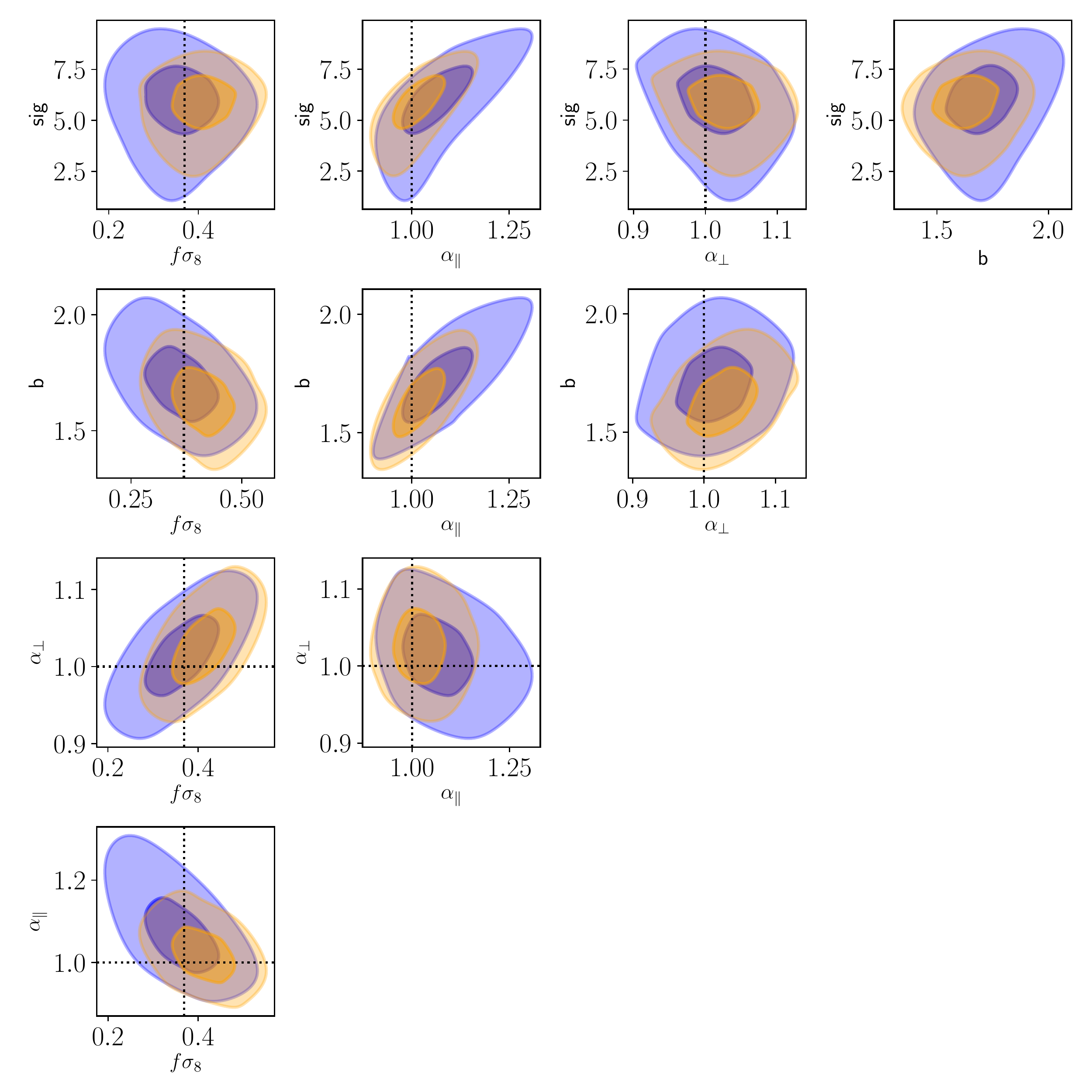}
\vskip -0.25cm
\caption{Likelihood contours, showing the 68\% and 95\% confidence intervals for various combinations of the parameters obtained from the anisotropic fit using 3-multipole (orange) and 3-wedge (blue).}
\label{fig:contour3M-3W}
\end{figure}

Figure~\ref{fig:contour3M-3W} displays the likelihood contours of the reference results for the two analyses using 3-multipole and 3-wedge. The differences observed between the two methods are consistently explained by the expected statistics.
We can see an important correlation between $\alpha_{\parallel}$ and $b\sigma_{8}$ which is consistent with the findings on the OuterRim catalogs and on the data when performing tests on the bias prescription. We also see a significant correlation between $\alpha_{\parallel}$ and $\sigma_{\rm tot}$ which is consistent with the fit on the data when fixing the redshift resolution improves the precision on  $\alpha_{\parallel}$. The degeneracy between $f\sigma_{8}$ and the AP parameters demonstrates the importance of fitting them jointly in order to provide a measurement of the growth rate of structure independent of the fiducial cosmology.

\label{lastpage}

\end{document}